\documentclass[11pt]{article}
\usepackage{epsfig}
\usepackage{amssymb,amsmath}
\usepackage{epigraph}
\usepackage{xcolor}

\usepackage[lined,boxed,commentsnumbered]{algorithm2e}

\usepackage{graphicx}				
\usepackage{amssymb}

\newcommand{\beq}{\begin{equation}}
\newcommand{\eeq}{\end{equation}}
\newcommand{\bea}{\begin{eqnarray}}
\newcommand{\eea}{\end{eqnarray}}

\begin{document}



\begin{center}

{\LARGE
``Quantum Equilibrium-Disequilibrium'': Asset Price Dynamics, 
\vskip0.5cm 
Symmetry Breaking, and Defaults as Dissipative Instantons
}

\vskip1.0cm
{\Large Igor Halperin\footnote{NYU Tandon School of Engineering. E-mail: igor.halperin@nyu.edu}  and Matthew Dixon\footnote{Department of Applied Math, Illinois Institute of Technology. Email: matthew.dixon@iit.edu}} \\
\vskip0.5cm
\today \\

\vskip1.0cm
{\Large Abstract:\\}
\end{center}
\parbox[t]{\textwidth}{
We propose a simple non-equilibrium model of a financial market as an open system with a possible exchange of money with an outside world and market frictions (trade impacts) incorporated into asset price dynamics via a feedback mechanism. Using a linear market impact model, this produces a non-linear 
two-parametric extension of the classical Geometric Brownian Motion (GBM) model, that we call the ``Quantum Equilibrium-Disequilibrium'' (QED) model. 

The QED model incorporates non-linear mean-reverting dynamics, broken scale invariance, and corporate defaults. 
In the simplest one-stock (1D) formulation, our parsimonious model has only one degree of freedom, yet calibrates to both equity returns and credit default swap spreads. Defaults and market crashes are associated with dissipative tunneling 
events, and correspond to instanton (saddle-point) solutions of the model.  
When market frictions and inflows/outflows of money are neglected altogether, ``classical'' GBM scale-invariant dynamics with an exponential 
asset growth and without defaults are formally recovered from the QED dynamics. 
However, we argue that this is only a {\it formal} mathematical limit, and in reality the GBM limit is {\it non-analytic} due to non-linear effects that produce both defaults and divergence of perturbation theory in a small market friction parameter.   

 }
 \newcounter{helpfootnote}
\setcounter{helpfootnote}{\thefootnote} 
\renewcommand{\thefootnote}{\fnsymbol{footnote}}
\setcounter{footnote}{0}
\footnotetext{
We would like to thank Eric Berger, Jean-Philippe Bouchaud, Sergei Esipov, Andrey Itkin, Andrei Lopatin, Sergey Malinin  and Rossen Roussev for useful comments.
}     

 \renewcommand{\thefootnote}{\arabic{footnote}}
\setcounter{footnote}{\thehelpfootnote} 

\newpage
 
\section{Introduction}

Since the groundbreaking work of Samuelson in 1965 \cite{Samuelson}, the log-normal asset return model, also known as the   
Geometric Brownian Motion (GBM) model
 \beq
 \label{log_norm}
 d X_t  =  
 (r_f  +  {\bf w}^T  {\bf z}_t) X_t dt    +  \sigma X_t  d W_t,
\eeq
remains the main work-horse of financial engineering. In Eq.(\ref{log_norm}), $ X_t $ is an asset price at time $ t $, $ r_f $ is a risk-free rate of return, 
$ {\bf z}_t $ are return predictors (``alpha''-signals), $ {\bf w} $ are their weights, and $ W_t $ is a standard Brownian motion. For what follows, we can view 
the GBM model as a model with {\it linear} drift $ f(X_t) =  (r_f  +  {\bf w}^T  {\bf z}_t) X_t  $ and {\it multiplicative} (i.e. proportional to $ X_t $) noise.

The  GBM model (\ref{log_norm}), which was in itself an improvement over an Arithmetic Brownian Motion (ABM) model suggested by Bachelier in 1900 \cite{Bachelier}, ensured non-negativity of model prices. The ABM model can also be viewed as a model with a {\it constant} drift and volatility terms.

Many classical results in finance are based on linear asset return models similar to Eq.(\ref{log_norm}). Most notable examples include the Capital Asset Pricing Model (CAPM) \cite{CAPM} and the Black-Scholes option pricing model \cite{BS}, \cite{Merton}.

On the other hand, these classical models are widely recognized to be in conflict with many empirical facts about financial markets, such as volatility clustering, market frictions, feedback effects, market crashes and ``6-sigma events'', corporate defaults etc. Countless models in the literature tried to improve the log-normal dynamics of 
Eq.(\ref{log_norm}) to make it more consistent with market data, essentially by modifying one or more elements in  Eq.(\ref{log_norm}):
(i) extend a set of predictors $ {\bf z}_t $ \cite{fama2012};
(ii) include non-linear dependencies on predictors $ {\bf z}_t $; and (iii) allow for a different state-dependent (and possibly stochastic) noise amplitude in (\ref{log_norm}) \cite{Heston93aclosed-form}.

A common feature of such extensions is that they all keep a linear (or constant) drift term of the GBM model. Such a linear drift term corresponds to a free diffusion, i.e. describes dynamics without interactions, as will be explained in more detail below.     
Generally, both the presence and importance of non-linear effects (in particular, due to market frictions) for market dynamics is widely recognized in the literature.
Yet at the same time it is also commonly believed that, even though non-linearities exist, the GBM model could still be used as a ``zero-order'' approximation to the 
``true'' behavior of financial markets\footnote{One of the rare exceptions is Dash \cite{Dash} who suggested that ``true'' dynamics cannot be as simple as the GBM model, and proposed Reggeon Field Theory (RFT) as a better candidate model for the stock price dynamics}.
Furthermore, another drawback of the GBM model and its extensions is that they are incapable of being simultaneously calibrated to both equity and credit markets without violation of modeling assumptions. One principal challenge is how to represent corporate bankruptcy and default in the GBM model - the point $X_t=0$ corresponding to such events is not attainable.

The classical Merton corporate default model \cite{Merton_74} belongs to a class of ``structural models'' that
simply bypass this problem by identifying a credit event as a first crossing event for some non-zero boundary 
level $ \bar{X} > 0 $ of an unobservable {\it firm value} process, rather than an observable stock price process\footnote{Structural models are {\it linear} models in the drift and are often formulated in the risk-neutral measure where it becomes redundant.}. 

This is problematic on multiple accounts. Assumptions of a well-defined default boundary at some fixed (and exactly known) value  $ \bar{X} > 0 $ are unrealistic. If such value exists, its 
{\it estimate} would be subject to noise. Uncertainty of the default boundary would lead to uncertainty of the default time itself. This means that if we strictly adhere to the model, at each moment in time we can't confidently determine the occurrence of default, as the exact location of the boundary is unknown.



The goal of this paper is to develop a two-parametric extension of the GBM model that describes markets as {\it interacting} and non-linear dynamical systems, rather than as 
non-interacting stochastic systems.  Our model is derived from the Langevin approach to stochastic dynamics that generalizes free diffusion dynamics to 
a diffusion of interacting particles in an external potential which is generally non-linear \cite{Langevin}. In our approach, such a potential has a specific quartic polynomial form.    
{\it Simultaneously}, non-linearity of the model produces a new mechanism of default, which ensures that stock prices mostly follow diffusion but sometimes may also crash to zero due to a corporate bankruptcy.  As a result, we obtain a model of a defaultable equity, with only one degree-of-freedom given by the equity price itself, that can be calibrated to both equity and credit default swap (CDS) markets. To the best of our knowledge, this is a property that is not shared by any other equity model. 

Such a non-linear extension of the GBM model, with two additional parameters $ \kappa $ and $ g $, controlling the amount of non-linearity, is obtained by modeling two aspects of real market dynamics that are absent in the GBM model and its descendants. We incorporate the fact that markets are open, rather than closed, systems and we include the market feedback effects of trading via market impact mechanisms. 

In this work, we focus on the simplest possible approach where these effects are treated using simple linear or quadratic functions. In particular, we use a simple linear model of market impact.  In choosing such function approximations, we 
rely on physics-inspired analyses of analytic properties and symmetries of resulting price dynamics. Such analyses suggest that other, more complex non-linear extensions of the GBM dynamics should converge to our model in a regime of ``mild'' non-linearities. Our model, that we call the ``Quantum Equilibrium-Disequilibrium'' (QED) model, thus offers a simple non-linear extension of the GBM model that captures the most essential characteristics of real markets such as non-linearities of dynamics due to market frictions, and the presence of corporate defaults. 

Importantly, a simple model {\it formulation} does {\it not} mean simple 
model {\it dynamics}. 
Dynamics of the QED model can be highly complex due to a joint effect of non-linearity and noise. As will be discussed later in the paper, the model captures a combination of non-linearities, noise {\it and} symmetries that play a key role in defining a dynamic behavior of stock prices.

In its reliance on analyticity and analysis of symmetries, our approach bears some similarity to the classical 
phenomenological   
Ginzburg-Landau (GL) approach to equilibrium phase transitions \cite{Landau}. As will be shown below, in the QED model, corporate defaults (or market crashes, in a multi-stock formulation of the model) can be described as phase transitions.

Phase transitions in our model occur as events of non-equilibrium noise-induced barrier crossing phenomena for ``particles'' representing firms, where a barrier is created, for certain values of model parameters, by a combination of a process of capital inflow/outflow in the market and a linear feedback mechanism. Such transitions are well studied in statistical and quantum physics, where 
they are known as {\it instantons}. In our model, corporate defaults and market crash events are described by instantons. 

As their name suggests, instanton transitions occur nearly instantaneously in time, therefore they might be good candidates to represent sudden large price moves such as those occurring during market crashes or corporate defaults. While such effects are often described by additional jump processes or by modifications of a diffusion term in a dynamic equation, in our framework they can be produced \emph{endogenously}, as a result of interactions of market non-linearities with random market noise using only one degree of freedom, and a white noise model.

By focusing on low-order non-linearities of stock price dynamics, our model appeal to physics-rooted ideas of universality of phase transitions, which suggest that qualitative features of phase transitions are determined by symmetries of a system, rather than details of a microscopic Hamiltonian \cite{Landau}.
Similarly to the classical Ginzburg-Landau theory 
\cite{Landau}, our model uses a quartic potential (see below) to capture dynamics of the phase transition. 

Differently from the GL theory, phase transitions in our model are in universality classes of non-equilibrium phase transitions to {\it absorbing states} \cite{Hinrichsen}.  Our model identifies these absorbing states with a zero-price level corresponding to a corporate bankruptcy or default. 

When two additional parameters  $ \kappa $ and $ g $ of the QED model are set to zero, it reduces to the original GBM model. This seems to conform to a  perception commonly held by practitioners and academics alike that while friction effects are important, in most practical cases\footnote{that is those who do not specifically target modeling market impact and other market friction effects, which is unavoidable for some applications, for example for optimal stock execution, or e.g. for portfolio optimization with transaction costs.}, they can be treated as ``first-order'' effects, while the original friction-free GBM model can still be used as a reasonable ``zero-order'' approximation to the dynamics of actual markets. The underlying idea here is that as a linear function can always be viewed as a linear approximation to a non-linear function, the GBM model 
could also be viewed as a result of such a Taylor-truncation of underlying non-linear dynamics. When frictions effects (e.g. due to market impact) become important, they could be computed using, say, perturbative methods.   

However, on closer inspection we observe that the limit  $ \kappa, g \rightarrow 0 $ needed to recover the GBM model may be problematic. Indeed, the model behavior in this limit is {\it qualitatively} different from a behavior obtained at non-zero values of these parameters. As will be shown in detail in Section ~\ref{sect:Model}, for certain values of parameters, the resulting dynamics are {\it metastable}, with a noise-induced barrier transition to a state of zero price associated with a corporate default. On the other hand,
the dynamics in the GBM limit  $ \kappa, g \rightarrow 0 $ are {\it unstable} (see in Sect.~\ref{sect:Model}), and do not allow for corporate defaults. 
This suggests that the GBM limit may be non-smooth (non-analytic).  

In statistical physics, non-analyticity often arises in phase transition phenomena. 
In particular, non-analyticity of a free energy of a statistical system leads to a second order phase transition, see e.g. Landau and Lifshitz \cite{Landau}.
The Lee-Yang theory of first order phase transitions relates them to zeros of a partition function in a complex plane of model parameters (see e.g. \cite{Itzykson_Druffe}, Sect. 3.2). In our model, corporate defaults are associated with noise-induced phase transitions of barrier crossing via an instanton mechanism. This suggests that 
our model may similarly be non-analytic in a complex plane of parameters $ \kappa, g $. Such non-analyticity would produce a non-analytic GBM limit. In other words, the exact GBM limit $ \kappa, g = 0 $ for computable quantities would not correspond to any continuous limit of a smooth function that would be obtained 
with non-zero values of this parameters in a fuller model such as the QED model.  

This might have important implications, as non-analyticity implies that certain measurable quantities, such as prices of derivative instruments, may have corrections that are non-analytic (non-perturbative) in parameter $ \kappa, g $. In particular, instantons in physics are {\it non-perturbative} phenomena: they cannot be seen at any finite order of a perturbative expansion around a non-interacting limit (which would correspond in our case to the limit $ \kappa, g = 0 $). 
In our model, corporate defaults are described by instanton solutions. The presence of instanton-induced 
corrections invalidates the idea of a Taylor truncation of the true dynamics, where the GBM model would become a ``zero-order'' approximation. Instead, this implies that the right ``zero-order'' approximation should already include non-zero friction parameters  $ \kappa, g $ (which can nevertheless be numerically small), and treat them non-perturbatively, rather than as small corrections to an ideal friction-free limit. 
     

The contribution of this paper is two-fold. On the theoretical side, we present the QED model and introduce multiple arguments in favor of non-analyticity of this model (or similar non-linear models) in friction parameters $ \kappa, g $. Sect.~\ref{sect:Verhulst_limit} identifies a qualitatively different behavior for positive and negative values of $ \kappa $, in the spirit of Dyson's argument about divergence of perturbative expansions in quantum electrodynamics \cite{Dyson}. In the same Sect.~\ref{sect:Verhulst_limit} we obtain an explicit solution of the noiseless ``Verhulst limit'' that demonstrates non-analytic behavior. 
Furthermore, we show qualitative differences in a long-term model behavior in the GBM limit and outside of this limit.
When noise is turned on, we show non-analyticity of a normalization coefficient of a stable solution of the Fokker-Planck equation, which is analogous to the Lee-Yang theory of phase transition (see Appendix~\ref{sect:Appendix_C}). 
And the last but not least, in Sect.~\ref{sect:Tunneling} we obtain explicit instanton solutions in our model, showing that they are non-analytic in the friction parameters.    

On the computational side, we develop numerical approaches for calibrating the QED model to univariate market data beyond using the path integral Langevin formulation.  We show how the Fokker Planck Equation (FPE) approach is used in the classical Kramers method to derive our model calibration technique.  We also describe the reduction to the Schrodinger equation as a potential technique for multi-variate calibration in finance and leave it to future work to implement this approach.

The remainder of this paper is organized  as follows. 
In Sect.~\ref{sect:Related_work}, we provide a short overview of related work.
Sect.~\ref{sect:Model} derives our QED model.  Sect.~\ref{sect:Model_solution} discusses methods for solving the model using the Langevin equation and its path integral formulation, in addition to introducing Langevin instantons. Sect.~\ref{sect:Tunneling} describes defaults as tunneling events in the equivalent quantum mechanical formulation of the model. Sect.~\ref{sect:Experiments} describes our 
numerical experiments. Finally, Sect.~\ref{sect:Summary} provides a short summary and directions for future research.

\section{Related work}
\label{sect:Related_work}

The GBM model (\ref{log_norm}) is often used within models based on the competitive market equilibrium paradigm, such as e.g. the CAPM and the Black-Scholes models \cite{Duffie}. 
Within this paradigm, a market 
is considered a closed system fluctuating around a state of perfect thermodynamic equilibrium, without any exchange of money and information with an outside world. 

An alternative to a closed-market concept of competitive market equilibrium paradigm is proposed by 
Amihud {\it et. al} \cite{Amihud_2005}. Instead of competitive market equilibrium, the authors argue that a better paradigm should 
acknowledge the very existence of financial markets. Indeed, markets are facilitated by market makers who process outside information and provide liquidity in the market in an amount that is optimal for {\it them}. This has an impact on market prices, which in turn impacts decisions of investors to inject or withdraw capital in the market. Their trades impact the market via a feedback mechanism.  

Under ``normal'' market conditions, such a scenario implies a dynamically stable state of a market, which 
Amihud {\it et. al} referred to as an ``Equilibrium-Disequilibrium''  \cite{Amihud_2005}. In physics, such states of physical systems are usually referred to as non-equilibrium steady states, see e.g. \cite{Landau}. Our ``Quantum Equilibrium-Disequilibrium'' model implements the market dynamics in accord to the approach of 
Amihud {\it et. al}. 
For a special case when we exclude signals and set $ g = 0 $, our model produces the 
so-called Geometric Mean Reversion (GMR) process which was used in the literature for modeling commodities and real options, see e.g. Dixit and Pindyck \cite{Dixit}. 
Its properties for $ \kappa > 0 $ were further studied by Merton \cite{Merton_growth} and Ewald and Yang \cite{EY}.

Our approach is based on a convex optimization approach of Boyd {\it et. al.} \cite{Boyd_2017}, that we extend here by adding linear market impact, and applying it to a {\it market portfolio}, instead of an individual investor portfolio, following the approach of \cite{IHIF}. Our previous model in \cite{IHIF} was a single-agent model where a bounded-rational agent\footnote{Such an agent embodies a ``collective wisdom'' of the market, that can also be related to various versions of an ``Invisible Hand''-type market mechanisms popular in economics since Adam Smith who coined the term itself.} aggregates {\it all} traders in the market. 

The Reinforcement Learning (RL) - based model of Ref.~\cite{IHIF} provides a mesoscopic description of market dynamics when viewed from the prospective of a market agent.
Here we present an alternative, and perhaps more ``physically'' clear view of the same dynamics, but seen this time from {\it outside} of the market.

The reason for adopting such a view is as follows. In the approach of \cite{IHIF}, agents' actions $u_t$ are adjustments of all positions (by all traders in the market) on a given stock at the beginning of the interval $ [t, t+\Delta t] $. Because now we look at {\it all} traders in the market at once, unlike the individual-investor setting of Boyd {\it et. al.}, it is natural to interpret $ u_t $ as an amount of {\it new capital} injected (or withdrawn, if it is negative) in the market by outside investors at the beginning of the interval  $ [t, t+\Delta t] $. As we will show later, such a ``dual'' view generalizes the simple market dynamics model suggested in 
\cite{IHIF} to describe not only a stable ``growth'' market phase (as implicitly assumed in both \cite{Boyd_2017} and \cite{IHIF}), but also more realistic market 
regimes with corporate defaults and market crashes. 

A model obtained in this way amounts to non-linear Langevin dynamics with multiplicative noise that contains a white noise and colored noise components. The  
colored noise term describes signals used by investors. Such dynamics and related phase transitions are well studied in physics, see e.g. \cite{Schmittmann}, \cite{Van_Broeck_1997}, \cite{Munoz_1998}, \cite{Hinrichsen}.

Approaches to modeling stock markets based on Langevin dynamics were previously considered in the econophysics literature, in particular by Bouchaud and 
Cont \cite{BC}, Bouchaud and Potters \cite{Bouchaud_book}, and Sornette \cite{Sornette, Sornette_book}. 
In particular, Bouchaud and Cont used the Kramers escape rate formula (which also appears in our model, see Sect.~\ref{sect:Kramers_escape} below) to describe market crashes. While we similarly use the non-linear Langevin equation for dynamics of stock prices, here we focus on phenomena that were not the focus in these previous works.

More specifically, we concentrate on (i) comparison of analytical properties of the GBM and QED models; (ii) analysis of their different {\it symmetries}; and (iii) the existence of a possible phase transition that differentiates these
two models. 
Unlike the GBM model, which does not admit defaults (in the sense of transitions into a truly absorbing state), our QED model leads to defaults described as phase transitions into absorbing states at the zero price level 
$ X = 0 $. 
 
 \section{``Quantum Equilibrium-Disequilibrium'' (QED) model}
 \label{sect:Model}
 
 \subsection{Linear impact and non-linear mean reversion}
 
 
 Let $ X_t $ be a total capitalization of a firm at time $ t $, rescaled to a dimensionless
 quantity of the order of one $ X_t \sim 1 $, e.g. by dividing by a mean capitalization over the observation period. 
We consider discrete-time dynamics described, in general form, by the following equations: 
\bea
\label{r_t_one_more}
&& X_{t+ \Delta t} = (1 + r_t \Delta t) (  X_t  - c X_t \Delta t +  u_t  \Delta t )  \nonumber, \\
&& r_t   = r_f + {\bf w}^T {\bf z}_t -  \mu  u_t + \frac{\sigma}{ \sqrt{ \Delta t}} \varepsilon_t, 
\eea
where $ \Delta t $ is a time step, $ r_f $ is a risk-free rate, $ c $ is a dividend rate (assumed constant here),  $ {\bf z}_t $ is a vector of predictors with weights $ {\bf w} $, $ \mu $ is a market impact parameter with a linear impact specification, $u_t\equiv u_t(X_t, {\bf z}_t)$ is a
cash inflow/outflow from outside investors, and $ \varepsilon_t \sim \mathcal{N} (\cdot| 0, 1) $ is white noise.
Here the first equation defines the change of the total market cap\footnote{or, equivalently, the stock price, if the number of outstanding shares is kept constant.} in the time step $ [t, t+\Delta t] $ as a composition of two changes to its time-$t$ value $ X_t $. First, at the beginning of the interval, 
a dividend $ c X_t \Delta t $ is paid to the investors, while  
they also may inject  the amount  $ u_t \Delta t  $ of capital in the stock.
After that, the new capital value $ X_t  - c X_t \Delta t + u_t \Delta t $ grows at rate $ r_t $. The latter is given by the second of Eqs.(\ref{r_t_one_more}), where the term $ \mu u_t $ describes a linear trade impact effect. Note that $ u_t $ can be either zero or non-zero.

The reason that the same quantity $ u_t $ appears in both equations in (\ref{r_t_one_more}) is simple.
In the first equation, $ u_t $ enters as a capital injection $ u_t \Delta t $, while in the second equation it enters via the market impact term $ \mu u_t $ because 
adding capital  $ u_t \Delta t $ means trading a quantity of the stock that is proportional to $ u_t $. Using a linear impact approximation, this produces the impact
term  $ \mu u_t $.
As will be shown below, this term is critical even for very small values of $ \mu $ because the limit $ \mu \rightarrow 0 $ of the resulting model is {\it non-analytic}.

In general, the amount of capital $ u_t \Delta t $ injected by investors in the market at time $ t $ should depend on the current market capitalization 
$ X_t $, plus possibly other factors (e.g. alpha signals). We consider a simplest possible functional form of $u_t$, without signals,
\beq
\label{determ_u}
u_t =  \phi X_t + \lambda X_t^2,
\eeq
with two parameters $ \phi $ and $ \lambda $. Note the absence of a constant term in this expression, which ensures that no investor would invest in a stock with a strictly zero price. Also note that the Eq.(\ref{determ_u})
can always be viewed as a leading-order Taylor expansion of a more general nonlinear
``capital supply'' function $ u( X_t, {\bf z}_t ) $
that can depend on both $ X_t $ and signals $ {\bf z}_t $. Respectively, parameters $  \phi $ and $ \lambda $ could be slowly varying functions of signals 
$ {\bf z}_t $. Here we consider a limiting case when they are treated as fixed parameters, which may be a reasonable assumption when an economic regime does not change too much for an observational period in data. 

In what follows, we assume that the second parameter is positive and very small, i.e. $ 0 < \lambda \ll 1 $.
This implies that the second term in (\ref{determ_u}) provides only a small correction to the first term in a parametrically wide region $ \left| X_t \right| \ll | \phi / \lambda | $. As will be more clear below, the main role of the quadratic term in (\ref{determ_u}) is to regularize large fields  $ \left| X_t \right| \sim | \phi / \lambda | $.

Thus, in the parametric region  $ \left| X_t \right| \ll | \phi / \lambda | $ the capital supply is mostly determined by the value of parameter 
$ \phi $ in Eq.(\ref{determ_u}), that can be either positive or negative. If $ \phi > 0 $, capital is injected in the market (a growth period), while if $ \phi < 0 $, capital is withdrawn (i.e. the market contracts).  

As will be clear below, while 
for positive values of $ \phi $ a contribution the second term $ \sim X^2 $ could be neglected when $ \lambda \rightarrow 0 $, a non-zero value $ \lambda > 0 $ is needed to have well defined dynamics  with {\it negative} values of $ \phi $. This implies that when $ \phi $ is negative (more accurately, when 
$\phi < \lambda/ \mu $), large fluctuations becomes critically important.
 


In a model proposed in \cite{IHIF}, the decision variable $ u_t $ is considered an action of a {\it goal-directed} bounded-rational agent that 
aggregates {\it all} traders in the market. Such agent can be viewed as an embodiment  of an ``Invisible Hand''-type market mechanism.
 Assuming that such market-agent sequentially optimizes risk-adjusted returns of a market portfolio, the model developed in \cite{IHIF} suggests that  an optimal action $ u_t $ is a constrained Gaussian random variable whose mean is linear in the state $ X_t $. 
If we take a deterministic limit in this approach and omit a possible dependence on signals $ {\bf z}_t $, this leads to Eq.(\ref{determ_u}) for a specific choice $ \lambda = 0 $. 

The reason that the optimal control of such market-wide agent implementing an ``Invisible Hand'' market mechanism in \cite{IHIF} is obtained 
as a {\it linear} control of the form 
(\ref{determ_u}) 
is that the model developed in \cite{IHIF}  uses a {\it quadratic} (Markowitz) utility 
function and assumes that the market is in a ``normal'' regime. For non-quadratic utilities, a non-linear optimal control, instead of a linear specification, would be obtained. These adjustments could be 
interpreted as higher-order terms of a Taylor expansion of a non-linear ``optimal capital supply'' function $ u(X_t, {\bf z}_t) $. 

As will be clear below, the strict limit $ \lambda \rightarrow 0 $ is well defined only for a growing market with $ \phi \geq \lambda/ \mu $. In this case, a non-vanishing value of $ \lambda $ would simply produce a small proportional shift in physical quantities computed with the model such as future price distribution or correlation functions. 
In other words, the quadratic term in (\ref{determ_u}) is {\it irrelevant} in the growing phase with $ \phi \geq \lambda/ \mu $. The model of \cite{IHIF} implicitly assumes that the market is in a stable ``growth'' mode, therefore it can be understood as a strict limit $ \lambda \rightarrow 0, \, \phi \geq 0 $ of Eq.(\ref{determ_u}).

On the other hand, as we will see below, a non-vanishing value of $ \lambda > 0 $ is needed to have well-defined dynamics for a regime when $ 
\phi < \lambda/ \mu $ that can describe {\it meta-stable} rather than stable market dynamics. 

Note that all this assumes that market friction $ \mu > 0 $, as it creates a feedback loop of impact of added or subtracted equity on market returns. 
As we will see next,  a non-zero friction $ \mu > 0 $  changes the dynamics from linear 
to {\it non-linear}.

Substituting Eq.(\ref{determ_u}) into Eqs.(\ref{r_t_one_more}), neglecting terms $ O(\Delta t)^2 $ and taking the continuous time limit $ \Delta t \rightarrow dt $
we obtain the ``Quantum Equilibrium-Disequilibrium" (QED) model: 
 \beq
 \label{QED}
 d X_t =\kappa  X_t \left( \frac{\theta}{\kappa}  -  X_t - \frac{g}{\kappa} X_t^2  \right) dt +  \sigma   X_t  \left( d W_t + {\bf w}^T {\bf z}_t \right),
\eeq
where we introduced parameters 
\beq
\label{params}
g  = \mu \lambda , \; \; \kappa   =   \mu  \phi - \lambda  , \; \;  \theta  =   r_f - c + \phi.  
\eeq   

If we keep $ \mu > 0 $ fixed, the mean reversion parameter $ \kappa $ can be of either sign, depending on the values of $ \phi $ and $ \lambda $. 
If $ \phi < \lambda/ \mu $, then $ \kappa < 0 $, otherwise one for $ \phi \geq \lambda/ \mu $ we get $ \kappa \geq 0 $. 

Eq.(\ref{QED}) with $ g = 0 $ is known in physics and biology as the Verhulst population growth model with a multiplicative noise \cite{Horsthemke}, where it is usually written in an equivalent form that can be obtained by a linear rescaling of the dependent variable $ X_t $ that makes the coefficient in front of term $ X_t^2 $ equal one. 

 Note that the higher order terms in the drift in (\ref{QED}) are responsible for a possible saturation of the process. In population dynamics, this corresponds to a population competing for a bounded food resource. In a financial context, this spells a limited total wealth in a market without an injection of capital from the outside world.  

\subsection{Classical potential}

Eq.(\ref{QED}) is a special case of the Langevin equation \cite{Langevin} 
\beq
\label{Langevin_0}
d \, x_t = -  U'(x) dt + \sigma x_t dW_t,
\eeq
which describes an overdamped Brownian particle in a potential $ U(x) $ whose negative gradient gives a drift term in the equation, in the presence of a multiplicative noise. 

We start with an analysis of a deterministic (noiseless) limit of Eq.(\ref{QED}) with  $ \sigma \rightarrow 0 $ and $ {\bf z}_t = 0 $. 
Comparing with Eq.(\ref{Langevin_0}), this is the same as a deterministic limit of the Langevin equation:
\beq
\label{QED_determ}
d X_t = \kappa X_t \left( \frac{\theta}{\kappa} - X_t  - \frac{g}{\kappa} X_t^2  \right) dt \equiv - \frac{\partial U(X_t)}{\partial X_t} dt.
\eeq
where the potential $ U(x) $ is a quartic polynomial:
\beq
\label{pot}
U(x) = - \frac{1}{2} \theta x^2 + \frac{1}{3} \kappa x^3 + \frac{1}{4} g x^4,
\eeq
In what follows, we will refer to 
(\ref{pot}) as the classical potential, to emphasize the fact that it gets ``dressed'' by noise, similarly to how classical potentials in quantum physics become modified by quantum corrections proportional to powers of the Planck constant $ \hbar $. In our case, the noise variance $ \sigma^2 $ will play the role of $ \hbar $ in quantum physics.

The quartic potential (\ref{pot}) can describe a number of different dynamic scenarios depending on the signs and values of parameters involved.
Note a cubic term in this expression, which appears because the dynamics in our model is strongly asymmetric in time (see below).
Such asymmetry is caused by the presence of an absorbing state at $ x = 0 $ in the dynamics. This is quite different from the Ginzburg-Landau theory of 
equilibrium second-order phase transitions where a time reversal symmetry prohibits odd powers in a polynomial expansion of a free energy \cite{Landau}. 


The classic potential (\ref{pot})  can also be parametrized differently in terms of zero level points $ a $ and $ b $:
\beq
\label{pot_ab}
U(x) = - \frac{1}{2} \theta x^2 \left( 1 - \frac{x}{a} \right) \left(1 - \frac{x}{b} \right),
\eeq
where parameters in Eq.(\ref{pot}) and (\ref{pot_ab}) are related as follows:
\beq
\label{relat_param}
\frac{\kappa}{\theta} = \frac{3}{2} \frac{a+b}{ab} , \; \; \; \frac{g}{\theta} = - \frac{2}{ab}.
\eeq
The potential (\ref{pot}) or (\ref{pot_ab})  
has a degenerate extremum $ \bar{x}_0  $ at zero and  two other extrema $  \bar{x}_{1,2}  $ at non-zero values 
\beq
\label{extrema_V}
\bar{x}_0 = 0, \; \; \; \bar{x}_{1,2} = \frac{- \kappa \pm \sqrt{ \kappa^2 + 4 g \theta}}{ 2 g } = \frac{3}{8} \left( a + b \pm \sqrt{ (a+b)^2 - \frac{32}{9} ab} \right),
\eeq
where $ \bar{x}_1 $ and $ \bar{x}_2 $ correspond to the plus and minus signs, respectively. 

\begin{figure}[ht]
\begin{center}
\includegraphics[
width=180mm,
height=55mm]{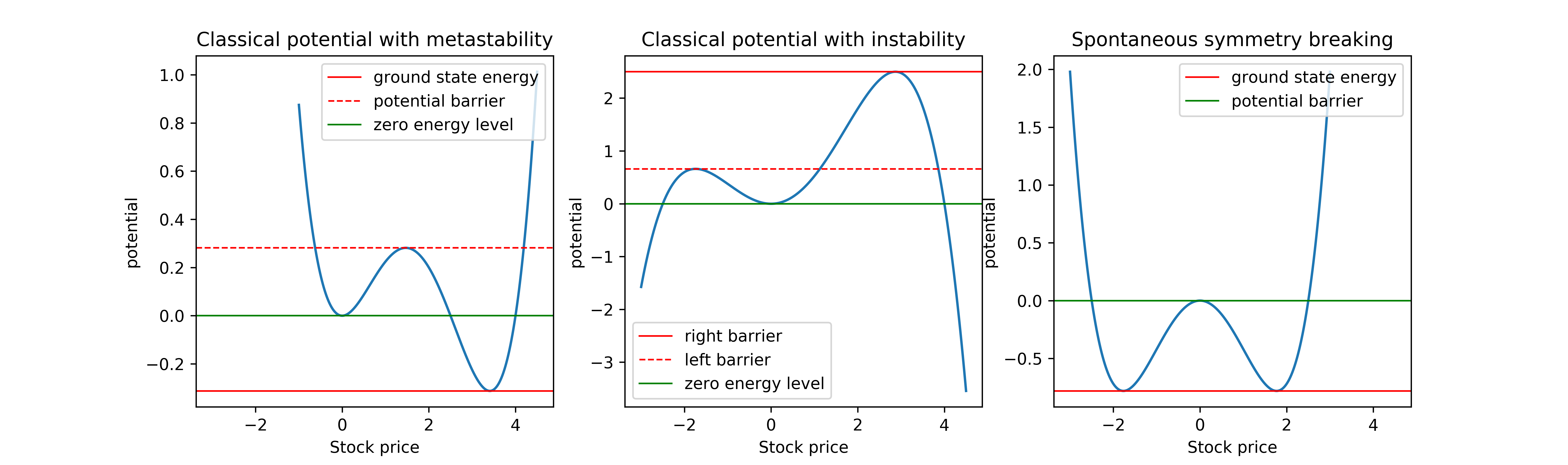}
\caption{Classical potentials for different choices of parameters in (\ref{pot_ab}). The graph on the left is obtained with $ \theta = -1,  \, a = 2.5 $ and 
$ b = 4.0 $. This describes a bimodal system with two metastable states and possible absorption at zero. The graph in the center corresponds to  
$ \theta = -1,  \, a = -2.4 $ and 
$ b = 4.0 $, and describes instability with a metastable state at $ x = 0 $. The graph on the right corresponds $ \theta = 1,  \, a = -2.5 $ and 
$ b = 2.5 $ (i.e. $ \kappa = 0 $), and describes a potential that leads to spontaneous symmetry breaking, where the point $ x = 0 $ becomes a maximum rather than a local minimum. 
} 
\label{fig:Classical_potentials}
\end{center}
\end{figure}

The roots are real-valued if the discriminant is non-negative, which produces a constraint
\beq
\label{positive_discr}
 \theta \geq - \frac{\kappa^2}{4g}.    
\eeq 
Different shapes of the classical potential that can be obtained with different parameters in Eq.(\ref{pot_ab}) are illustrated in Fig.~\ref{fig:Classical_potentials}.

For small values $ g \rightarrow 0 $, the roots (\ref{extrema_V})
can be approximated as follows:
\bea
\label{x_12}
&&  \bar{x}_{1} = \frac{\theta}{\kappa} \left( 1 - \frac{g \theta}{ \kappa^2} \right) + O (g^2) \nonumber, \\
&&  \bar{x}_{2}  = - \frac{\kappa}{g} -  \frac{\theta}{\kappa} \left( 1 - \frac{g \theta}{ \kappa^2} \right) + O (g^2). 
\eea
Note that the first root $ \bar{x}_{1} $ is non-perturbative in $ \kappa $ and perturbative in $ g $, while the second root in non-perturbative in both $
\kappa $ and $ g $. 

When $ \theta > 0 $ and $ g > 0 $ is small, we $  \bar{x}_{1} > 0 $ and $  \bar{x}_{2} < 0 $ for $ \kappa > 0 $. 
As the second root $  \bar{x}_{2} < 0 $, we can safely take the limit $ g \rightarrow 0 $ in this expression. As a result, 
$  \bar{x}_{2} < 0 $ will move to a negative infinity, without substantially affecting the (classical, i.e. without noise) dynamics for positive values $ x \geq 0 $.

On the other hand, when  $ \kappa < 0 $, we have $  \bar{x}_{1} < 0 $ and $  \bar{x}_{2} > 0 $. In this case, taking the limit 
$ g \rightarrow  0 $ means that $  \bar{x}_{2} $ approaches a {\it positive} infinity in the physical region.
This behavior suggests that in this regime, the behavior of the system is sensitive to the specific choice of value of $ g $. 

Depending on the values of parameters, potentials shown in Fig.~\ref{fig:Classical_potentials} may therefore lead to quite different dynamic scenarios. The most insightful scenario is presented on the left graph.
In this case, there exists a global minimum of the potential at a non-zero value of $ X $, while the origin $ X = 0 $ is a local minimum. If a particle hops to the well centered at the origin from the right well, this can be interpreted as a large drop of the stock price. Following this, the stock price can continue to diffuse to the left towards the origin at zero. The even of reaching the zero level can
be identified as a default event. 

Note that in the 
noiseless limit considered in this section, the first 
jump to the left well is not possible (it requires adding noise, see below), but the process of sliding to the minimum at zero {\it while} in the left-most well has a well-defined zero-noise limit given by
Eq.(\ref{QED_determ}). In particular, it shows that the stock price cannot become negative if we start with a positive initial value: as the gradient of the potential is zero at $ x = 0$, the price will stay at this level forever once it is reached. Note for what follows that such irreversibility of defaults persists in the noisy case as well, as long as noise is multiplicative as in
Eq.(\ref{QED}).

 Therefore, a potential such as shown on the left graph in
 Fig.~\ref{fig:Classical_potentials} can simultaneously describe small diffusion in the right well, and large drops and defaults via thermally-induced tunneling. Reintroducing the signals $ {\bf z}_t $ will result in a fluctuating barrier.

On the other hand, two other scenarios shown in the middle and the right figure are of marginal interest as they do not provide insight into defaults. The figure on the right describes a scenario with spontaneous symmetry breaking, where the zero-level point becomes the repelling boundary, as the potential in the vicinity of this point can be approximated by the inverted harmonic oscillator potential:
\beq
\label{inverted_harm_pot}
U(x) = - \frac{1}{2} \theta x^2,  \, \; \; \theta > 0. 
\eeq
Such scenario can only be realized in our setting if $ \kappa = 0 $, which means a market with zero net influx of capital. Using this as an approximation to the potential around zero, we obtain an approximation to Eq.(\ref{QED}) for small values of $ X_t $ (we omit signals $ {\bf z}_t $ 
here):
\beq
\label{QED_GBM_limit}
d X_t =  \theta X_t  dt +  \sigma   X_t  d W_t, 
\eeq
which is the same as the GBM model (\ref{log_norm}).   Therefore, the latter can be formally considered as 
either the formal limit of $ \kappa, \lambda  \rightarrow 0 $ in the potential $ U(x) $, or as an approximation for $ X_t \ll \frac{\theta}{\kappa}, \, \lambda \ll 1  $ of the QED model. 


Therefore, with the spontaneous symmetry breaking potential the GBM model (\ref{log_norm}) can be thought of as a small-field approximation to the QED dynamics of Eq.(\ref{QED})\footnote{A potential with spontaneous symmetry breaking as a model for asset prices was discussed by Sornette \cite{Sornette}.}. However, this scenario loses defaults: the point $ x = 0 $ is now repelling rather than attracting, as indicated by 
Eq.(\ref{inverted_harm_pot}). 

This can be compared with the scenario in the left figure. In this case, the point $ x = 0 $ is a local minimum of the potential, therefore it becomes an attractive point for a particle, once it hops to the left well from the right well. 

On the other hand, the same GBM approximation (\ref{QED_GBM_limit}) can also be obtained in this scenario as well. From the perspective of local dynamics for large prices, the only difference from the figure on the right is that now the potential is locally quadratic around a non-zero value. But a big difference from the spontaneous symmetry breaking scenario is that defaults are still both possible, and also have a chance to be modeled. The latter is because in the figure on the left, they are associated with exponentially small tunneling probabilities, also known as escape probabilities. 

Integrating Eq.(\ref{QED_GBM_limit}) in the limit $ \sigma \rightarrow 0 $, we obtain the conventional exponential asset growth, but this time only for sufficiently small deviations from a minimum of the potential $ U(X) $:
\beq
\label{exp_growth_appr}
X_t = X_0 e^{\theta t} , \; \; \; X_t \ll 1.  
\eeq
Therefore, the exponential growth (\ref{exp_growth_appr}) that is often used in classical ``financial equilibrium'' models, i.e. equilibrium in a financial rather than a physical context, is associated within the QED model (\ref{QED}) with an {\it initial exponential growth} during the process of relaxation to a true vacuum\footnote{In cosmological models used in physics, such behavior is often referred to as an initial inflation period.}.
In the next section we will return to this scenario, after we analyze a  
time-dependent solution of the model in the deterministic limit.

The same exponential growth (\ref{exp_growth_appr}) is also obtained with a potential with spontaneous symmetry breaking, though in this case it corresponds to an expansion around a maximum rather than a minimum at $ X = 0 $, and defaults would not be allowed classically,

To summarize so far, out of many possible scenarios for dynamics that can be obtained with the quartic potential (\ref{pot}) or (\ref{pot_ab}), the graph on the left in Fig.~\ref{fig:Classical_potentials} is the most interesting scenario. In what follows, we will be mostly concerned with the type of dynamics corresponding to such a scenario. 


\subsection{The $ \mathcal{CPT} $-symmetry}

Consider the classical Newtonian's second law of dynamics for a particle in the potential (\ref{pot}):
\beq
\label{Newton}
\frac{d^2 x}{dt^2}  = - \frac{\partial U}{\partial x} = \theta x - \kappa x^2 - g x^3.
\eeq
It is instructive to consider symmetries of Eq.(\ref{Newton}) under the following set of transformations that 
we will refer to as the $\mathcal{CPT} $ transformations, using an analogy with physics:
\bea
\label{CPT}
& & \mbox{ $\mathcal{C}$-parity:}   \hskip0.3cm \kappa \rightarrow - \kappa \nonumber \\
& & \mbox{$ \mathcal{P}$-parity:}  \hskip0.3cm x \rightarrow - x  \\
& & \mbox{$\mathcal{T} $-parity (Time reversal):} \hskip0.3cm    t \rightarrow - t.   \nonumber \nonumber
\eea
It is easy to see that the Newtonian second law equation (\ref{Newton}) is separately symmetric with respect to the time reversal $\mathcal{T} $
and the joint $\mathcal{CP} $-inversion (while the latter is also equivalent to flipping the sign of the potential). As a consequence, it is also invariant with respect to a simultaneous $\mathcal{CPT} $ transformation.

However, unlike the Langevin equation (\ref{Langevin_0}), Newton's second law (\ref{Newton}) contains the second derivative in time in the left-hand side. This difference is important because it implies that for classical mechanics we need both an initial and terminal conditions to define 
two arbitrary constants of integration, while for the Langevin dynamics we have only one. This results in the Langevin dynamics having different symmetries from the symmetries of classical mechanics.

It is therefore interesting to see how the Langevin dynamics can arise from the classical dynamics. To this end, we have to add in Eq.(\ref{Newton}) a 
dissipative force term that is proportional to particle velocity $ \dot{x} $, with a friction coefficient $ \gamma $, and in addition, make parameter $ \theta $ fluctuating:
\beq
\label{Newton_fric}
\frac{d^2 x}{dt^2}  = \left( \theta + \varepsilon_t \right) x - \gamma \frac{dx}{dt}  - \kappa x^2 - g x^3.
\eeq 
To obtain a limiting behavior of this system in the limit of large friction $ \gamma \rightarrow \infty $, note that we can re-write Eq.(\ref{Newton_fric})
by introducing dimensionless time $ \tau = t/ \gamma $:
\beq
\label{Newton_fric_2}
\frac{1}{\gamma^2} \frac{d^2 x}{d \tau^2}  = \left( \theta + \varepsilon_t \right) x - \frac{dx}{d \tau}  - \kappa x^2 - g x^3.
\eeq 
Taking here the formal limit $ \gamma \rightarrow \infty $ (called the overdamped limit), the term in the left-hand side vanishes, and the remaining terms exactly reproduce the  
Langevin equation (\ref{Langevin_eq}) in time $ \tau $:
\beq
\label{Langevin_tau}
\frac{dx}{d \tau}  =  - \frac{\partial U}{\partial x} + \sigma x d W_t, \; \; U(x) \equiv
- \frac{1}{2}\theta x^2 + \frac{1}{3} x^3 + \frac{1}{4} x^4. 
\eeq
The acceleration term in the left-hand side of Eq.(\ref{Newton_fric_2}) is dropped in the overdamped limit, which means that in the overdamped limit, we assume 
that the velocity is nearly constant, hence its fluctuations are neglected, resulting in zero acecleration.

The Langevin equation is no longer invariant under time reversal $\mathcal{T} $, though it maintains invariance with respect to the joint $\mathcal{CP} $-inversion.
As the time reversal symmetry is broken, the  simultaneous $\mathcal{CPT} $ invariance does {\it not} hold. An interesting property of Eq.(\ref{Langevin_tau}) is that 
it implies the a  {\it backward} motion with $ \tau \rightarrow - \tau $ in a potential $ U $ is mathematically equivalent to a {\it forward} motion in an inverted potential $ - U $. We will use this observation below. 

%
%
%
%
%

\subsection{Second order phase transition in the deterministic Verhulst limit}
\label{sect:Verhulst_limit}

In this section, we temporarily set $ \lambda = 0 $, and hence $ g = 0 $, to simplify the analysis of the deterministic limit.
The solution of Eq.(\ref{QED_determ}) in this limit is
\beq
\label{Verhulst_determ}
X_t = X_0 \frac{e^{\theta t}}{ 1 + \frac{\kappa}{\theta} X_0 \left( e^{\theta t} - 1 \right)}  
= \frac{\theta}{\kappa} \frac{1}{1 - \left( 1 - \frac{\theta}{\kappa X_0} \right) e^{- \theta t} }.
\eeq 
This solution is strongly asymmetric in time. Indeed, for an arbitrarily small starting point $ X_0 > 0 $, the process (\ref{QED_determ}) becomes singular 
in a finite time $ t_{\infty} $ given by the following expression:
\beq
\label{t_infty}
t_{\infty} = \frac{1}{\theta} \log \left( 1 - \frac{\theta}{\kappa X_0 } \right).
\eeq 
Note, however, that in an ‘`expanding'' regime with $ \kappa > 0 $, $ t_{\infty} $ is actually {\bf negative}. It can be interpreted as an emergence from 
a negative singularity at  time $ t_{\infty} < 0 $ in the past. (and being at a {\it positive} singularity immediately prior to that). For any given initial condition, $ X_0 $, Eq.(\ref{t_infty}) specifies a time in the past when the current market emerged from a negative singularity. For positive times, the solution remains positive and bounded for $  \kappa > 0 $ and $ X_0 > 0 $.

\begin{figure}[ht]
\begin{center}
\includegraphics[
width=160mm,
height=60mm]{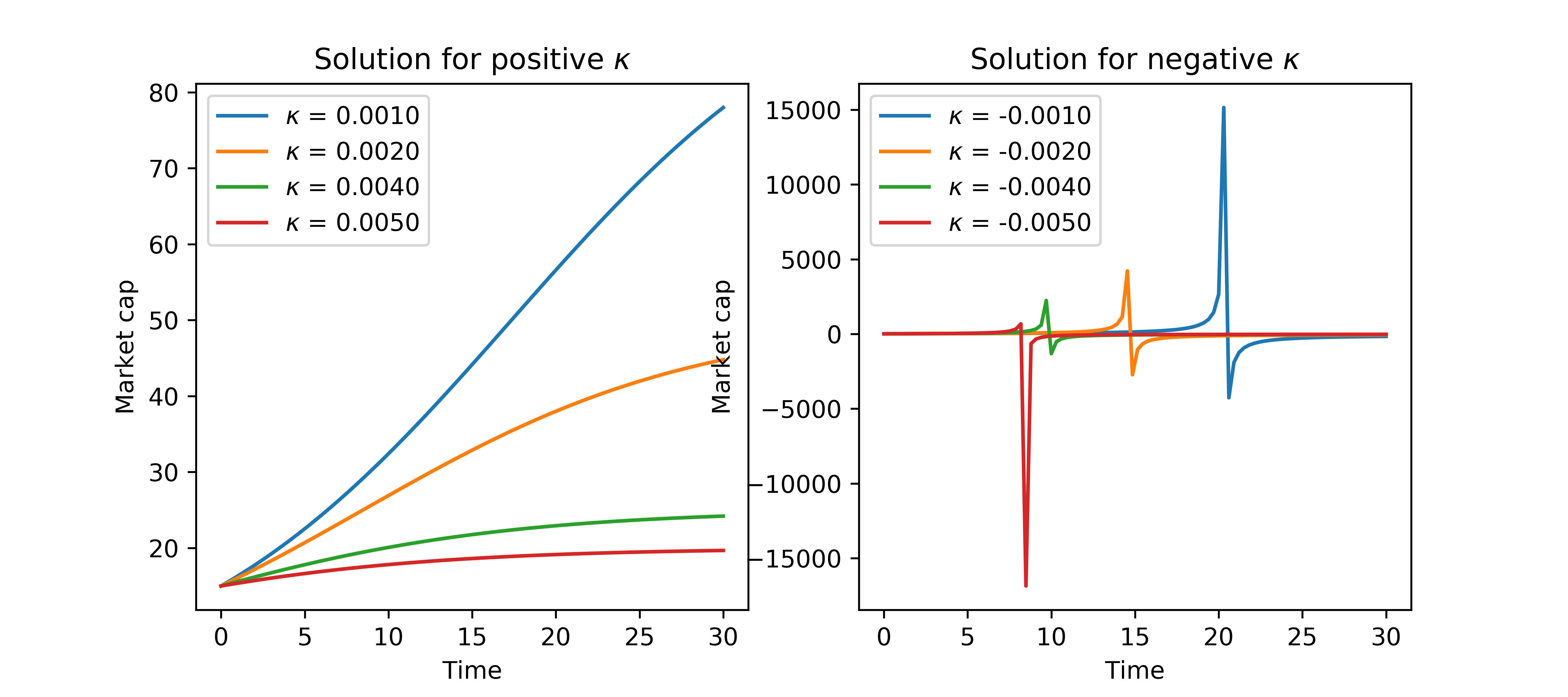}
\caption{Solutions of the ``Verhulst limit'' $ g = 0 $ of the QED model for positive and negative values of $ \kappa $. The following parameters are used:
$ r_f = 0.015, c = 0.01,  \mu = 0.001 $.} 
\label{fig:Classical_Verhulst_solution}
\end{center}
\end{figure}

On the other hand, consider the case $ \kappa < 0 $, i.e. capital is being withdrawn from the market, so that $ \phi < 0 $. In this case, $ t_{\infty} $ is {\bf positive}, i.e. the 
Verhulst dynamics (\ref{Verhulst_determ}) exhibits a {\it positive} singularity in a {\it finite} time  $ 0 \leq  t_{\infty} < \infty $ starting from an arbitrarily small value $ X_0 > 0 $. After that, it re-emerges from a {\it negative} singularity and remains on a ``non-physical'' solution branch with $ X_t < 0 $ by asymptotically decaying to $ 0 $.
The presence of two singular solutions, with singularities in the past or the future for $ \kappa > 0 $ and $ \kappa < 0 $, respectively, is a consequence of the 
$ \mathcal{CPT} $-symmetry (\ref{CPT}) of the dynamics.  Examples of solution of the ``Verhulst limit'' $ g \rightarrow 0 $ of the classical dynamics with $ \sigma \rightarrow 0 $ are shown in Fig.~\ref{fig:Classical_Verhulst_solution}. 

To preclude such catastrophic scenarios from happening is precisely the reason we keep an additional {\it cubic} term in Eq.(\ref{QED}) with a {\it positive} coefficient 
$ g \ll 1 $. However small the value of $ g $ is, it regulates the model for very large fields $ | X | \sim 1/ g $, while producing only very small corrections to the behavior for small fields $ | X | \ll 1/ g $. A non-zero value of $ g $ smoothes out the singularity as it becomes the second root $ \bar{x}_2 $ in 
Eqs.(\ref{x_12}) when $ g \neq 0 $.

Keeping $ g > 0 $ (or equivalently $ \lambda > 0 $) arbitrarily small but finite, ensures that the limit $ \kappa \rightarrow 0 $ (when $ \kappa $ itself is adjusted by 
$ \lambda $ as $ k = \mu \phi - \lambda $) is {\it non-analytic} in the product $ \mu \phi $.  if we keep $ \mu > 0 $ fixed, the analytic structure of the model can equivalently be formulated in 
terms of either $ \kappa $  or $ \phi $. We prefer to use the parameter $ \kappa $ to describe analyticity as it directly enters the QED diffusion (\ref{QED}).  

Physical quantities can only be defined in a complex $ \kappa $ plane with a branch cut singularity along the negative $ \kappa $ axis. Therefore, any perturbative expansion
developed for $ \kappa > 0 $ around the singularity $ \kappa = 0 $ would be a divergent expansions\footnote{This is similar to the famous argument by Dyson about divergence of perturbation theory in Quantum Electro-Dynamics (QED) \cite{Dyson}. In our case, negative values of $ \kappa $ would produce a {\it positive} feedback loop from investing in the market as long as $ u_t > 0 $, leading directly to an explosion at a finite time for a negative $ \kappa $.}.
   
Therefore, when $ \kappa < 0 $, the only stable steady solution of Eq.(\ref{QED_determ}) is $ \bar{X} = 0 $.
At $ \kappa = 0 $, this solution becomes unstable, and a new stable steady state solution 
\beq
\label{np_solution}
 \bar{X} = \frac{\theta}{ \kappa} , \; \; \; \kappa > 0,
\eeq 
bifurcates at $ \kappa \geq 0 $. Recall that in this section we temporarily set $ g = 0 $. When $ g > 0 $, the steady state solution (\ref{np_solution}) is
corrected by terms of $ \sim O(g) $, see Eqs.(\ref{x_12}). 

This solution (or (\ref{x_12})) is {\it non-perturbative} in $ \kappa $, i.e. this branch emerges in a continuous but non-differentiable way. This means that in the deterministic limit $ \sigma \rightarrow 0 $ without signals $ {\bf z}_t = 0 $, the system undergoes a {\it second-order phase transition} at $ \kappa = 0 $, with a non-vanishing order parameter (\ref{np_solution}) that emerges for $ \kappa > 0 $  \cite{Horsthemke}.

Of course, the discussion of an actual second-order phase transition in the classical system with the potential (\ref{pot}) is somewhat contrived, as the dynamics of such a classical system are smooth. The notion of a phase transition is applied here to the phase diagram of steady states, that does exhibit a 
discontinuous behavior at $ \kappa = 0 $. As we will see below, the noisy problem is mathematically equivalent to a problem of quantum mechanical tunneling, where a concept of a dynamic absorbing phase transition, as a special kind of a second order phase transition, becomes less formal and more interesting. 


For small values of $ \kappa $ and short times such that $ \theta t \ll 1 $, we can approximate the solution (\ref{Verhulst_determ}) as follows:
\beq
\label{QED_det_appr}
X_t \simeq X_0  e^{\theta t} \left( 1 - \frac{k}{\theta} X_0 \left( e^{\theta t} - 1 \right) \right) \simeq X_0  e^{\theta t} \left( 1 - \kappa X_0 t \right) 
\simeq X_0 e^{ (\theta - \kappa X_0 ) t } = X_0 e^{ (r_f - c + \phi - \kappa X_0 ) t }. 
\eeq  
Therefore, the return on equity $ R_e $ is given by the following expression:
\beq
\label{cost_of_capital}
R_e = r_f -c + \phi - \kappa X_0 \simeq r_f - c + \phi.
\eeq  
(the last relation holds as long as $ \kappa X_0 \ll 1 $).
But importantly, the exponential growth with such rate $ R_e $ is correct only for sufficiently short times, where approximations leading 
to Eq.(\ref{QED_det_appr}) are justified. 
In a {\it long-term} limit, the asset price converges to the stable solution (\ref{np_solution}):
\beq
\label{long_term}
\lim_{t \rightarrow \infty} X_t = \frac{\theta}{\kappa} \left[ 1 - \left( \frac{\theta}{ \kappa X_0} - 1 \right) e^{-\theta t} \right].  
\eeq
This is one of the key differences between the QED model and the GBM model, where an exponential asset growth continues indefinitely. 
As the steady state solution  (\ref{np_solution}) is {\it non-perturbative} (non-analytic) in $ \kappa $, it could {\it not} be recovered with either the GBM model or any perturbative scheme of handling frictions that would be still based on the GBM model (\ref{log_norm}).

Assume for the moment that the QED model (\ref{QED}) is the ``true'' model of the market, but an observer is unaware of this, and uses instead the GBM model 
(\ref{log_norm}). These dynamics can be viewed as fluctuations around the {\it unstable} steady state solution $ \langle X \rangle = 0 $. Such fluctuations can only 
be about a local equilibrium at best, simply because their ``ground state'',  $ \langle X \rangle = 0 $, itself becomes unstable for $ \kappa > 0 $. If one starts with such an unstable state, it will relax into the stable state (\ref{np_solution}). 

This suggests that the meaning of ``equilibrium vs non-equilibrium'' in our model is exactly {\it opposite} to its meaning in classical 
``competitive market equilibrium'' models \cite{Duffie}. The exponential asset growth, assumed to be an attribute of equilibrium dynamics in these models, 
 turns out to be just  
an initial (non-equilibrium) exponential growth stage of a relaxation into a stable state in our model.
 
Thus, both the GBM and QED models produce an exponential asset growth in a short term, but strongly differ in their long-term predictions. 
Yet, depending on the values of model parameters, and given their possible non-stationarity, testing for boundedness  vs unboundedness of an 
exponential asset growth may not be an easy task.  Fortunately, the two models (\ref{log_norm}) and (\ref{QED}) can be discerned by their behavior. More specifically, while the GBM model does not admit defaults as a process of ``falling to the origin'', $ X = 0 $, in the QED model such events become possible as noise-activated barrier penetration transitions. 
The following sections consider these phenomena in more details.


\section{Noise-induced metastability:  Langevin instantons}
\label{sect:Model_solution}

\subsection{Langevin dynamics in the log-space}

When we keep the white noise (the Brownian motion $ W_t $) in Eq.(\ref{QED}), but omit (or take averages of) signals $ 
{\bf z}_t $, it can be viewed as a Langevin equation with multiplicative noise:
\beq
\label{Langevin_eq}
d \, x_t = -  U'(x) dt + \sigma x_t dW_t,
\eeq
where the classical quartic potential $ U(x) $ is defined in Eq.(\ref{pot}), and $  U'(x) $ stands for the derivative of the potential with respect to $ x $.  

Instead of working with multiplicative noise, Eq.(\ref{Langevin_eq}) can be transformed into a Langevin equation with additive noise by defining a log-price variable $ y_t  =  \log x_t $. Using It\^{o}'s lemma, we obtain 
\beq
\label{Ito}
dy_t =  \frac{\partial y_t}{\partial x_t} d x_t + \frac{1}{2}   \frac{\partial^2 y_t}{\partial x_t^2} \left(d x_t \right)^2 = \left( - \frac{U'(x)}{x} - \frac{\sigma^2}{2} 
\right) dt + \sigma d W_t. 
\eeq 
This produces an equivalent Langevin equation for the log-price (more accurately, the log-cap):
\beq
\label{Langevin_y_0}
d y_t = - V'(y) dt + \sigma d W_t, \; \; \; V'(y) \equiv \frac{\partial V(y)}{\partial y} = -  \left( \theta - \frac{\sigma^2}{2} \right) + \kappa e^{y} + g e^{2y},
\eeq
where the potential $ V(y) $ in the log-space is defined as follows:
\beq
\label{pot_y}
V(y) 
= - \bar{\theta} y + \kappa e^{y} + \frac{1}{2} g e^{2y}, \; \; \;\bar{\theta} \equiv \theta - \frac{\sigma^2}{2}.
\eeq  
Note that this is no longer a classical potential, as now it acquires a ``quantum'' It\^{o}'s correction $ \sim \sigma^2 $ in the linear term $ \bar{\theta} y $. 

\begin{figure}[ht]
\begin{center}
\includegraphics[
width=180mm,
height=55mm]{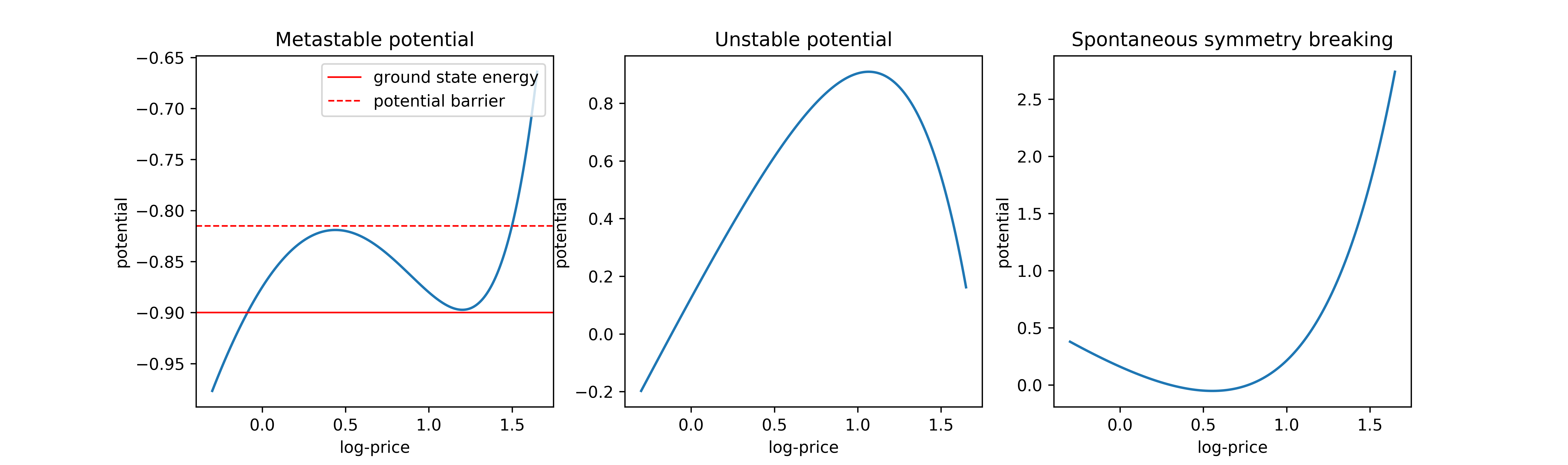}
\caption{Potentials $ V(y) $ in the log-variables $ y = \log x $ for the same choices of parameters in (\ref{pot_ab}) as used in 
Fig.~\ref{fig:Classical_potentials}, with $ \sigma = 0.25 $. The graph on the left describes metastability corresponding to a bimodal system with two metastable states 
in Fig.~\ref{fig:Classical_potentials}. Absorption at zero is replaced by escape to negative infinity in the $ y $-space.} 
\label{fig:Classic_potentials_log_space}
\end{center}
\end{figure}

Different shapes of the $y$-space potential that can be obtained with different parameters in Eq.(\ref{pot_ab}) are shown in 
Fig.~\ref{fig:Classic_potentials_log_space}.

The Langevin equation (\ref{Langevin_y_0}) in the log-space $ y = \log x $, 
can also be written in a form more conventional in physics (crudely, by dividing both sides of  (\ref{Langevin_y_0}) by $ dt $):
\beq
\label{Langevin_y}
\frac{dy}{dt} = - \frac{\partial V}{\partial y}  + \xi_t\, \; \;  \; \langle \xi_{t_1} \xi_{t_2} \rangle =  \sigma^2 \delta(t_1 - t_2), 
\eeq
where $ \xi_t $ is Gaussian white noise. 
Note that the form of noise correlation in Eq.(\ref{Langevin_y}) suggests that the combination $ \frac{\sigma^2}{2} $ can be identified with an effective temperature 
 $ T $ of the system, i.e. $ T = \frac{\sigma^2}{2} $. 
 
 Another important point to note is the behavior of the potential $ V(y) $ at positive and negative infinities in the $ y $-space. For $ y \rightarrow \infty $, we obtain 
 $ V \rightarrow + \infty $ if $ g \geq 0 $. On the other hand, the behavior at a negative infinity depends on the sign of $ \bar{\theta} $: $ V = 
 \mbox{sign}(\bar{\theta}) \infty $. 
 
 Therefore, the potential $ V(y) $ leads to a bounded motion on the real $ y $-axis only for $ \theta \geq  \frac{\sigma^2}{2} $.
 On the other hand, when $ \theta \leq  \frac{\sigma^2}{2} $ (i.e. $ \bar{\theta} \leq 0 $), the potential {\it decreases} as $ y \rightarrow - \infty $. This means that a particle placed in such potential will escape to a negative infinity. 
 
 The escape to $ y = - \infty $ is the same as a ``fall to the origin'' $ x = 0 $ in the original $ x $-space, which can be interpreted as corporate defaults.
 The latter are therefore identified in our model with a decay of a metastable state and escape to $ y = - \infty $ in the log-space.
 
Metastability of an initial state is ensured by the presence of a barrier separating this initial state and a ‘`free fall'' regime for $ y \rightarrow - \infty $.
This happens when parameters are in a certain range that will be specified momentarily. For other values of parameters (but such that the constraint  $ \theta \leq  \frac{\sigma^2}{2} $ still holds), there would be no barrier, and escape to $ y \rightarrow - \infty $ 
would correspond to a decay of an {\it unstable} initial state. 

Two extrema of the potential (\ref{pot_y}) are given by
\beq
\label{bar_y}
\bar{y}_{1,2} = \log \frac{ - \kappa \pm \sqrt{\kappa^2 + 4 g  \bar{\theta} }}{2g}.
\eeq
We are interested in the range of parameters when both extrema $ \bar{y}_{1,2} $ are real-valued. This requires that $ \kappa < 0 $ and 
$ \bar{\theta} < 0 $. Using the definitions of $ \kappa $ and $ \theta $ in terms of $ \phi $, this produces the following 
range of parameter $ \phi $ for a metastable decay to occur in the model:
\beq
\label{decay_params}
\phi_{1} \leq \phi \leq \bar{\phi}, \; \; \;  \bar{\phi}  \equiv \frac{\sigma^2}{2} - r_f + c, \; \; \;     
\phi_1 \equiv \frac{- \lambda + \sqrt{ 4 \mu \lambda \bar{\phi}}}{\mu}. 
\eeq
For yet smaller values $ \phi < \phi_1 $, the barrier disappears, and the process becomes unstable, rather than metastable. This can describe a regime of a ``free fall'' to default, or a total market collapse in a multivariate context.

For the height of the barrier $ E_b  \equiv V(\bar{y}_2) - V(\bar{y}_1) $, we obtain
\beq
\label{dV}
E_b  =  \bar{\theta} \log 
\frac{ 2g  \bar{\theta}+ \kappa^2 - \kappa \sqrt{\kappa^2 + 4 g\bar{\theta}}}{
 2g \bar{\theta} }  
 + \frac{\kappa}{2g} 
 \sqrt{\kappa^2 + 4 g\bar{\theta}}.
\eeq
For small values of $ \bar{\theta} $, this yields 
\beq
\label{non_anal}
E_b = \bar{\theta} \log  \frac{g \bar{\theta}}{\kappa^2} + 
O \left( \bar{\theta}^2 \right). 
\eeq
Note non-analyticity of this expression in the It\^{o}-adjusted drift $ \bar{\theta} $. If we view this expression in a complex plane of $ z = \theta $, it has a branch cut singularity on the negative semi-axis $ z \in (-\infty, 0] $. 
 
On the other hand, for small values of $ \kappa $ and fixed $ \theta $, Eq.(\ref{dV}) gives rise to another approximate formula:
\beq
\label{dV_appr}
E_b =  \frac{\kappa^2}{2g} + O \left( \kappa^2  
 \sqrt{\kappa^2 + 4 g \left( \theta - \frac{\sigma^2}{2} \right)} \right).
\eeq
These formula shows that the barrier has a non-perturbative origin (it depends on $ 1/\sqrt{g} $), and its height vanishes in the strict limit $ \kappa = 0 $. 
Also, note the effect of noise volatility $ \sigma $ on the barrier: as $ \sigma $ increases, the barrier becomes smaller.

An additional point to note in Eq.(\ref{dV_appr}) (or the full formula (\ref{dV})) is that it is {\it non-analytic} at $ \theta = \frac{\sigma^2}{2} $. As is  
shown in Appendix C, this non-analyticity is related to the fact that a stationary state of the Fokker-Planck equation exists only for 
$ \theta \geq \frac{\sigma^2}{2} $ (for the It\^{o} prescription), or for $\theta \geq \sigma^2 $ (for the Stratonovich interpretation). 

\subsection{Default as an escape from a metastable state}
\label{sect:Kramers_escape}

A potential such as shown in Fig.~\ref{fig:Classic_potentials_log_space} enables a noise-activated escape of a particle initially near a bottom 
of a local minimum of the potential
(the point $ \bar{y} \simeq 1.4 $ on the left graph) through a potential barrier. This is a classical problem studied in statistical physics, see e.g. 
\cite{Hanggi_1986}.

The classical transition state theory \cite{Hanggi_1986} gives the probability of a particle jumping over a barrier as a product of two factors:
\beq
\label{P_jump}
P \left( \mbox{jump} \right) = \Gamma \exp \left( - \frac{E_b}{T} \right)  \equiv \Gamma B,
\eeq
where $\Gamma $ is a pre-factor and $ B $ the Arrhenius factor 
\beq
\label{Arrhenius}
B = \exp \left( - E_b/T \right), 
\eeq
where $ E_b $ is the barrier height, and $ T = \frac{\sigma^2}{2} $ is the temperature. The pre-factor $ \Gamma $ can be interpreted the probability per unit time of finding the particle hitting the barrier while oscillating in the right well. 
For a potential with one minimum at $ y_{\star} $ and 
one maximum at $ y^{\star} $, such as in Fig.~\ref{fig:Classic_potentials_log_space}, the escape rate is of the form of Eq.(\ref{P_jump}), and is given by the famous 
Kramers formula \cite{Hanggi_1986} (see Appendix A for a derivation):
\beq
\label{Kramers_rate}
r = \frac{\sqrt{ V''(y_{\star}) \left| V''(y^{\star})) \right| }}{2 \pi} \exp \left[ - \frac{2}{\sigma^2} (V(y^{\star})) - V(y_{\star}) ) \right].
\eeq
The Kramers escape rate (\ref{Kramers_rate}) applies as long as the barrier height $ E_b \equiv V(y^{\star})) - V(y_{\star}) \gg \frac{\sigma^2}{2} $. As in Eq.(\ref{Arrhenius}), the escape rate $ r $ is exponentially small in the height  $ E_b $.

While the Kramers relation (\ref{Kramers_rate}) is obtained with classical stochastic dynamics, the same exponentially suppressed probabilities of under-barrier transitions are obtained in quantum mechanics \cite{Landau_QM} within the semi-classical (WKB) approximation. This is of course not incidental, and is due to the fact that the Langevin dynamics can be mapped onto quantum mechanics in Euclidean time 
\cite{Feigelman_Tsvelik,Zinn-Justin-QFT}. 

Such mapping of classical stochastic dynamics onto 
Euclidean quantum dynamics is very useful as it allows one to utilize the power of quantum mechanics and quantum field theory (QFT) for solving problems of classical stochastic dynamics that are conventionally ad'' using the Langevin or Fokker-Planck equation. In particular, these methods enable efficient computational methods for tunneling in multivariate (and non-linear) setting in dimensions $ D > 1 $\footnote{In the financial context, $ D > 1 $ corresponds to a multivariate setting with $ D $ different stocks.}, where the Kramers relation (corresponding
to $ D = 1 $) needs modifications.
We will now consider a modern alternative QFT-based approach to computing such hopping (tunneling) probabilities that is called the instanton approach.    


\subsection{Langevin instantons: the fate of a 
``false price''}
\label{sect:Langevin_instantons}

Let's start with the Langevin equation (\ref{Langevin_y}). Following \cite{LI}, we use the path integral formulation of the Langevin dynamics. 
It is given by the following generating functional (see 
Eq.(\ref{Lagrangian_MSR_2}) 
in Appendix B):
\beq
\label{Gener_Z}
Z[h]  = \int D y \, D \hat{p} \mathcal{J}(y) \exp\left( -  \frac{1}{\sigma^2} \mathcal{A}(y, \hat{p} ) + \int h(t) y(t) dt \right),  
\eeq
where $  \int D y \, D \hat{p} $ stands for integration over trajectories, $ \hat{p} $ is the so-called response (Martin-Siggia-Rose) dynamic field,
 $  \mathcal{A}(y, \hat{p} )  $  is the Euclidean action of fields $ y , \hat{p} $ with the Lagrangian $ \mathcal{L} $:
\beq
\label{Lagrange_MSR}
 \mathcal{A}(y, \hat{p} ) =\int \mathcal{L} (y, \dot{y}, \hat{p} ) dt, \, \; \; \; 
\mathcal{L}  (y, \dot{y}, \hat{p} ) =  i \hat{p} \sigma^2 \left( \frac{d y}{dt}  +  \frac{d V}{dy} \right) + \frac{\sigma^4}{2} \hat{p}^2. 
\eeq
 $ h(t) $ is an external source field, and
$  \mathcal{J}(y) $ is a Jacobian
\beq
\label{Jacobian}
 \mathcal{J}(y)  = \mbox{det} \left( \frac{d}{dt} + \frac{d^2 V(y)}{dy^2} \right). 
 \eeq
In the weak-noise limit $ \sigma^2 \rightarrow 0 $ (which is formally the same as a low-temperature limit, if we identify temperature $ T $ as $ T = \sigma^2/2 $), the 
Jacobian  $  \mathcal{J}(y) $ and the source field $ h $ can be neglected \cite{LI} (see Eqs.(\ref{Jacobian_A}) and (\ref{Lagrangian_MSR})).  
The generating functional is dominated in this limit by solutions of classical Lagrange equations \cite{Landau_mechanics} (here $ \dot{y} = dy(t)/dt $) 
\beq
\label{Lagrange_equation}  
\frac{d}{d t} \frac{\partial \mathcal{L}}{\partial \dot{y}} - \frac{\partial \mathcal{L} }{\partial y} = 0 , \; \; 
\frac{d}{d t} \frac{\partial \mathcal{L}}{\partial \dot{\hat{p}}} - \frac{\partial \mathcal{L} }{\partial \hat{p}} = 0,
\eeq
that are obtained by varying the action $  \mathcal{A}(y, \hat{p} ) $ with respect to $ y  $ and $ \hat{p} $, respectively.
For the QED Lagrangian (\ref{Lagrange_MSR}), this yields
\beq
\label{classical_motion}
\frac{d\hat{p}}{dt} - \frac{d^2 V}{d y^2} \hat{p} = 0 , \; \; \; 
\frac{d y}{dt} + \frac{d V}{d y}  =  i \sigma^2 \hat{p}.
\eeq
Note that the second equation has the same form as the original Langevin equation (\ref{Langevin_y}) where the role of noise is played by the 
rescaled response field $ i \hat{p} $.

The classical equations of motion (\ref{classical_motion}) admit two drastically different solutions. The first solution reads
\beq
\label{classical_sol}
\frac{dy}{dt} = -  \frac{d V}{d y} , \; \; \hat{p} = 0 , \; \;  \mathcal{A}(y, \hat{p} )  = 0.
\eeq
This is a ``normal'' solution, where the response field $ \hat{p} $ vanishes, and a particle goes ``downhill'', against the gradient of the potential. This solution has zero action and corresponds to the noiseless (``classical'') limit of the dissipative Langevin dynamics (\ref{Langevin_y}). 

There is however another solution
\beq
\label{instanton}
\frac{dy}{dt} =   \frac{d V}{d y}, \; \;   i \hat{p} =  \frac{2}{\sigma^2} \frac{dy}{dt} =  \frac{2}{\sigma^2}  \frac{d V}{d y}.
\eeq
It is easy to see that as long as Eqs. (\ref{instanton}) hold, the first of Eqs.(\ref{classical_motion}) is obtained from the second equation by differentiating both sides with respect to time. 

The second solution (\ref{instanton}) is very different from the 
first, normal solution (\ref{classical_sol}). As suggested by the second equation in (\ref{instanton}), the response field 
$ \hat{p} $ for this solution is purely imaginary. The first equation shows that this solution describes a noiseless limit of the Langevin equation with an {\it inverted} potential $ V(y) $, or equivalently under time reversal $ t \rightarrow - t $. Such inversion of a potential when describing tunneling also occurs in quantum mechanics (see below), and defines the essence of the instanton approach: classically forbidden transitions become classically allowed with an inverted potential. 

The solution (\ref{instanton}) is a {\it vacuum} ({\it zero energy}) solution, as the Hamiltonian (\ref{H_QED}) vanishes for this solution:
\beq
\label{H_insta}
 \mathcal{H} =  
 - \frac{\sigma^4}{2} \left( i \hat{p} \right)^2  +  i \sigma^2 \hat{p} \frac{d V}{dy}  =   - 2  \left( \frac{d V}{d y} \right)^2 
 + 2  \left( \frac{d V}{d y} \right)^2  = 0.
 \eeq
On the other hand, the action for this solution for a transition from $ y_R $ at $ t_i $ to $ y_0 $ at $ t_f $ is non-zero: 
\beq
\label{action_instanton}
 \mathcal{A}(y, \hat{p} )  = \frac{2}{\sigma^2} \int_{t_i}^{t_f} \left( \frac{dy}{dt} \right)^2 dt = \frac{2}{\sigma^2} \int_{y_R}^{y_0}   \frac{dy}{dt} dy
 =  \frac{2}{\sigma^2} \int_{y_R}^{y_0}   \frac{dV}{dy} dy =   \frac{2}{\sigma^2} \left( V(y_0) - V(y_{R}) \right) \equiv \frac{E_b}{T},
 \eeq
where we replaced $ \frac{\sigma^2}{2} $ by the effective temperature $ T $ at the last step. The probability of such transition is therefore
\beq
\label{prob_instanton}
P(\mbox{escape})  \sim \exp \left( -  \mathcal{A}(y, \hat{p}) \right) =   \exp \left( -   \frac{E_b}{T} \right),
\eeq
which coincides, up to a pre-exponential factor, with Eq.(\ref{Arrhenius}), as well as with the conventional WKB approximation of quantum mechanics for under-barrier 
tunneling \cite{Landau_QM}. 


It turns out that in quantum mechanics (QM), tunneling can be described using a similar approach. Classical Euclidean equations of motion arise in the QM semi-classical limit $ \hbar \rightarrow 0 $ upon the so-called Wick rotation of time to {\it imaginary time} $ \tau = i t $ (see Appendix B for a crash introduction).
Solutions of classical Euclidean (imaginary time) equations of motion describing tunneling phenomena are 
known in quantum field theory and statistical physics as {\it instantons}, see e.g. \cite{Zinn-Justin-QFT}, \cite{Weinberg_book}, or \cite{Morano}. A pre-exponential factor $ \Gamma $ can be computed from the analysis of fluctuations around instanton solutions. 

\begin{figure}[ht]
\begin{center}
\includegraphics[
width=140mm,
height=50mm]{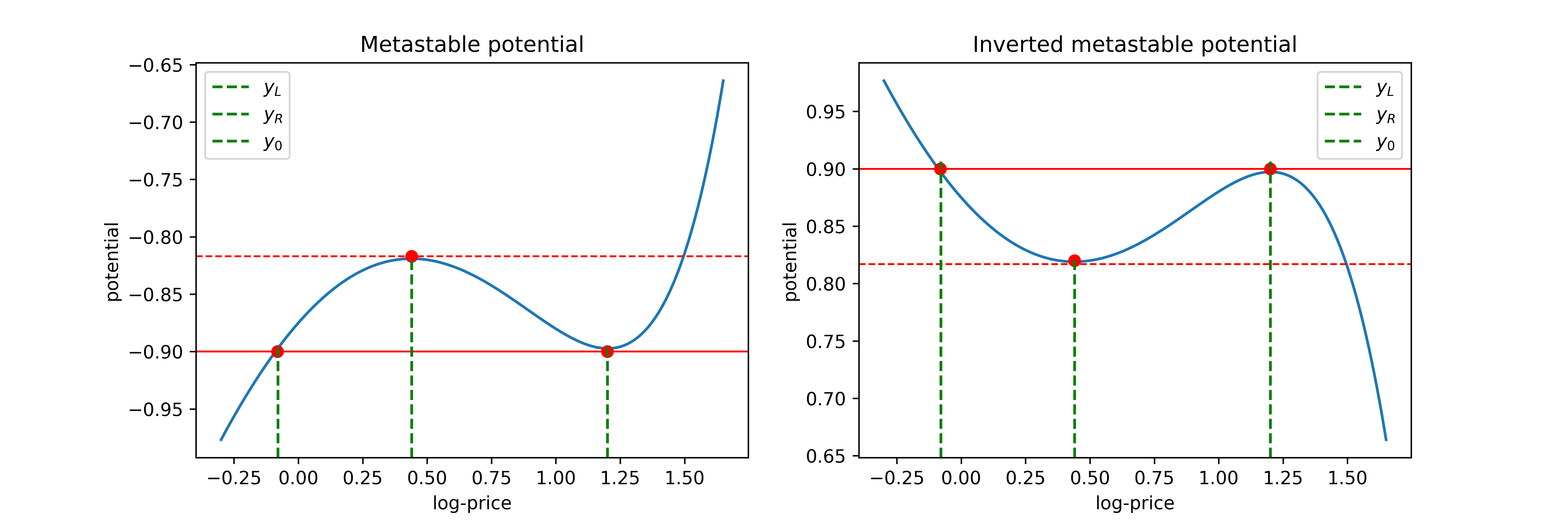}
\caption{A metastable potential in the log-space $ y = \log x $ for the same choices of parameters in (\ref{pot_ab}) as used in 
Fig.~\ref{fig:Classical_potentials}. The graph on the right shows the inverted potential. 
} 
\label{fig:Inverted_potential}
\end{center}
\end{figure}

The instanton solution is illustrated for a metastable potential 
on the left 
in Fig.~\ref{fig:Inverted_potential}.
An instanton starts at time $ t = - T $ (where $ T \rightarrow \infty $) at the point $ y_R $ of the right local minimum (or equivalently of the top of the inverted potential on the right graph). At time $ t = T $, it arrives at the return point $ y_L $.
Note that $ y_L $ is a reflection point rather than a local minimum of a potential. Therefore, our instantons are different from 
instantons that arise in theories with vacuum tunneling between degenerate potentials. In these models, initial and terminal conditions for an instanton correspond to degenerate vacua connected by the instanton.  

In the present problem, we deal with a different case of a decay of a metastable state. Corresponding classical solutions for such scenarios are called bounces (see \cite{Coleman_Callan} or \cite{Coleman_book}, Chapt. 7). Unlike instantons, bounces are period solutions, where the final state coincides with the initial one, which in its turn is chosen to be a particle located at or near the bottom of the local metastable minimum of the potential. 
As we will show below, the bounce is made of an instanton-anti-instanton pair.


The instanton solution (\ref{instanton}) on the left of 
Fig.~\ref{fig:Inverted_potential} 
describes a particle that starts at the point $ y_R $ at the bottom of right-most well of the original potential $ V(y) $ at time $ t = - T $, and then climbs the hill at $ y_0 $, following Eq.(\ref{instanton}). After that, the first solution (\ref{classical_sol}) kicks in, and describes the classical downhill move from $ y_0 $ to $ y_L $, achieving the point $ y_L $ at time 
$ T $. An anti-instanton follows the same trajectory reversely in time. By stacking together the instanton and anti-instanton, we obtain a bounce that starts at $ y_R $ at $ t = - T $, and ends at the same point $ y_R $ at time $ t = T $.
Note that the parts of this trajectory that go from $ y_R $ to $ y_0 $ and then from $ y_L $ to $ y_0 $ are classically forbidden as they go uphill the classical potential $ V(y) $.

On the other hand, on the right graph in Fig.~\ref{fig:Inverted_potential}, the same solution is interpreted as a purely classical motion in the inverted potential $ -V(y) $ (alternatively, one can reverse time $ t \rightarrow - t $). In this case, the particle starts on the top of the hill at $ y_R $, then relaxes to $ y_0 $ and climbs to $ y_L $, after which it ``bounce'' (hence justifying its name) back to $ y_R $ by reverting the trajectory. The explicit form of the instanton solution will be given below in 
Sect.~\ref{sect:Instantons_and_WKB}.

To summarize, the instanton-anti-instanton pairs (bounces) produce exponentially small probabilities (\ref{prob_instanton}) suppressed by the height of a potential barrier, as in the 
Kramers escape rate (\ref{Kramers_rate}).
The pre-exponential factor can be obtained by computing fluctuations around the bounce \cite{Coleman_Callan,Coleman_book}. Instead of completing this calculation (that produces again the Kramers relation), in the next section we present an equivalent approach to quantum mechanical tunneling based on the Fokker-Planck equation. 

\section{Defaults and the Fokker-Planck dynamics}
\label{sect:Tunneling}

 \subsection{The Fokker-Planck equation}
\label{sect:FPE}

\begin{figure}[ht]
\begin{center}
\includegraphics[
width=160mm,
height=60mm]{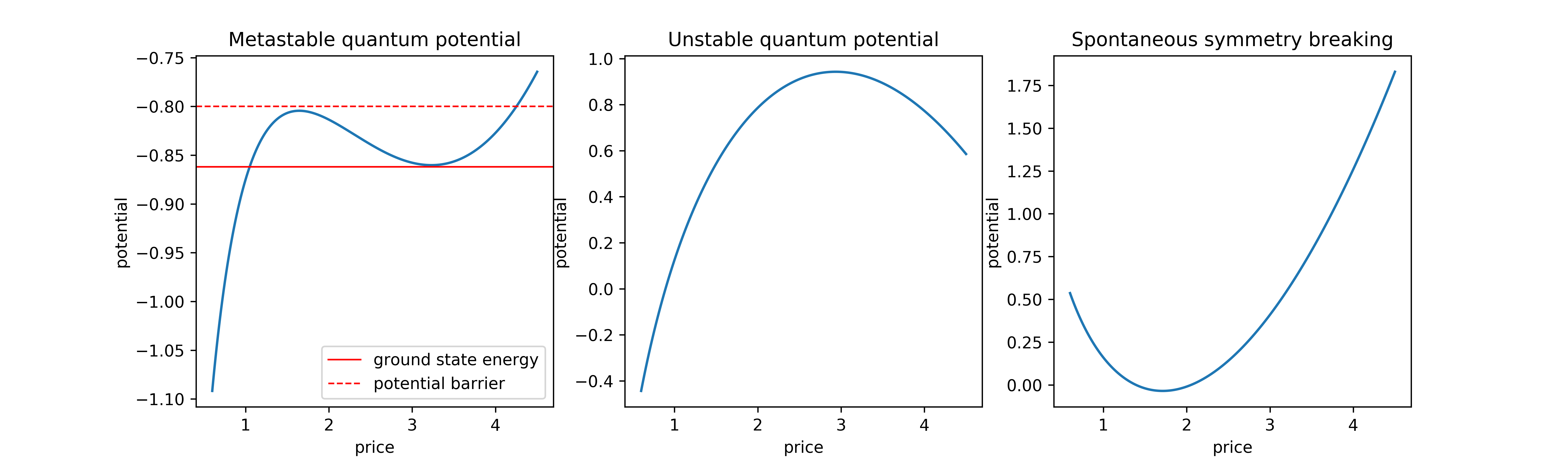}
\caption{Effective potentials with metastable states for the QED model. 
The same values of parameters as in Fig.~\ref{fig:Classical_potentials} are used.
The figure on the right shows extrema of the effective potential and a potential barrier between the local minimum and an essential singularity at zero.} 
\label{fig:Effectiive_potentials}
\end{center}
\end{figure}

Let $ p(x,t) $ be a probability density of the non-linear diffusion of Eq.(\ref{Langevin_eq}). Its behavior depends on rules for stochastic integration,
namely the Stratonovich vs It\^{o} prescriptions, see e.g.  \cite{Horsthemke}. The Fokker-Planck equation (FPE) reads
\beq
\label{FPE}
\partial_t p(x,t) = - \partial_x \left[ f(x) p(x,t) \right] + \frac{\sigma^2}{2} \partial_{xx} \left[ g(x) p(x,t) \right],
\eeq
where
\beq 
\label{drift_diff}
f(x) =  \left(\theta x - \kappa x^2  - g x^3 + (2 - \nu) \frac{\sigma^2}{2} x  \right), \; \; \; g(x) = x^2,
\eeq
and $ \nu = 2 $ or $ \nu = 1 $ for the It\^{o} or Stratonovich interpretation of the SDE (\ref{Langevin_eq}), respectively.
Similar FPEs with polynomial drifts and a multiplicative noise (\ref{drift_diff}) were studied in the physics literature, see e.g. \cite{Munoz_1998}, 
\cite{Van_Broeck_1997}. The steady-state solution of this equation is
\beq
\label{steady_state_Str}
p_s(x) = \frac{1}{Z} \frac{1}{g(x)} \exp\left\{ \int_{0}^{x} \frac{f(y) }{ \frac{\sigma^2}{2} g(y)} dy \right\} 
= \frac{1}{Z}  \exp \left[ - U(x) \right],
\eeq
where $ Z $ is a normalization factor (also known as a partition funciton), and $ U(x) $ is the effective potential:
\beq
\label{eff_pot}
U(x) =  - \frac{2}{\sigma^2} \int_{0}^{x} \frac{f(y) }{ g(y)} dy + \log g(x) 
= \frac{2}{\sigma^2} \left[  \kappa x  + \frac{g}{2} x^2 -  \left(\theta - \frac{\sigma^2}{2} (4-\nu) \right) \log x  \right].
\eeq
This is a potential energy of a ``particle'' with mass $ g $ in an external field given by a combination of a quadratic and logarithmic potentials.
This potential leads to {\it metastability} that was already mentioned above, for certain values of  
model parameters, as illustrated in Fig.~\ref{fig:Effectiive_potentials}. 

The extremal points of the effective potential 
(\ref{eff_pot}) are (we set $ \nu = 2 $ here) 
\beq
\label{extremum}
 \bar{x}_{1,2} = \frac{- \kappa \pm \sqrt{ \kappa^2 + 4 g \left( \theta - \sigma^2 \right) }}{ 2 g }.
\eeq
In addition, the effective potential (\ref{eff_pot}) has an essential singularity at $ x = 0 $.
Comparing  Eq.(\ref{extremum}) with Eq.(\ref{extrema_V}), we find that that only effect of noise on the extrema $ \bar{x}_{1,2} $ of the classical potential is a shift of parameter $ \theta $ by an amount $ \sim \sigma^2 $. 

On the other hand, the trivial vacuum $ x = 0 $ is impacted by noise in a more drastic way, and becomes an essential singularity. 
Consider first the scenario where 
the coefficient in front of the logarithmic term in Eq.(\ref{eff_pot}) is positive. In this case, the state $  \bar{x}_0 = 0 $ becomes a {\it repelling boundary} and 
absorption at $ x = 0 $ is impossible. Such regime would be similar to a default-free GBM world in the sense that in both cases the boundary at $ x = 0 $ becomes unattainable.

On the other hand, if the coefficient in front of the logarithm in Eq.(\ref{eff_pot}) is negative, than a particle can ``fall to the origin'', $ x= 0 $, in a potential such as shown in Fig.~\ref{fig:Effectiive_potentials}.  If a particle is initially placed to the right of a local maximum of the effective potential, it will be in a {\it meta-stable} state that will decay into the absorbing state in a finite time due to thermal fluctuations. A ``fall to the origin'' (a corporate default) from a meta-stable initial state may occur when the coefficient of the logarithm  in Eq.(\ref{eff_pot}) is negative.

To summarize, adding white noise 
induces a new  ``effective potential'' (\ref{eff_pot}) that leads to considerable changes of the analysis of the phase structure of the classical model (\ref{pot}) without noise.
For certain values of parameters, there is a non-zero probability of a ``collapse to the origin'' at $ X = 0 $, that can describe a corporate default. Such events would be examples of the so-called 
noise-induced phase transitions \cite{Horsthemke,Van_Broeck_1997}.



\subsection{Transformation to  the Fokker-Planck equation with an additive noise}

Let's consider the FPE (\ref{FPE}) where we introduce a new independent variable $ y = \log x $, with an initial condition 
$ y(t_0) = y_0 = \log x_0 $ for some time $ t_0 $.  Let $ \tilde{p}(y,t|y_0) $ is be a probability density for $ y $. For clarity, we display the initial point $ y_0 $ here, or its respective value $ x_0 $ in the $ x$-space
in this section. The two densities $ p(x,t|x_0)$ and $ \tilde{p}(y,t) $ are related as follows:
\beq
\label{density_y}
p(x,t | x_0) dx = \tilde{p}(y,t | y_0) dy \; \;  \Leftrightarrow \; \; \;  p(x(y),t | x_0) = e^{-y} \tilde{p} (y,t | y_0)  \; \;  \left( y = \log x , \, 
\frac{\partial}{\partial x} = e^{-y} \frac{\partial}{\partial y} \right).
\eeq
This substitution transforms the FPE (\ref{FPE}) into a new FPE for the density $ \tilde{p} (y,t) $ that contains an {\it additive} rather than a multiplicative noise:
\beq
\label{FPE_tilde}
\frac{\partial \tilde{p}(y,t | y_0)}{\partial t} = - \frac{\partial}{\partial y} \left[ \tilde{f}(y) \tilde{p}(y,t | y_0) \right] + \frac{\sigma^2}{2} \frac{\partial^2}{\partial y^2} 
\tilde{p}(y,t | y_0),
\eeq
where 
\beq
\label{drift_tilde}
\tilde{f}(y) = \theta - (\nu -1 ) \frac{\sigma^2}{2} - \kappa e^y - g e^{2y} \equiv - \frac{ \partial V(y)}{\partial y}.
\eeq  
The steady state solution for this equation reads (compare with Eq.(\ref{steady_state_Str}))
\beq
\label{p_s_y}
\tilde{p}_s(y) = \frac{1}{\tilde{Z}} e^{ -  \frac{2}{\sigma^2} V(y)}, \; \; \; V(y) \equiv - \left( \theta - (\nu -1 ) \frac{\sigma^2}{2} \right) y + 
\kappa e^y + \frac{1}{2} g e^{2y},
\eeq
where $ \tilde{Z} $ is a normalization constant. This solution exists for parameter values that give rise to a normalizable steady state solution. 
For parameters that produce metastability, the normalization constant $ Z $ in Eq.(\ref{p_s_y}) diverges, and therefore for such parameters  Eq.(\ref{p_s_y})
ceases to be a valid solution. For mathematical details, see Appendix C.


\subsection{Schr{\"o}dinger equation}  
\label{sect:SE}

In what follows, we set $ \nu = 2 $, i.e. choose the It\^{o} interpretation of the FPE (\ref{FPE}), and work with the FPE
(\ref{FPE_tilde}) in the log-space. 
We use the following ansatz to solve Eq.(\ref{FPE_tilde}) (see e.g. \cite{Junker} or \cite{vanKampen}):
\beq
\label{ansatz}
 \tilde{p}(y,t | y_0) =  e^{ - \frac{1}{\sigma^2} \left(V(y) - V(y_0) \right)}  K_{-}(y,t | y_0).
\eeq
Using this in Eq.(\ref{FPE_tilde}) produces an imaginary time Schr{\"o}dinger equation (SE) for the forward Feynman propagator
$ K_{-}(y,t | y_0) $. As discussed in Appendix A, it is useful to analyze simultaneously both the forward and backward 
propagators, resp. $ K_{-}(y,t | y_0) $ and $ K_{+}(y,t | y_0) $,
where $ K_{+}(y,t | y_0) $ describes the dynamics with a flipped potential (or equivalently the dynamics under time-reversal).
A pair of SEs for $ K_{\pm}(y,t | y_0) $ 
reads
\beq
\label{SE}
- \sigma^2 \frac{\partial K_{\pm}(y,t | y_0)}{\partial t} = \mathcal{H}_{\pm} K_{\pm} (y,t | y_0),
\eeq
where $ \mathcal{H}_{\pm} $ are the Hamiltonians 
\beq
\label{Hamiltonian}
\mathcal{H}_{\pm} = -  \frac{\sigma^4}{2} \frac{\partial^2}{\partial y^2} +  \frac{1}{2} \left( \frac{ \partial V}{\partial y} \right)^2 
\pm \frac{1}{2}  \sigma^2 \frac{\partial^2 V}{\partial y^2}  \equiv -  \frac{\sigma^4}{2} \frac{\partial^2}{\partial y^2} + U_{\pm}(y). 
\eeq
The solution of the Schr{\"o}dinger equation can be formally represented by the heat kernel
\beq
\label{heat_kernel}
 K_{\pm} (y,t | y_0) = \langle y | e^{ - \frac{1}{\sigma^2} t \mathcal{H}_{\pm} } | y_0 \rangle.
\eeq
Let $ \left\{ \Psi_n^{-} \right\} $ be a complete set of eigenstates of the Hamiltonian $ H_{-} $ with eigenvalues 
$ E_n^{-} $:
\beq
\label{Eigen_values} 
\mathcal{H}_{-} \Psi_n^{-} = E_n^{-} \Psi_n^{-}, \; \; \; 
\sum_{n} \Psi_n^{-}(x) \Psi_n^{-}(y) = \delta(x-y).
\eeq
We assume here a discrete spectrum corresponding to a 
motion in a bounded domain, so that the set of values 
$ E_{n}^{(-)} $ is enumerable by integer values $ n = 0, 1, \ldots $. 
Substituting this into Eqs.(\ref{heat_kernel}) and (\ref{ansatz}), we obtain the spectral decomposition of the original FPE (see e.g. \cite{vanKampen}):
\beq
\label{spectral_FPE}
 \tilde{p}(y,t | y_0) =  e^{ - \frac{1}{\sigma^2} \left(V(y) - V(y_0) \right)}  \sum_{n=0}^{\infty} e^{ - \frac{1}{\sigma^2} E_n^{-} t} 
 \Psi_n^{-}(y) \Psi_n^{-}(0).
 \eeq
When $ t \rightarrow \infty $, only one term with the lowest energy 
survives in the sum in Eq.(\ref{spectral_FPE}). If such lowest energy $ E_1^{(-)} $ is larger than zero for a particle located in a potential well, such as shown in 
Fig.~\ref{fig:Inverted_potential}, this means that such an initial state $ y_0 $ is metastable with the decay rate 
$ E_1^{(-)}/\sigma^2 $. Therefore, in the quantum mechanical (QM) approach to barrier transitions in the Langevin dynamics computing rate transitions amounts to computing the eigenvalue spectra of the QM 
Hamiltonians $ H_{\pm} $. The transform
(\ref{ansatz}) and the resulting first factor in 
Eq.(\ref{spectral_FPE}) take care of the non-equilibrium component of the dynamics, and reduce the problem of metastability in the Langevin dynamics to a stationary or quasi-stationary quantum mechanical problem \cite{Feigelman_Tsvelik}.

\begin{figure}[ht]
\begin{center}
\includegraphics[
width=160mm,
height=60mm]{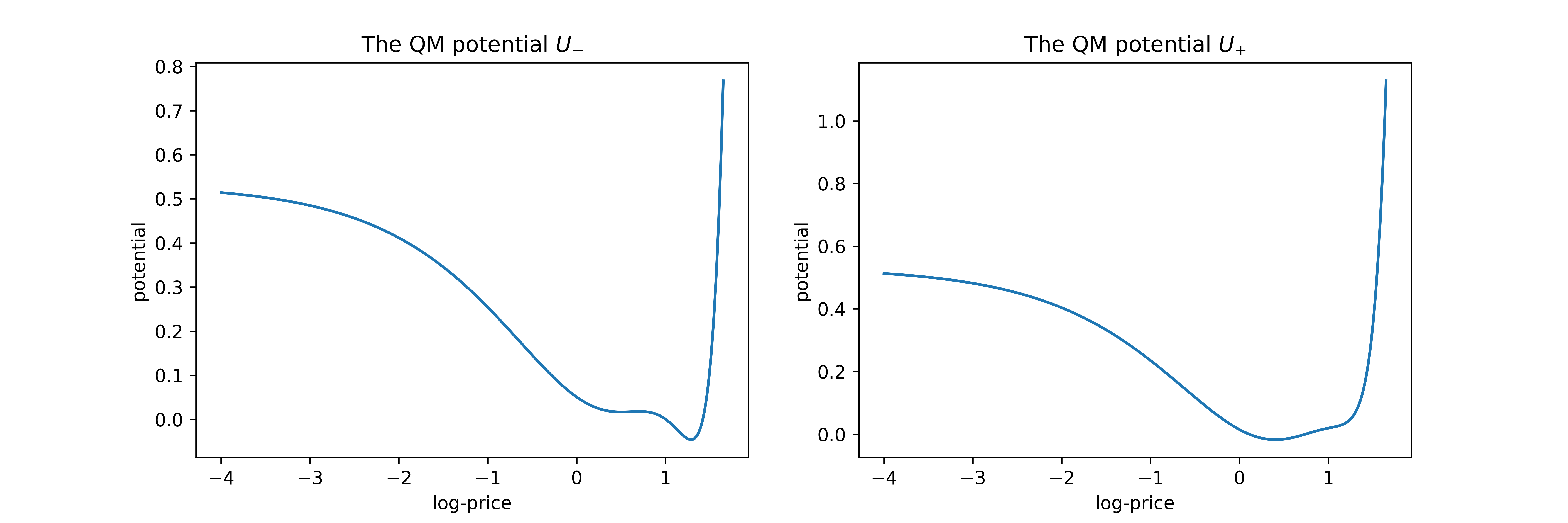}
\caption{Quantum mechanical potentials $ U_{\pm} (y) $ for the QED model.} 
\label{fig:QM_SUSY_potentials}
\end{center}
\end{figure}

Using function $ V(y) $ from Eq.(\ref{p_s_y}), we obtain (recall
that $ \bar{\theta} = \theta - \sigma^2/2 $)
\bea
\label{W_y}
&&  \frac{1}{2} \left( \frac{ \partial V}{\partial y} \right)^2 = \frac{\bar{\theta}^2}{2} -\kappa \bar{\theta}  e^y + 
\left( \frac{\kappa^2}{2} - g\bar{\theta} \right) e^{2y} 
+  \kappa g e^{3y} + \frac{g^2}{2} e^{4y}, 
\nonumber \\
&& \frac{\sigma^2}{2} \frac{\partial^2 V}{\partial y^2} = 
\kappa \frac{\sigma^2}{2}  e^{y} + g \sigma^2 e^{2y}. 
\eea
The QM potentials $ U_{\pm} (y) $ are therefore as follows:
\beq
\label{V-}
U_{\pm} (y) = \frac{\bar{\theta}^2}{2} - \kappa \left( \bar{\theta}
 \mp \frac{\sigma^2}{2} \right) e^y + 
\left( \frac{\kappa^2}{2} - g\bar{\theta} 
 \pm g \sigma^2 \right) e^{2y} 
+  \kappa g e^{3y} + \frac{g^2}{2} e^{4y}.
\eeq
Therefore, both potentials $ U_{\pm} (y) $ are quartic polynomials
in $ x = e^{y} $, and moreover differ from one another only in 
coefficients in front of $ x $ and $ x^2 $. As $ y \rightarrow
- \infty $, both potentials approach the same constant level 
$ \bar{\theta}^2/2 $, see Fig.~\ref{fig:QM_SUSY_potentials}.

Extrema of $ U_{\pm} (y) $ are determined by the equation (here $ x = e^{y} $)
\beq
\label{extrema_U_pm}
  x \left(x^3 + a x^2 + b_{\pm} x + c_{\pm} \right) = 0,
 \eeq
 where 
 \beq
 \label{coeffs_cubic}
 a =  \frac{3 \kappa}{2 g} , \; \; 
 b_{\pm} = \frac{1}{g^2} \left( \frac{\kappa^2}{2}  - 
 g \bar{\theta}  \pm  g {\sigma^2}  \right) , 
 \; \; c_{\pm} =  - \frac{\kappa}{2 g^2} \left( \bar{\theta} 
 \mp \frac{\sigma^2}{2}  \right). 
 \eeq
 Eq.(\ref{extrema_U_pm}) has a trivial root $ x = 0 $ ($ y = - \infty $) corresponding to a particle absorbed at the origin $ X = 0 $. In addition, the cubic polynomial 
 arising in Eq.(\ref{extrema_U_pm}) can have either one real root and two complex roots, or three real-valued roots. The roots are found by noticing that by a substitution
 $ x = u - a/3 $, the cubic equation $ x^3 + a x^2 + b_{\pm} x 
 + c_{\pm}  = 0 $ reduces to the standard Cardano form 
 \beq
 \label{Cubic_eq}
 u^3 + p_{\pm} u + q_{\pm} = 0 , \; \; p_{\pm} = - \frac{a^2}{3} + b_{\pm}, \; \; q_{\pm} = 2 \left( \frac{a}{3} \right)^3 
 - \frac{ab_{\pm}}{3} + c_{\pm}.
 \eeq
 This equation has three roots, that we write here in terms of the original variable  $x $:
 \beq
 \label{roots_Cardano}
 x_1 = A_{\pm} + B_{\pm} - \frac{a}{3}, \; \; \; 
 x_{2,3} = - \frac{A_{\pm} + B_{\pm}}{2} \pm i 
 \sqrt{3} \frac{A_{\pm} - B_{\pm}}{2}  - \frac{a}{3}, 
 \eeq
 where 
 \beq
 \label{Cardano_aux}
 A_{\pm} = \sqrt[\leftroot{-1}\uproot{2}\scriptstyle 3]{ - \frac{q_{\pm}}{2} + \sqrt{Q_{\pm}} } , \; \; 
 B_{\pm} = \sqrt[\leftroot{-1}\uproot{2}\scriptstyle 3]{ - \frac{q_{\pm}}{2} - \sqrt{Q_{\pm}} }, \; \; Q_{\pm} = \left( \frac{p_{\pm}}{3} \right)^3 + \left( \frac{q_{\pm}}{2} \right)^2. 
 \eeq
 If the discriminant $ Q_{\pm} $ in (\ref{Cardano_aux}) is positive, Eq.(\ref{Cubic_eq}) has one real root and two complex roots. 
If $ Q_{\pm} = 0 $, then there are tree roots such that two of them are equal each other. Finally, when $ Q_{\pm} < 0 $, there are three distinct real roots.

Fig.~\ref{fig:QM_SUSY_potentials} shows that while the potential $ U_{-} $ has two local minima and a local maximum (so that
$ Q_{-} < 0 $), the partner potential $ U_{+} $ has only one global real-valued minimum (corresponding to $ Q_{+} < 0 $). This implies that it might be easier to compute the eigenvalues of the Hamiltonian $ \mathcal{H}_{+} $ than of $ \mathcal{H}_{-} $.
It turns out that the spectra (and eigenfunctions) of the Hamiltonians $ \mathcal{H}_{\pm} $ are 
elated by a symmetry known in physics as {\it supersymmetry}.  Analysis of supersymmetry of the FPE and a supersymmetric derivation of the Kramers relation are presented in Appendix D.

\subsection{Instantons and the modified WKB approximation}
\label{sect:Instantons_and_WKB}

While reproducing the Kramers relation along with calculable corrections, the quantum-mechanical SUSY-based calculation of the decay rate 
given in  in Appendix D does not provide an explicit link to  
noise-induced barrier transitions that we described by Langevin instantons in Sect.~\ref{sect:Langevin_instantons}. Here we show that the same instantons arise in the Schr{\"o}dinger equation
(\ref{SE}).

To this end, consider the stationary Schr{\"o}dinger equation for the lowest energy $ E $, which is obtained from the time-dependent Schr{\"o}dinger equation (\ref{SE}) by the substitution $ K_{-}(y,t | y_0) \rightarrow e^{- Et} \Psi(y) $:  
\beq
\label{SE_2}
\mathcal{H}_{-} \Psi = E \Psi, \; \; \;  \mathcal{H}_{-} = 
 -  \frac{\sigma^4}{2} \frac{\partial^2}{\partial y^2} +  \frac{1}{2} \left( \frac{ \partial V}{\partial y} \right)^2 
- \frac{1}{2}  \sigma^2 \frac{\partial^2 V}{\partial y^2}.
\eeq

 We look for a solution of the SE (\ref{SE_2}) in the form
\beq
\label{WKB}
\Psi(y) = \exp\left[ - \frac{1}{\sigma^2} S_0(y) - S_1(y) \right],
\eeq
which replaces one wave function $ \Psi $ by two unknown functions $ S_0(y) $ and $ S_1(y) $.
The change of variables (\ref{WKB}) is the same as in the 
conventional WKB approximation of quantum mechanics (also known as the Eikonal approximation, see e.g. \cite{Landau_QM}). The QM WKB approximation computes functions 
$ S_0 $, $ S_1 $ (and possibly higher-order corrections in $ \sigma^2 $) by assuming a perturbative expansion for the energy $ E $ of the ground state as follows:
\beq 
\label{ground_state_energy}
E = E^{(0)} + \sigma^2 E^{(1)} + \sigma^4 E^{(2)} + \ldots.
\eeq 
The standard WKB approximation for functions $ S_0 $, $ S_1 $ is obtained by 
substituting Eqs.(\ref{WKB}) and (\ref{ground_state_energy}) into 
the SE (\ref{SE_2}), and matching coefficients in front of different powers of $ \hbar = \sigma^2 $.

Instead of the conventional WKB method, we use a {\it modified} WKB method in which $ E_0 = 0 $, so that any difference of the ground state energy from zero is a pure quantum effect (i.e. it vanishes in the limit $ \sigma^2 \rightarrow 0 $) 
\cite{Milnikov_2008, Milnikov_book}. While the modified WKB approach was developed in \cite{Milnikov_2008, Milnikov_book} in the context on non-supersymmetric quantum mechanics, here we apply it to SUSY QM obtained from our initial Langevin formulation of classical stochastic dynamics. Note that the classical vacuum energy in SUSY QM is zero as can be seen from (\ref{SE_2}). Therefore, setting  
$ E^{(0)} = 0 $ is perfectly justified in our case.   

Substituting (\ref{WKB}) and (\ref{ground_state_energy}) into Eq.(\ref{SE_2}) and equating expressions in front of like powers of 
$ \sigma^2 $, we obtain, to the zeroth order, the following equation:
 \beq
 \label{WKB_eq_0}
 - \frac{1}{2} \left( \frac{\partial  S_0}{\partial y} \right)^2 +
 \frac{1}{2} \left( \frac{\partial  V}{\partial y} \right)^2 = 0.
 \eeq 
 The first-order equation reads
 \beq 
 \label{WKB_eq_1}
 \frac{1}{2} \frac{\partial^2  S_0}{\partial y^2} -
 \frac{\partial  S_0}{\partial y} \frac{\partial S_1}{\partial y}-
\frac{1}{2} \frac{\partial^2  V}{\partial y^2} =  E^{(1)}. 
\eeq  
The first equation (\ref{WKB_eq_0}) is the Euclidean Hamilton-Jacobi (HJ) equation for the shortened (or Maupertuis) action 
$ S_0 = S_0 (y| y_0) $ as a function of a final position $ y $ starting with an initial position $ y_0 $ \cite{Landau_mechanics}, with zero energy $ E = 0 $. 
Its difference from the real-time HJ equation is in the flipped sign of the kinetic energy. 
Replacing the partial derivative $ \partial  S_0 /\partial y $ with 
the canonical momentum $ p = \partial  S_0 /\partial y = \dot{y} $,
Eq.(\ref{WKB_eq_0}) is solved by a solution of one of the following two equations:
\beq 
\label{Instanton_QM}
\frac{dy}{dt} = \pm \frac{\partial V}{\partial y} 
\; \; \; \Leftrightarrow \; \; \; S_0(y|y_0) = \pm \left(
V(y) - V(y_0) \right)
\eeq
As the action should be non-negative and we assume that 
$ V(y) > V(y_0) $,
we should select the plus sign. 
This is the same instanton solution that was obtained in Eq.(\ref{instanton}) in the path integral Langevin dynamics in Sect.~\ref{sect:Langevin_instantons}. Choosing the minus sign produces anti-instanton that can be interpreted as an instanton that propagates backward in time. The action for a bounce is twice the instanton action (\ref{Instanton_QM}).

Therefore, we see that Langevin instantons re-appear as quantum-mechanical instantons in the equivalent QM formulation. We view three different derivations (FPE, SUSY, and QM instantons) of the same Kramers escape rate (\ref{Kramers_rate}) presented here not as a useless mathematical exercise but rather as a demonstration that powerful methods of quantum mechanics and quantum field theory can be applied to stochastic nonlinear dynamics of stock prices both for a single stock (as shown in this paper) and in a multivariate setting (which is left here for future research). 

The inverse of the instanton solution can be obtained by 
integrating Eq.(\ref{Instanton_QM}):
\bea  
\label{instanton_solution_inv}
t - t_0 
= && \hskip-0.5cm \pm \int_{y_0}^{y} \frac{dy'}{\partial V / \partial y'} = 
\pm \frac{1}{g} \int_{x_0}^{x} \frac{dx'}{x' (x'-x_1) (x'-x_2)} 
\nonumber \\
= && \hskip-0.5cm
\pm \frac{1}{g} \frac{1}{x_2 - x_1}  
\left( \frac{1}{x_2} \log \frac{| x - x_2|}{|x_0 - x_2|} - 
\frac{1}{x_1}  \log \frac{| x - x_1|}{ | x_0 - x_1|} 
\right)
\pm \frac{1}{g}\frac{1 }{x_1 x_2}  \log \frac{x}{x_0},
\eea
where plus or minus signs correspond to the instanton and anti-instanton, respectively, and $ x = e^{y} $ with $ x_{1,2} = e^{y_{1,2}} $ where $ y_{1,2} $ are defined 
in Eq.(\ref{bar_y}). Given the inverse of the instanton (or anti-instanton) solution, the instanton itself can be obtained by a numerical inversion of Eq.(\ref{instanton_solution_inv}).

\begin{figure}[ht]
\begin{center}
\includegraphics[
width=180mm,
height=55mm]{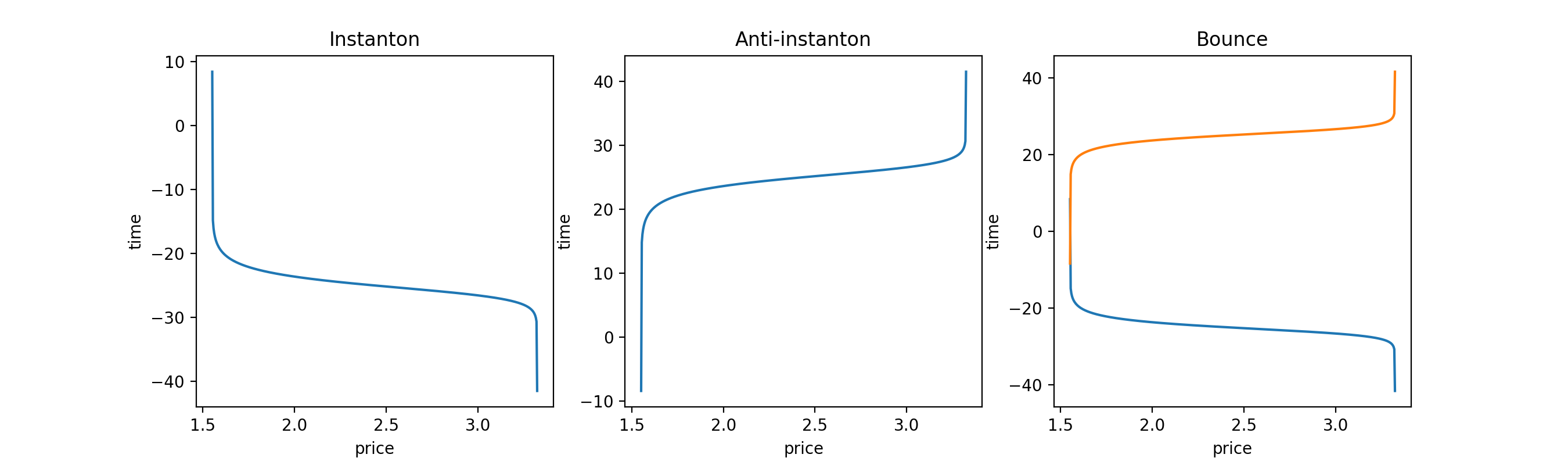}
\caption{Instanton, anti-instanton, and bounce solutions.} 
\label{fig:Instanton_and_bounce}
\end{center}
\end{figure}

The instanton solution starts at $ x_1 $ at time 
$ t = - T $ (where $ T \rightarrow \infty $), and arrives at point 
$ x_2 $ at time $ t = T $. 
The anti-instanton solution is the same in the reverse order: it starts at time $ t = - T $
at the reflection point $ x_2 $, and arrives at point 
 $ x_1 $ at  $ t = T $. A bounce is a periodic solution with period $ 2 T $ obtained with an instanton followed by a time-shifted anti-instanton that starts at time $ t = 0 $, instead of $ t = - T $, see Fig.~\ref{fig:Instanton_and_bounce}. Note that the instanton and anti-instanton are well localized in time, and are different from a constant solution only for a relatively short time interval. Therefore, in the limit $ T \rightarrow \infty $ typically assumed
in instanton-based calculations, a bounce can be thought of as a pair of well-separated instanton and anti-instanton. This implies, in particular, that the action for a bounce is twice the instanton action\footnote{As was mentioned above, only  periodic {\it bounce} solutions that start and end at same local minimum can be classical extrema for metastable potentials \cite{Coleman_Callan,Coleman_book} such as obtained in the QED model for $ \bar{\theta} < 0 $. We represent bounces as instanton-anti-instanton pairs, but our instantons are {\it not} transitions between classical local minima.}. 
 
Note that in our problem, the instanton action can be directly integrated as in Eq.(\ref{Instanton_QM}) because the quantum mechanical potential in the SE (\ref{SE_2}) is $ W(y) \equiv 1/2 
\left( \partial V / \partial y \right)^2 $ (plus the quantum correction proportional to 
$ \partial^2 V / \partial y^2 $.) Therefore,
Eq.(\ref{Instanton_QM}) coincides with the regular WBK expression for the instanton action, as an integral of the momentum corresponding to the zero total energy $ E $ in the potential 
$ W(y) $ and the canonical momentum $ p_0 \equiv \sqrt{ 2 W(y)} = 
\partial V / \partial y $ \cite{Landau_QM}:
\beq 
\label{action_instanton_WKB}
S_0 = \int_{y_0}^{y} p_0(y') dy' =  \int_{y_0}^{y} \sqrt{ 2 W(y')} dy' = 
\int_{y_0}^{y}  \frac{ \partial V}{ \partial y'} dy' = 
V(y) - V(y_0). 
\eeq 
Therefore, the Euclidean HJ equation (\ref{WKB_eq_0}) in our setting is the same as the leading-order HJ equation in the (non-supersymmetric) modified WKB 
approach of \cite{Milnikov_2008,Milnikov_book} provided we 
use it for the QM potential $ W(y) \equiv 1/2 
\left( \partial V / \partial y \right)^2 $. This particular form of the QM potential allows one to compute the instanton action explicitly as in Eq.(\ref{Instanton_QM}). The bounce action exactly reproduces the exponential factor in the Kramers relation 
(\ref{Kramers_rate}).

The pre-exponential factor can be obtained from 
the second equation (\ref{WKB_eq_1}). Note the difference of this equation relatively to its regular QM version in \cite{Milnikov_2008,Milnikov_book}: the instanton solution of Eq.(\ref{WKB_eq_1}) leads to a cancellation of the first and third terms in Eq.(\ref{WKB_eq_1}). Rearranging terms, we obtain
\beq 
\label{WKB_2_integrated}
\frac{\partial S_1}{\partial y} = - \frac{E^{(1)}}{
\partial S_0 / \partial y} 
\eeq 
that can be easily integrated:
\beq 
\label{second_WKB_integrated}
S_1(y|y_0) = - E^{(1)} \int_{y_0}^{y} \frac{dy'}{
\partial S_0 / \partial y'} = 
- E^{(1)} \int_{y_0}^{y} \frac{dy'}{
\partial V / \partial y'} 
= - E^{(1)} (t - t_0),
\eeq 
where $ t = t(y) $ as determined by its dependence on $ x = e^y $
in Eq.(\ref{instanton_solution_inv}). Further approximations of this expression reproduce the pre-exponential factor in the Kramers relation (\ref{Kramers_rate}). 

\section{Experiments}
\label{sect:Experiments}

Assuming that data correspond to a stable (quasi-stationary) market regime, the FPE equation in the $y$-variable (\ref{FPE_tilde}) can be used to estimate the model parameter using maximum likelihood 
estimation (additional constraints will be introduced below). To this end, note that the FPE (\ref{FPE_tilde}) corresponds to the following diffusion law in the 
$ y $-space (where we re-install the drivers $ {\bf z}_t $):
\beq
\label{MLE_y}
d y_t = - \frac{ \partial V(y)}{\partial y} dt + \sigma dW_t , \; \; \; V(y) \equiv - \left( \theta -  \frac{\sigma^2}{2} + {\bf w} {\bf z}_t \right) y  + \kappa e^y 
+ \frac{1}{2}  g e^{2y},
\eeq
where $ W_t $ is a standard Brownian motion.
In terms of variables $ y = \log x $, the negative log-likelihood of data is therefore (assuming a small time step $ \Delta t \rightarrow 0)$)
\beq
\label{log_like_market}
LL_M (\Theta) = - \log \prod_{t=0}^{T-1} 
\frac{1}{ \sqrt{ 2 \pi  \sigma^2 \Delta t}  } 
\exp \left\{  - \frac{1}{2 \sigma^2 \Delta t} \left(   \frac{ y_{t+ \Delta t} -   y_{t}}{ \Delta t} +  \frac{ \partial V(y)}{\partial y}   
\right)^2
\right\} , 
\eeq
where $ {\bf y}_t  = \log x_t $  now stands for observed values of log-cap. Note that because the model is Markov, the product over $ t = 0, \ldots, T-1 $ does not 
necessarily mean a product of transitions along the same trajectory. The negative log-likelihood should be minimized to estimate parameters $ 
\sigma $, $ \theta $,  $ \kappa $, $ g $ and $ {\bf w} $. 

To guide the numerical optimization, one typically imposes regularization. In our case, we use a regularization that ensures the presence of a barrier between a ``diffusive'' and ``default'' regions of stock log-cap value $ y_t $ (or equivalently of market prices). These regions would be concentrated around the metastable minimum of the effective potential and the reflection point, respectively. In such a regime, defaults via tunneling become possible, and therefore we can approximate model-implied hazard rates by the Kramers escape rate formula (\ref{Kramers_rate})\footnote{Alternatively, we could perform unconstrained optimization with model parameters that do not necessarily ensure the existence of a potential barrier. This may provide a better in-sample fit if we are only interested in fit to equity returns but not to credit spreads. Out-of-sample performance may however be worse with such approach, presumably depending on a fraction of large market moves in an out-of-sample dataset.}. 

By fitting to CDS data referencing stocks in our dataset, we get approximate calibration to two sorts of data: stock returns and CDS spreads. Note that from the point of view of calibration to equity returns, calibration to the Kramers formula with a fit to CDS spreads amounts to what can be called the ``Kramers regularization''. With this regularization, we simply compute the Kramers rate as a function of model parameters, and then add 
a penalty to the negative log-likelihood that is proportional to the squared difference of this quantity and a CDS-implied hazard rate.
The total penalty (regularization) is made of this contribution plus the penalty ensuring the presence of a barrier.

We used 7 years of daily returns and daily CDS spreads for 16 stocks that make just over a half of all names in the Dow Jones Industrial index. All results are reported for the simplest case with no signals (so that 
$ {\bf z}_t = 0 $ or equivalently $ {\bf w} = 0 $).  

We modify the loss function to fit the QED model to the observed CDS spreads. For each year of daily spread history, we estimate the sample mean $\bar{r}_{obs}$ and then add a penalty term of the form $\lambda_1(\bar{r}_{obs}-r)^2$, where $r$ is the Kramers rate. 



Furthermore, we have to enforce the condition that all transitions 
included in the likelihood are within the diffusive region. It is convenient to add this as a soft penalty for the initial value $ y_{t_0} $ for each transition to be to the right of the maximum $ y_2 $ of the potential. This can be enforced e.g. by adding a penalty 
$ \lambda_2 max(y_2 - y_{t_0} , 0 ) $, or its smoothed-out version.  Unless otherwise stated, $\lambda_2=1\times 10^5$.



To ensure positivity of the discriminant defining the roots 
$ y_{1,2} $ (and hence that they are both real-valued) and the presence of a barrier, we use the following simple numerical trick. At each iteration step, for any given previous value of 
$ \kappa $, we define a new variable 
$\kappa' = -(2\sqrt{g|\bar{\theta}|} + |\kappa|)$, which is negative by construction, as well as it automatically satisfies the condition 
$(\kappa')^2\geq 4.0g\bar{\theta}$. Using such recomputed value
$ \kappa \rightarrow \kappa' $ to reinforce negativity of $ \kappa $ at each iteration step of calibration, we eventually converge to a stable solution with both a negative $ \kappa $ and  a barrier. We use 
this trick instead of a more brute-force approach that would enforce the real-valued root and a barrier at a cost of two additional Lagrange multipliers.



\begin{figure}[ht!]
\begin{center}
\includegraphics[
width=100mm,
height=70mm]{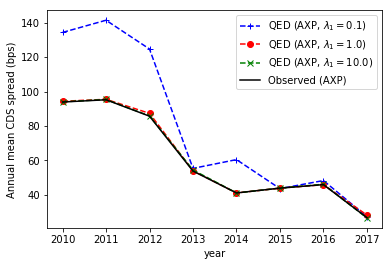}
\caption{A comparison of the observed annual mean spread (bps) of the AXP CDS (black solid) with the calibrated Kramers spread: hazard $\times~ (1-R)$. The calibrated spreads are shown for three different values of $\lambda_1$ corresponding to Table \ref{tab:gbm_vs_qed}. When $\lambda_1=10$ (green dashed), the model and observed annual mean CDS spreads are practically identical.} 
\label{fig:axp_hazard}
\end{center}
\end{figure}

Figure \ref{fig:axp_hazard} compares the annual mean spread (bps) of the AXP CDS spread with the Kramers hazard rate. 

Using the same symbol, Table \ref{tab:gbm_vs_qed} compares the maximum log likelihood function under GBM with the stock price calibrated QED model - without using the constraints described above and excluding CDS spreads. The maximum log likelihood function is also shown for the calibrated QED model with the constraints and CDS data for various values of $\lambda_1$. The maximum log likelihoods are observed to be consistently higher under the QED model (without constraints) than the GBM model. With increasing values of $\lambda_1$, we observe a general trend of decreasing log likelihood values, due to greater emphasis placed on the CDS spread history. 

For completeness, the calibrated model parameters, using both prices and CDS spreads, are reported for all symbols in 
Tables \ref{table:theta} to \ref{table:g} using $\lambda_1=10$. Additionally, the calibrated Kramers escape rate and annual CDS spread history is given in Tables \ref{table:Kramers_r} and \ref{table:CDS} respectively.  All numerical experiments are performed using \verb|TensorFlow| version 1.3.

\begin{table}[h!]

\resizebox{\columnwidth}{!}{%
\begin{tabular}{|c|c|c|c|c|c|c|c|c|}
\hline
Model &  2010 & 2011 & 2012 & 2013 & 2014 & 2015 & 2016 & 2017\\
\hline
GBM &  878.739 & 895.622 & 991.374 & 983.511 & 1007.410 & 946.753 & 1011.842 & 1008.518 \\
\hline
QED (unconstrained) & 882.301 & 900.213 & 997.0149 & 984.307 & 1008.076 & 947.164 &  1012.872 & 1010.485 \\
\hline
QED (constrained, $\lambda_1=0.1$) & 875.025 & 891.0852 & 989.552 &  977.831 &   1008.032 & 946.543 & 994.030 & 1005.150 \\
\hline
QED (constrained, $\lambda_1=1.0$) & 866.31 &879.7991 &980.2564 &961.6697 &1008.0321 &946.7632 &1010.5832 &989.9231 \\
\hline
QED (constrained, $\lambda_1=10.0$) & 838.477 & 832.561 & 850.736 & 938.622 & 1005.481 & 946.821 & 994.871 & 1004.654 \\
\hline
\end{tabular}
}
\caption{Comparison of the maximum log likelihood function under GBM and QED models for AXP price history between 2010 to 2017. Note that the second row shows the values of the maximum log likelihood function for the QED model calibrated without constraints to price history only. The bottom three rows show the maximum log likelihood functions for the QED model calibrated, with constraints, to price and CDS spread history, for different values of the regularization parameter.}
\label{tab:gbm_vs_qed}
\end{table}

\begin{table}[!ht]

\resizebox{\columnwidth}{!}{%
\begin{tabular}{|c||cccccccc|}
\hline
&2010&2011&2012&2013&2014&2015&2016&2017\\
\hline
AXP&93.844&95.2148&85.4901&53.4911&41.0538&43.8309&45.9352&26.8229\\
BA&152.9697&240.5589&232.8387&109.406&70.0338&72.3545&88.2742&54.3214\\
CAT&77.0086&85.8795&95.7828&71.005&43.4454&61.1637&79.0823&41.6516\\
CSCO&57.6397&80.3585&79.1721&43.3382&36.8152&30.7575&34.0111&27.4797\\
DIS&48.913&37.9151&29.388&22.2677&19.3259&18.4489&25.1561&30.2328\\
GS&144.1252&193.7553&241.3186&125.7899&84.7&89.1107&101.7708&71.5498\\
HD&67.5283&63.3773&49.3592&35.1921&25.4508&21.9486&26.1475&25.2557\\
IBM&39.3108&41.4169&36.4958&34.752&41.2068&48.8432&57.4105&37.2366\\
JNJ&42.302&40.0684&36.3307&23.9545&14.4517&15.2582&17.4625&17.8959\\
JPM&87.7228&104.4257&119.2014&83.2466&59.6081&71.0212&68.7511&49.8118\\
MCD&44.9489&34.8205&25.3095&18.6306&21.2441&39.4674&34.741&26.4765\\
PFE&54.2214&67.408&62.2827&35.3363&23.3582&21.4857&26.4003&28.567\\
PG&43.9813&44.7094&50.3334&34.7521&26.4204&18.1977&20.7467&20.9639\\
UNH&132.8105&100.0228&93.4989&50.3178&37.7885&28.2273&34.7879&24.5506\\
VZ&76.6715&67.0173&60.6482&65.1896&52.368&64.9342&62.631&71.1288\\
WMT&44.973&48.2212&42.4751&30.3341&17.5377&19.7762&39.3315&36.0448\\
\hline
\end{tabular}
}
\caption{The calibrated Kramers hazard (default) rate $r$ without a signal. Note that the hazard rate has been scaled here by the factor $(1-R)$, where $R=0.4$ is the recovery rate,  to facilitate a comparison with data expressed in terms of CDS spreads (in bps). 
This relation between hazard rates and credit spreads is commonly known in the industry as the ``credit triangle'' relation.
}

\label{table:Kramers_r}
\end{table}

\begin{table}[!ht]
\resizebox{\columnwidth}{!}{%
\begin{tabular}{|c||cccccccc|}
\hline
&2010&2011&2012&2013&2014&2015&2016&2017\\
\hline
AXP&93.883&95.2114&85.6554&53.7578&41.0243&43.8309&45.8682&26.6747\\
BA&153.0882&240.7485&233.2632&109.254&70.1769&72.35&88.3443&54.7185\\
CAT&77.3004&86.0846&95.8098&71.0846&43.445&61.1695&79.0182&41.6533\\
CSCO&57.6367&80.4074&79.2337&43.3009&36.6438&30.779&33.8902&27.9398\\
DIS&48.8364&37.9136&29.4179&22.3007&19.17&18.4568&25.1843&30.2362\\
GS&144.1496&193.7821&241.147&125.7654&84.6226&89.0994&101.9467&71.4695\\
HD&67.4886&63.4055&49.4462&35.2651&25.8489&21.9501&26.1376&25.1368\\
IBM&39.3434&41.366&36.4887&34.7548&41.2087&48.844&57.4953&37.2389\\
JNJ&42.3546&40.0621&36.3065&23.8851&14.8077&15.2604&17.336&17.9215\\
JPM&87.7208&104.4329&119.1014&83.2168&59.6392&71.2143&69.01&49.9987\\
MCD&44.8073&35.0409&25.3138&18.6199&21.1898&39.2849&34.6896&26.5041\\
PFE&54.2365&67.482&62.3983&35.3102&23.3617&21.5014&26.3561&28.6097\\
PG&43.9865&44.7123&50.3846&34.717&26.2986&18.182&20.763&20.9647\\
UNH&132.7989&99.9259&93.4899&50.2713&37.7979&28.2098&34.6348&24.5253\\
VZ&76.6777&66.9812&60.5316&65.1253&52.2397&64.9528&62.5775&71.2035\\
WMT&45.0123&48.1945&42.4534&30.3034&17.5621&19.7762&39.115&35.9252\\
\hline
\end{tabular}
}
\caption{Observed annual mean CDS spreads (in bps).}
\label{table:CDS}
\end{table}

\section{Summary and outlook}
\label{sect:Summary}

This paper presents the ``Quantum Equilibrium Disequilibrium'' (QED) model of a financial market, which is inspired by reinforcement learning and physics. The initial formulation of the QED model for a special case $ g = 0 $ was proposed in \cite{IHIF} based on inverse reinforcement learning (IRL) applied to the modeling of investment portfolios and a market as a whole. A generalization of the resulting dynamics to the case
 $ g > 0 $ developed in this paper using physics-motivated analyses of symmetries and analytic properties  enabled incorporating defaults and large market moves in the dynamics of stock prices.

Our model produces stock price dynamics with defaults using only one degree of freedom. This is very different from most of the existing financial models that either model stocks without defaults, or introduce additional variables such as hazard rates or credit spreads to describe defaults. Our model can be simultaneously calibrated to credit default swap (CDS) data and equity return data, despite 
having very modest data requirements with only 4 
free parameters $ \theta, \, \sigma, \, \kappa $ and $ g $ (plus the weights $ {\bf w} $, if any).

For practical applications with more stringent requirements for quality of fit to data, one might include time-dependent predictors (alpha-signals) $ {\bf z}_t $. In addition to producing calibrated weights $ \bf{w} $, calibrated model parameters
$ \theta, \, \sigma, \, \kappa $ will also be different in the presence of signals. We focused in our examples on the basic benchmark case with no signals. 

Calibration of our model to CDS data can also be viewed as a proxy to calibration to (unavailable) large market moves data. Unlike the GBM model and other models of ``small'' marker fluctuations, our non-linear model is both capable of modeling large market moves, and offers an approximate way to calibrate probabilities of such events to available CDS market data. 

With strong practical motivation, we showed in this paper non-linearities of resulting price dynamics are non-perturbative (non-analytic) in friction parameters. Consequently, the GBM model turns out to be only 
a {\it formal} mathematical limit of the QED model, or any other non-linear model where defaults arise as noise-induced transitions into an absorbing state. 

In reality, this limit is {\it non-analytic},  which shows up, depending on a parameter regime, as a divergence of perturbation theory, or as a non-equilibrium and non-perturbative noise-activated phase transition into an absorbing state, associated in our model with a corporate default.  
As a result, the QED and GBM model are {\bf not} related by a {\it physically-meaningful} smooth transition. In the language of physics, they cannot be in the same universality class - which is easy to understand as the GBM model altogether misses defaults, in the first place.

The formal limit of zero market frictions thus leads to both  
divergence of a formal perturbation theory around a zero-friction limit, and missing defaults. Such rare events become a mathematical impossibility in the GBM and other linear models, simply because they cease to exists in this formal limit.   
Similarly to tunneling phenomena in statistical physics, quantum mechanics and quantum field theory, such exponentially rare events cannot be caught at any finite order of perturbation theory around a zero-friction limit. Therefore such events are not attainable either in the GBM model or any other model that treats market frictions as a perturbation around a ``friction-free'' market. On the other hand, such market does {\it not} exist anyway, at least not as any meaningful physical limit.

To describe defaults and other large market moves, we need different tools that do {\it not} rely on perturbative expansions around such non-existing ``friction-free'' markets. 
Fortunately, such tools are readily available in physics. In this paper, we presented three inter-related, non-perturbative, physics-based approaches. 

The first approach is based on the analysis of the Fokker-Planck equation. In a one-dimensional setting, this leads to the celebrated Kramers escape rate formula as a theoretical prediction for hazard (default) rates in the QED model. Such analysis of defaults as tunneling phenomena is simple in 1D, but in higher dimensions the Kramers relation requires modifications.

Such modifications for $ D > 1 $ (multi-stock or portfolio) 
settings are feasible under two other approaches for computing the hazard rate in the QED model: the supersymmetric quantum mechanics (SUSY QM) approach, and the instanton approach. In this paper, we showed how these methods work in 1D and reproduce the Kramers formula. Multi-dimensional extensions are left for a future research.

In addition to multidimensional extensions, the model presented here can be extended along other directions. The most important one is incorporating signals $ {\bf z}_t $ into the dynamics. This requires modifications to calculations we presented above for the hazard rates. For example, if signals $ 
\bf{z}_t $ are modeled as colored (correlated) noise, this gives rise to a fluctuating barrier separating the diffusive and default regions. Such analysis will be presented elsewhere. 

In our approach, both large market moves and defaults are described 
by instanton solutions whose distinguished property is their locality in time: the instanton is well approximated by a smoothed step function, see Fig.~\ref{fig:Instanton_and_bounce}. The instanton transition happens very fast in time. Therefore, even though the noise term in a simplest version of our model can be just white noise, nonlinearity and instantons provide a mechanism explaining large market jumps without the need for introducing jump-diffusion or stochastic time changes that are sometimes introduced in mathematical finance models to include jumps in prices.

Our model does not claim to provide an accurate time-dependent (inter-day, within each year) calibration to {\it daily} CDS spreads. This is of course because without time-dependent signals $ {\bf z}_t $, the model is stationary and does not exhibit any explicit time-dependence. In addition, incorporating non-zero trading signals $ {\bf z}_t $ produces fluctuating barriers. Consequently, predictions for CDS spreads based on the 
Kramers relation will be affected by time-varying corrections that might be considerably sizable.

Another interesting problem is to model CDS spreads at different tenors. Instanton-based or SUSY QM-based approaches are only able to model the simplest term structure of CDS spreads. Namely, only a simplest CDS curve with a constant hazard rate (equal to the Kramers escape rate) can be modeled based on instantons or SUSY QM for the ground state. This is because both these methods are asymptotic in the transition time $ T \rightarrow 
\infty $. Producing non-flat CDS curves requires computing corrections to this asymptotic behavior, which is another problem that we leave for future research.

\def\thesection{A}	
\setcounter{equation}{0}
\def\theequation{\thesection.\arabic{equation}}

\section*{Appendix A: Escape from a metastable minimum}
\label{sect:Appendix_A}

Here we provide a brief summary of the derivation of the probability to escape from a metastable minimum using the Fokker-Planck equation.
For more details, see e.g. \cite{Gardiner}, Chapter 9.

Consider the FPE equation 
\beq
\label{FPE_c}
\partial_t p(y,t) = \partial_t \left[ V'(y) p(y,t) \right] + D \partial_y^2 p(y,t) \equiv - \partial_y J(y) , \; \; \; D \equiv \frac{\sigma^2}{2},
\eeq
where a potential $ V(y) $ has a metastable minimum at point $ a $, a stable minimum at point $ c $, and a local maximum between these points at point $ b $, 
and $ J(y) $ is the probability current:

%
 \beq
\label{J_curr}
J = - \frac{\partial V}{\partial y} P(y,t) - \frac{\sigma^2}{2} \frac{\partial P(y,t)}{\partial y} = -  \frac{\sigma^2}{2} e^{ - \frac{2}{\sigma^2} V} 
\frac{\partial}{\partial y} \left(  e^{  \frac{2}{\sigma^2} V}  P(y,t) \right).
\eeq
In equilibrium with a stationary distribution $ P_0(x) $, the left hand side of the FPE vanishes. The right hand side  trivially vanishes as well, if the probability current is zero. This can be achieved if the product $  e^{  \frac{2}{\sigma^2} V}  P_0(x) $ is independent of $ x $. This immediately produces the equilibrium density:
\beq
\label{eq_density}
P_0(y) = \frac{1}{Z} e^{ - \frac{2}{\sigma^2} V(y)},
\eeq
where $ Z $ is a normalization factor. When the potential $ V(y) $ is metastable, Eq.(\ref{eq_density}) leads to a diverging normalization constant $ Z $ and hence to non-existence of the stationary solution  Eq.(\ref{eq_density}).

A proper treatment for a metastable scenario is obtained as follows. First, we assume a constant non-vanishing current $ J $ throughout the system, so that we have 
\beq
\label{J_constr}
 -  \frac{\sigma^2}{2} e^{ - \frac{2}{\sigma^2} V} 
\frac{\partial}{\partial y} \left(  e^{  \frac{2}{\sigma^2} V}  P \right) =  J  \; \; 
\Leftrightarrow 
\frac{\partial}{\partial y} \left(  e^{  \frac{2}{\sigma^2} V}  P \right)  = - J \frac{2}{\sigma^2} e^{  \frac{2}{\sigma^2} V}. 
\eeq
Let's integrate this equation between points $ a $ and $ c $:
\beq
\label{integral_a_c}
\left.  e^{  \frac{2}{\sigma^2} V}  P \right|_{a}^{c} = - \frac{2 J}{\sigma^2} \int_{a}^{c} e^{  \frac{2}{\sigma^2} V(y') } dy'.
\eeq
Neglecting the term proportional to $ P(c) $ in comparison to $ P(a) $, we obtain 
\beq
\label{integral_a_c_1}
 e^{  \frac{2}{\sigma^2} V(a)}  P(y=a) =  - \frac{2 J}{\sigma^2} \int_{a}^{c} e^{  \frac{2}{\sigma^2} V(y') } dy' \; \; \; \Rightarrow 
 J = \frac{\sigma^2}{2} \frac{P(a) \exp\left[ 2 V(a) / \sigma^2 \right] }{
 \int_{a}^{c}  e^{  \frac{2}{\sigma^2} V(y') } dy'}.
 \eeq
 The quantity $ J $ gives the total probability of a particle to leave the metastable well given that it is initially placed in this well. We can therefore write $ J = p r $ where  $ p $ is the probability to be in the well, and $ r $ is the escape rate. For the former, we can use use the fact that probabilities at $ y $ and $ y = a $ are approximately related by the equilibrium relation
\beq
\label{equilib_Prob}
P(y) = P(a) \exp\left[ - \frac{ V(y) - V(a)}{T} \right], \; \; \; T \equiv \frac{\sigma^2}{2}.
\eeq
Integrating this relation in the limits $ [a - \Delta, a + \Delta] $ where $ \Delta $ is the size of the well, and expanding the integrand on the right hand side around the point $ y = a $ to perform a Gaussian integration, we obtain
\beq
\label{p}
p = \int_{a - \Delta}^{a + \Delta} P(y) dy = P(a)  \int_{a - \Delta}^{a + \Delta}  
\exp\left[  - \frac{V''(a)(y-a)^2}{2T} \right] = P(a) \sqrt{ \frac{ 2 \pi T}{ \left| V''(a) \right| } }.
\eeq
For the denominator in Eq.(\ref{integral_a_c_1}), we can expand around the peak of the potential at $ y = b $ to obtain
\beq
\label{peak_b}
 \int_{a}^{c}  e^{  \frac{2}{\sigma^2} V(y') } dy' =  \sqrt{ \frac{ 2 \pi T}{ \left| V''(b) \right| } } e^{  \frac{2}{\sigma^2} V(b) }.
\eeq
Using this together with the relation $ r = J/p $ and Eq.(\ref{p}), we finally obtain the Kramers escape rate formula:
\beq
\label{Kramers_rate_C}
r = \frac{\sqrt{ V''(a) \left| V''(b) \right| }}{2 \pi} \exp \left[ - \frac{2}{\sigma^2} (V(b) - V(a) ) \right].
\eeq
This formula applies as long as the barrier height $ \Delta E \equiv  (V(b) - V(a) \gg \frac{\sigma^2}{2} $. 

%
%

\def\thesection{B}	
\setcounter{equation}{0}
\def\theequation{\thesection.\arabic{equation}}

\section*{Appendix B: Path integrals for the Langevin dynamics}
\label{sect:Appendix_B}

Here we describe steps needed to re-formulate the Langevin dynamics in terms of path integrals. For more details, see e.g. \cite{Zinn-Justin-QFT}, 
\cite{Junker}, or \cite{Gozzi_83}.

The method of path integrals for the Langevin dynamics is based on using the following identity valid for an arbitrary function $ f(x) $ acting in a real-valued space $ \mathbb{R}^n$:
\beq
\label{identity_x}
\int dx \delta (f(x)) \left| \mbox{det} \; \frac{\partial f}{\partial x} \right| = \int dx \frac{1}{\left| \mbox{det} \; \frac{\partial f}{\partial x} \right|}  \delta (x) 
\left| \mbox{det} \; \frac{\partial f}{\partial x} \right|  = \int dx \delta(x) =  1.
\eeq
In quantum mechanics and quantum field theory, a similar identity is used for infinite-dimensional path integrals.

We introduce {\it two} conjugated (``Nicolai'') maps defined by the following pair of relations for two independent Gaussian white noise processes 
$ \xi_t^{\pm} $:
\beq
\label{Nicolai}
\xi_t^{\pm} = \frac{dy}{dt} \mp \frac{\partial V}{\partial y}.
\eeq
If we take the minus sign in this relation, we recover the Langevin equation  (\ref{Langevin_y}) which we repeat here for convenience:
\beq
\label{Langevin_y_2}
\frac{dy}{dt} = -  \frac{\partial V}{\partial y}  + \xi_t^{-} \, \; \;  \; \langle \xi_{t_1}^{-} \xi_{t_2}^{-} \rangle =  \sigma^2 \delta(t_1 - t_2). 
\eeq
On the other hand, picking the plus sign in (\ref{Nicolai}), we obtain a similar Langevin equation to (\ref{Langevin_y_2}) with the independent noise 
$ \xi_t^{+} $, and with an {\it inverted} potential, or equivalently written in the backward time $ t \rightarrow - t $.  As we will see shortly, we need both the forward and backward solutions if we want to describe periodic solutions that propagate in both directions.

Now consider a transition probability $ m_t^{\pm} (y, y_0) $ to some state $ y $ given an initial state $ y_0 $ at time $ t_0 $
under the influence of noise $ \xi_t^{\pm} $. It can be represented as an integral of a delta-function 
with respect to all realizations of noise $ \xi_t^{\pm} $, where $ D \xi_{\pm} $ stands for integration over values of $ \xi_t^{\pm} $ for all times $ t $, see e.g. \cite{Zinn-Justin-QFT} or 
\cite{Junker}:
\beq
\label{Langevin_means}
m_t^{\pm} (y, y_0) = 
\langle \delta \left( y(t) - y \right) \rangle_{\pm} = \int D \xi_{\pm} \exp \left( - \frac{1}{2 \sigma^2} \int_{t_0}^{t} \left( \xi_t^{\pm} \right)^2  dt \right) \delta \left( y(t) - y \right). 
\eeq
This is a Gaussian path integral for the noise $ \xi_{\pm}(t) $. We can convert it into a Wiener path integral for $ y_t $ 
using the Nicolai maps Eqs.(\ref{Nicolai}) as the change of measure 
$ \xi_t^{\pm} \rightarrow y(t) $. Note that this involves the Jacobian of this transformation 
\beq
\label{Jac_Nicolai}
 \mathcal{J}  \equiv \left| \mbox{det} \, \left[ \frac{\delta \xi^{\pm}(t)}{\delta y(t') } \right] \right| = 
  \left| \mbox{det} \, \left[ \left( \frac{\partial}{\partial t} \mp \frac{\partial V}{\partial y} \right) \delta( t - t') \right] \right|. 
 \eeq 
 As was shown in \cite{EK}, \cite{Junker} (see also below), the value of the determinant depends on the definition of the stochastic calculus. For the It\^{o} and Stratonovich 
 prescriptions, we obtain, respectively
 \beq
 \label{Jacob_Ito_Strat}
 \mathcal{J}   = 
 \left\{ \begin{array}{l}
\text{const} \times 1,  \; \; \; \text{It\^{o}},  \\
\text{const} \times \exp \left\{ \mp \frac{1}{2} \int_{0}^{t} d \tau \frac{ \partial^2 V(y(\tau))}{\partial y^2} \right\}, \; \; \; \text{Stratonovich}.   \\
\end{array} \right.
\eeq  
The fact that with the It\^{o} prescription, the Jacobian is equal to an unessential constant factor can be seen directly in our initial discrete-time 
equations (\ref{r_t_one_more}). When viewed ``backwards'' as a particular discretization of a continuous-time diffusion, Eq.(\ref{r_t_one_more}) correspond to 
the most conventional forward Euler discretization of an It\^{o}'s SDE. The Jacobian of transition from the noise $ \xi_t $ to the next-value price $ x_{t+1} $ is clearly a constant factor here. Such constant can be ignored as it does not impact the dynamics, and is eventually re-absorbed in a denominator of the path integral.

Now we replace the path integration wrt $ \xi_{\pm}(t) $ in the Gaussian path integral (\ref{Langevin_means}) by integration with respect to paths of the state variable 
$ y_t $ using the Nicolai maps (\ref{Nicolai}) (with either the plus or minus sign) as a definition of the mapping, and keeping the Jacobian (\ref{Jac_Nicolai}).
This yields
\bea
\label{Langevin_y_path_int}
m_t^{\pm} (y, y_0) = 
&&\hskip-0.3cm
\int D y  \mathcal{J}   \exp \left( - \frac{1}{2 \sigma^2} \int_{t_0}^{t} \left( \frac{dy}{du} \mp \frac{\partial V}{\partial y} \right)^2  du \right) \delta \left( y(t) - y \right) 
\nonumber \\
= &&\hskip-0.3cm \int_{y_0}^{y_t} Dy   \mathcal{J}  \exp \left( - \frac{1}{2 \sigma^2} \int_{t_0}^{t} \left( \frac{dy}{du} \mp \frac{\partial V}{\partial y} \right)^2  du \right) 
\\
=  &&\hskip-0.3cm \int_{y_0}^{y_t} Dy   \mathcal{J}  \exp \left[ 
- \frac{1}{2 \sigma^2} \int_{t_0}^{t} 
\left( \left(\frac{dy}{du} \right)^2 
+ \left( \frac{\partial V}{\partial y} \right)^2 
\right) du 
\pm  \frac{1}{ \sigma^2}  \int_{t_0}^{t}  \frac{\partial V}{\partial y} \frac{dy}{du} du 
\right]. \nonumber 
\eea
The last integral entering this expression requires a proper interpretation, as its value depends on whether we use It\^{o}'s or Stratonovich's calculus \cite{Junker}:
\bea
\label{surface_term}
\int_{t_0}^{t}  \frac{\partial V}{\partial y} \frac{dy}{du} du 
= 
 \left\{ \begin{array}{l}
  \left[ V(y_t) - V(y_0) \right] - \frac{1}{2} \sigma^2 \int_{0}^{t} d \tau \frac{\partial^2 V}{\partial y^2},  \; \; \; \text{It\^{o}},  \\
 \left[ V(y_t) - V (y_0) \right],   \; \; \; \text{Stratonovich}.   \\
\end{array} \right.
\eea  
If we now substitute this expression along with the Jacobian (\ref{Jacob_Ito_Strat}) into Eq.(\ref{Langevin_y_path_int}), we obtain a final Lagrangian path integral representation for the transition probability. Importantly, this final representation is {\it independent} of the rule of stochastic calculus:
\beq
\label{Langevin_PI}
m_t^{\pm} (y, y_0) =  e^{ \pm \frac{ V(y_t) - V(y_0)}{\sigma^2}} \times
 \int_{y_0}^{y_t} Dy  \, 
 e^{ - \frac{1}{\sigma^2} \int_{t_0}^{t} \mathcal{L}_{\pm} du }
 = 
 e^{ \pm \frac{ V(y_t) - V(y_0)}{\sigma^2}} \times
 \int_{y_0}^{y_t} Dy  \, 
 e^{ - \frac{S_{\pm}}{\sigma^2} },
\eeq
where $ \mathcal{S}_{\pm} = \int \mathcal{L}_{\pm} dt $ is the Euclidean action with the Euclidean Lagrangian 
\beq
\label{Lagrangians}
\mathcal{L}_{\pm} =  \frac{1}{2} \left(\frac{dy}{dt} \right)^2 
+ \frac{1}{2} \left( \frac{\partial V}{\partial y} \right)^2 
\pm  \frac{\sigma^2}{2} \frac{\partial^2 V}{\partial y^2}.
\eeq
Therefore, once we obtained the path integral (\ref{Langevin_PI}) using It\^{o}'s calculus, from that point on we can assume the Stratonovich rules in for integrals
appearing in what follows within the path integral formulation of stochastic dynamics. This is convenient because it allows one to use conventional rules of calculus such as e.g. integration by parts.

The path integral entering Eq.(\ref{Langevin_PI}) is equivalent to an Euclidean quantum mechanical path integral with the Planck constant $  \sigma^2 $.
We recall that 
in quantum mechanics, an Euclidean Lagrangian in imaginary time $ \tau $ is obtained by making the so-called Wick rotation from the physical time $ t $ to 
the imaginary time $ \tau \equiv  i t $ in the action:
\beq
\label{QM_Euclead}
i \frac{S}{\sigma^2} = i \frac{1}{\sigma^2} \int_{0}^{t} du \left[ \frac{1}{2} \left( \frac{dy}{du} \right)^2 - U(y) \right] \Rightarrow
 - \frac{1}{\sigma^2} \int_{0}^{it} d \tau \left[ 
\frac{1}{2} \left( \frac{dy}{d \tau} \right)^2 + U(y) \right].
\eeq
As a result, the Euclidean Lagrangian is given by the kinetic energy {\it plus} the potential energy, see e.g. \cite{Weinberg_book} or 
\cite{Zinn-Justin-QFT}. In our case, a pair of Euclidean Lagrangians corresponding, respectively, to the transition probabilities $ m_t^{\pm} (y, y_0) $, are 
given by Eq.(\ref{Lagrangians}).

 We can also construct a corresponding Euclidean Hamiltonian from the Euclidean Lagrangian (\ref{Langevin_PI}) by making the Legendre transform from 
 the velocity $ \dot{y} $  
 to the canonical momentum $ p $:
 \beq
 \label{HamiltonFP}
 \mathcal{H}_{\pm} = - p \frac{dy}{dt} + \mathcal{L}, \; \; \; p \equiv \frac{\partial \mathcal{L}}{\partial \dot{y}}.
 \eeq
 For the Lagrangian (\ref{Langevin_PI}), this produces
 \beq
 \label{Hamiltonian_FP_QED}
  \mathcal{H}_{\pm} =  - \frac{1}{2} p^2 +  \frac{1}{2}\left( \frac{\partial V}{\partial y} \right)^2 
\pm \frac{\sigma^2}{2} \frac{\partial^2 V}{\partial y^2}. 
 \eeq
 Note the peculiar properties of these Euclidean Hamiltonians. While the last two terms 
 in $ \mathcal{H}_{\pm} $ can be identified with a potential energy term $ U_{\pm} $, the first term is a kinetic energy. In real-time classical mechanics,
 the kinetic energy of a particle of mass $ m $ and velocity $ v = \dot{y} $ is $ \frac{1}{2} m v^2 $,
 However,  in the Euclidean Hamiltonians  $ \mathcal{H}_{\pm} $, the coefficient in from of this term is {\it negative} rather than positive. This has important implications, in particular, it lets us to describe noise-induced tunneling as a classical motion in Euclidean time with zero energy.
  
 The ``quantum-mechanical'' Fokker-Planck Hamiltonians are obtained from these classical Hamiltonians by the regular quantum mechanical 
 replacement of the canonical momentum by a derivative $ p \rightarrow - \sigma^2 \partial / \partial y $. This gives a pair of FP Hamiltonians
 \beq
 \label{Hamilton_FP}
  \mathcal{H}_{\pm}^{(FP)} =  - \frac{\sigma^4}{2} \frac{\partial^2}{\partial y^2} +  \frac{1}{2}\left( \frac{\partial V}{\partial y} \right)^2 
\pm \frac{\sigma^2}{2} \frac{\partial^2 V}{\partial y^2}. 
 \eeq
The Fokker-Planck Hamiltonians (\ref{Hamilton_FP}) can also be obtained directly from the Fokker-Plank equations (FPE) for the transition probabilities 
$ m_t^{\pm} (y, y_0) $ \cite{Zinn-Justin-QFT, Junker}:
\beq
\label{trans_probs_FPE}
\frac{\partial}{\partial t} m_t^{\pm} (y, y_0) = \mp \frac{\partial}{\partial y} \left[ \frac{\partial V}{\partial y} m_t^{\pm} (y, y_0) \right] + 
\frac{1}{2} \sigma^2 \frac{\partial^2}{\partial y^2} m_t^{\pm} (y, y_0). 
\eeq
Note that the conventional FPE corresponds to the equation for $  m_t^{-} (y, y_0) $. The second equation for  $  m_t^{+} (y, y_0) $ corresponds to the forward dynamics in an inverted potential, or equivalently the backward in time dynamics with the original potential. Define the change of the dependent variable in Eq.(\ref{trans_probs_FPE}) as follows:
\beq
\label{change_variables_FPE}
 m_t^{\pm} (y, y_0)  = e^{ \pm \frac{ V(y_t) - V(y_0)}{\sigma^2}} K_{\pm} ( y, t | y_0).
 \eeq 
Substituting this ansatz in the FPE (\ref{trans_probs_FPE}), the latter is transformed into the Schr{\"o}dinger equation in imaginary time with the Planck constant $ 
\hbar = \sigma^2 $ and the Hamiltonian $ \mathcal{H}_{\pm}^{(FP)}  $ defined in 
Eq.(\ref{Hamilton_FP}):
\beq
\label{SE_imaginary_time}
- \sigma^2 \frac{\partial}{\partial t} K_{\pm}  ( y, t | y_0) =  \mathcal{H}_{\pm}^{(FP)}   K_{\pm}  ( y, t | y_0) , \; \; \;  K_{\pm}  ( y, t_0 | y_0) = \delta(y - y_0 ). 
\eeq
Using position eigenstates of the Hamiltonian $  \mathcal{H}_{-}^{(FP)} $ that we denote as $ | y_t \rangle $,  the formal solution of this imaginary-time Schr{\"o}dinger equation is given by the heat kernel
\beq
\label{heat_kernel_A}
  K_{\pm}  ( y, t | y_0) = \langle y | \exp \left( - \frac{1}{\sigma^2} t \mathcal{H}_{\pm}^{(FP)} / \sigma^2 \right) | y_0 \rangle.
  \eeq
Combining Eqs.(\ref{heat_kernel_A}) and (\ref{Langevin_PI}) which produces the Euclidean path integral representation of transition probability
\begin{equation}
\label{path_integral_Euclid}
 \langle y | \exp \left( -  t \mathcal{H}_{\pm}^{(FP)} / \sigma^2 \right) | y_0 \rangle = 
 \int_{y_0}^{y_t} Dy  \, 
 e^{ - \frac{1}{\sigma^2}  \int \mathcal{L}_{\pm} dt },
\end{equation} 
 where the Euclidean Lagrangians $ \mathcal{L}_{\pm} $ are defined in Eq.(\ref{Lagrangians}).
 
Transition probabilities such as (\ref{path_integral_Euclid}) are conventionally obtained in the path integral method by adding a source term $ h(t) $ in the 
exponent, and defining the generating functional $ Z[h] $ as follows:
\beq
\label{Gener_Z_A}
Z[h]  = \int D y \,\exp\left( - \frac{1}{\sigma^2}  \int \mathcal{L}_{\pm} dt 
+ \int h(t) y(t) dt \right).  
\eeq
In particular, the propagator $ \langle y(t_1) y(t_2) \rangle $ can be obtained by twice differentiating $ Z[h] $ with respect to $ h(t_1) $ and 
$ h(t_2) $, and then setting $ h(t) = 0 $ in the final expression, see e.g. 
\cite{Zinn-Justin-QFT}.

\subsection{Tunneling and instantons in quantum mechanics}
  
A convenient formulation of quantum mechanics in general, and tunneling phenomena in particular, is provided by the Feynman path integral method \cite{Feynman_QM}. Feynman's path integral version of quantum mechanics is  
formulated in terms of classical paths. Weights of such paths are complex-valued, and are given by $ \exp(iS/\hbar) $ where 
$ i $ is the square root of $ - 1 $, $ \hbar $ is the Planck constant, and 
$ S $ is the action on a path
\beq
\label{Hamilton}
S = \int_{y_0}^{y_f}  \mathcal{L}(y, \dot{y}) d \tau =  \int_{y_0}^{y_f}  \left[  \frac{1}{2} \left( \frac{dy}{d \tau} \right)^2 - V(y) \right] d \tau. 
\eeq
 
Consider the classical action  $ S $ multiplied by $ i $ starting from the time $ \tau $ and making the Euclidean rotation (also known as the Wick rotation)
\beq
\label{Euclead_rotation}
iS = i \int_{\tau_0}^{\tau_1} \left[ \frac{1}{2} \left(\frac{d y}{d \tau} \right)^2 - V_{-} (y) \right] d \tau \xrightarrow[ \tau = - i t ]{}  - \int_{t_0}^{t_1} \left[ 
 \frac{1}{2} \left(\frac{d y}{d t} \right)^2 + V_{-} (y) \right] d t \equiv - S_E.
\eeq   
Minimization of the Euclidean action $ S_E $ is done by solving the classical Lagrange equation 
\beq
\label{Lagrange_eq_2}
\frac{d}{d \tau} \frac{\partial \mathcal{L}}{\partial \dot{y}} + \frac{\partial \mathcal{L} }{\partial y} = 0 \; \; \; \Rightarrow \; \; \;  \frac{d^2 y}{d \tau^2} =  \frac{\partial V}{\partial y}.
\eeq
The key observation is that the potential $ V(y) $ got flipped as a result of the Euclidean (Wick) rotation \cite{Weinberg_book}. What was a classically forbidden region in the original time $ \tau $ becomes a ``classically'' feasible region in the Euclidean time $ t $, as now the potential is turned upside down, and what was a forbidden region now becomes a ``classically'' allowed region.

\subsection{Path integral with the response field}  
 
Instead of the Euclidean Lagrangian path integral, we can introduce an equivalent phase-space form by introducing the so-called Habbard-Stratonovich
transform that expresses the quadratic term in the action 
in the path integral (\ref{Langevin_y_path_int}) as an integral over an auxiliary (real-valued) field $ \hat{p} $    
also known as a response or Martin-Siggia-Rose (MSR) 
field  (we omit an unessential constant factor in front of this integral, as it will be re-absorbed into the overall normalization factor $ \mathcal{N} $ of a 
path integral, see below):
\bea
\label{Langevin_y_path_int_2}
m_t^{\pm} (y, y_0) 
= &&\hskip-0.3cm \int_{y_0}^{y_t} Dy   \mathcal{J}  \exp \left( - \frac{1}{2 \sigma^2} \int_{t_0}^{t} \left( \frac{dy}{du} \mp \frac{\partial V}{\partial y} \right)^2  du \right) 
\nonumber \\
=  &&\hskip-0.3cm \int_{y_0}^{y_t} Dy \, D \hat{p}  \mathcal{J}
\exp \left[ - \int_{t_0}^{t} \left(  \frac{\sigma^2}{2} \hat{p}^2 +   i \hat{p} \left( \frac{dy}{du} \mp   \frac{\partial V}{\partial y} \right)  \right)   du \right].
\eea
In this formulation, we use the Stratonovich rules of stochastic calculus. The Jacobian $ \mathcal{J} $ is given by the determinant of differential operator $ M $ obtained by differentiating the Langevin equation with respect to 
$ y(t') $:
\bea
\label{M_det}
 \mathcal{J} && \hskip-0.4cm = \mbox{det} \; M \equiv \mbox{det} \; \left[ \left\{ \frac{\partial}{\partial t} + \frac{\partial^2 V}{\partial y(t) \partial y(t')} 
 \right\} \delta(t- t') \right]
=
 \exp \left[ \mbox{Tr} \; \log \left[ \left\{  \frac{\partial}{\partial t} + \frac{\partial^2 V}{\partial y(t) \partial y(t')}  \right\} \delta(t- t') \right]  \right]
 \nonumber \\
&&  \hskip-0.4cm =   
 \exp \left[ \mbox{Tr} \; \log \left[  \partial_{t}  \left\{  \delta (t - t') +  \partial_{t}^{-1}  \frac{\partial^2 V}{\partial y(t) \partial y(t')} \right\}   \right] \right].
\eea
Here we used the identity $ \log \mbox{Det} \; A = \mbox{Tr} \; \log A $, and $ \partial_{t}^{-1} $ stands for the Green's function $ G(t - t') $ that satisfies 
\beq
\label{Green_fun}
\partial_t G (t - t') = \delta(t-t').
\eeq
The solutions of this equation are $  G (t - t') = \theta( t - t') $ if we choose propagation forward in time, or $  G (t - t') = - \theta( t' - t) $ for propagation backward in time \cite{Gozzi_83}. If we choose propagation forward in time, we obtain
\bea
\label{Jacob_forward}
 \mathcal{J}  && \hskip-0.4cm =  \exp \left[ \mbox{Tr} \left\{ \log  \partial_{t}  + \log \left[ \delta(t - t') + \theta(t-t')  \frac{\partial^2 V}{\partial y(t) \partial y(t')} \right] \right\} \right]
  \nonumber \\
&& \hskip-0.4cm =   \exp \left[ \mbox{Tr} \log  \partial_{t}  \right] \, \exp \left[ \mbox{Tr} \log  
 \left\{ \log \left[ \delta(t - t') + \theta(t-t')  \frac{\partial^2 V}{\partial y(t) \partial y(t')} \right] \right\} \right].
 \eea
 Here the first term can be dropped, as it cancels out a similar term in an overall normalization factor  in Eq.(\ref{Langevin_y_path_int_2}). On the other hand, the second term can be evaluated using the Taylor expansion of the logarithm:
 \bea
 \label{expans_log}
 \mathcal{J}  && \hskip-0.4cm =   \exp \left[ \mbox{Tr}   
 \left\{ \theta(t-t')  \frac{\partial^2 V}{\partial y(t) \partial y(t')}  + 
 \theta(t-t')  \theta(t'-t)   \frac{\partial^2 V}{\partial y(t) \partial y(t')}  \frac{\partial^2 V}{\partial y(t') \partial y(t)} + \cdots \right\} \right]  
 \nonumber \\ 
&& \hskip-0.4cm =  
  \exp \left[  \int d t \theta(0)  \frac{\partial^2 V}{\partial y^2(t)} + \int d t' 
   \theta(t-t')  \theta(t'-t)   \frac{\partial^2 V}{\partial y(t) \partial y(t')}  \frac{\partial^2 V}{\partial y(t') \partial y(t)} + \cdots 
   \right].
 \eea 
The second term and all the subsequent terms here vanish because $ \theta(t-t') \theta(t'-t) = 0 $, therefore only the first term survives. Choosing $ \theta(0) = 
\frac{1}{2} $, one has \cite{Gozzi_83} (see also Sect.4.8.2 in \cite{Zinn-Justin-QFT})
\beq
\label{Jacobian_A}
\mathcal{J} = \mbox{det} \; M = \exp\left( \frac{1}{2} \int \frac{\partial^2 V}{\partial y^2} dt \right).
\eeq
On the other hand, if we choose backward propagation in Eq.(\ref{Green_fun}), the Jacobian (\ref{Jacobian_A}) gets a negative sign in the exponent.
Respectively, with either of this choices, we obtain a pair of a forward and backward 
Euclidean Lagrangians  $ \mathcal{L} $ and $ \mathcal{L}_{B} $ with actions $  \mathcal{A}(y, \hat{p} ) $ 
 and  $ \mathcal{A}_B(y, \hat{p} ) $:
\beq
\label{Lagrangian_MSR}
 \mathcal{A}_{(B)}(y, \hat{p} ) = \frac{1}{\sigma^2} \int   \mathcal{L}_{(B)}  dt, \, \; \; \; 
\mathcal{L}_{(B)} =  \frac{\sigma^4}{2} \hat{p}^2  +  i \hat{p} \sigma^2 \left( \frac{d y}{dt}  \mp  \frac{d V}{dy} \right) \mp \frac{1}{2}  \sigma^2 \frac{\partial^2 V}{\partial y^2}.
\eeq
To the leading order for the forward propagator $ 
m_t^{-} (y, y_0) $, the last term in Eq.(\ref{Lagrangian_MSR}) originating from the Jacobian can be neglected, producing the following Lagrangian:
\beq
\label{Lagrangian_MSR_2}
\mathcal{L} =  \frac{\sigma^4}{2} \hat{p}^2  +  i \hat{p} \sigma^2 \left( \frac{d y}{dt}  + \frac{d V}{dy} \right). 
\eeq
As an alternative to the Lagrangian mechanics, one can change variables from $ y,  \frac{d y}{dt} $ to Hamiltonian mechanics using the Legendre transform from
the velocity $ \dot{y} = dy/dt $ to the momentum 
\beq
\label{can_mom_Langevin}
p = \frac{\partial \mathcal{L}}{\partial \dot{y}} =  i \sigma^2 \hat{p}.
\eeq
This defines the classical Euclidean Hamiltonian 
\beq
\label{Hamiltonian_cl}
\mathcal{H} = - p \dot{y} + \mathcal{L} \left( y, \dot{y}, \hat{p} \right).
\eeq 
In our case with the specific Lagrangian (\ref{Lagrange_MSR}), we obtain the 
Hamiltonian 
\beq
\label{H_QED}
\mathcal{H} 
=   \frac{\sigma^4}{2} \hat{p}^2  +  i \sigma^2 \hat{p} \frac{d V}{dy} = - \frac{1}{2} p^2 + p \frac{d V}{dy}.
\eeq
The Hamilton form of dynamics is given by two first-order PDEs for $ y $ and $ p  $, instead of one second-order PDE of the Lagrangian formulation 
(\ref{Lagrange_equation}) \cite{Landau_mechanics}:
\beq
\label{Hamilton_eq}
\dot{y} = \frac{\partial  \mathcal{H} }{\partial p} , \; \; \; \dot{p} = - \frac{\partial  \mathcal{H} }{\partial y}.
\eeq
This of course produces the same equations of motion (\ref{classical_motion}) if we substitute here the QED Hamiltonian (\ref{H_QED}) with the canonical 
momentum (\ref{can_mom_Langevin}). 


\subsection{SUSY path integral}
\label{sect:Appendix_A_SUSY}

 As was discussed above, while the path integral representation (\ref{Langevin_y_path_int})
is independent of the rules of stochastic calculus (see
Eq.(\ref{Langevin_PI})), our derivation above relied on the It\^{o} calculus, under which the Jacobian $ \mathcal{J} $ is trivial, 
while the last term in Eq.(\ref{Langevin_y_path_int}) is given by the first expression in Eq.(\ref{surface_term}).

As the whole expression is independent of the stochastic calculus rules, we can instead use the Stratonovich rule, under which the 
Jacobian $ \mathcal{J} $ is non-trivial. We can now 
express this determinant as an integral over dynamic Grassmann (fermion) variables $ \psi_t , \, \bar{\psi}_t $ using the Grassmann integration rules (\ref{Grassmann_integration}). 

Unlike ordinary (boson) variables that commute, i.e. $ [y_1, y_2]
\equiv y_1 y_2 - y_2 y_1 = 0 $, Grassmann (fermion) variables
 $ \psi, \bar{\psi} $
{\it anti-commute}, i.e. $ \{ \psi_1, \psi_2 \} \equiv 
\psi_1 \psi_2 
+ \psi_2 \psi_1  = 0 $ (which means, in particular that they are 
{\it nilpotent}, i.e. $ \psi^2 = 0 $).  
Grassmann variables satisfy the following integration rules:
\beq 
\label{Grassmann_integration}
\int d \psi = 0, \; \; \int d \psi \, \psi  = 1, \; \; 
\int d \psi_1 d \bar{\psi}_1 \ldots d \psi_n d \bar{\psi}_n 
\exp \left\{ \sum_{i,j=1}^{n} \bar{\psi}_i A_{ij} \psi_j  \right\}
= \text{det} \left[ A_{ij} \right].
\eeq
These rules can be used in order to express the Jacobian 
$ \mathcal{J} $ in terms of an integral over Grassmann variables.
To this end, we introduce two dynamic Grassmann fields 
$ \psi $ and $ \bar{\psi} $, that satisfy the following anti-commuting relations:
\beq 
\label{anti_commuting}
\{ \psi(t), \psi(t) \} = \{ \bar{\psi}(t), \bar{\psi}(t) \} = 0, \; \; \; \{ \psi(t), \bar{\psi}(t) \} = 1.
\eeq 
This produces an 
equivalent representation of transition probabilities 
(\ref{Langevin_y_path_int}) in terms of a path integral that involves both the original (boson) variables $ y_t $ and Grassmann variables 
 $ \psi , \, \bar{\psi} $:
\beq
\label{Langevin_PI_SUSY}
m_t^{\pm} (y, y_0) =  e^{ \pm \frac{ V(y_t) - V(y_0)}{\sigma^2}} \times
 \int_{y_0}^{y_t} Dy D \psi D \bar{\psi} \, 
 e^{ - \frac{1}{\sigma^2} \int_{t_0}^{t} \mathcal{L}_{\pm} du }.
\eeq
where 
\beq
\label{Lagrangians_SUSY}
\mathcal{L}_{\pm} =  \frac{1}{2} \left(\frac{dy}{dt} \right)^2 
+ \frac{1}{2} \left( \frac{\partial V}{\partial y} \right)^2 
-  \frac{\sigma^2}{2}
\psi^{T} \left( \frac{d}{dt} \mp \frac{\partial^2 V}{\partial y^2}
\right) \psi.
\eeq 
This Lagrangian is invariant under the SUSY transformations
\beq 
\label{SUSY_transforms}
\delta y = \bar{\varepsilon} \psi - \bar{\psi} \varepsilon, \; \; 
\delta \psi = \varepsilon \left( \dot{y} \pm \frac{\partial V}{
\partial y} \right), \; \; 
\delta \bar{\psi} =  \left( \dot{y} \mp \frac{\partial V}{
\partial y} \right) \bar{\varepsilon}.
\eeq 
where $ \varepsilon, \bar{\varepsilon} $ are infinitesimal 
Grassmann parameters of the SUSY transformation.

We can relate the SUSY Lagrangians (\ref{Lagrangians_SUSY}) to 
the Fokker-Planck Hamiltonians (\ref{Hamilton_FP}) by introducing 
a matrix-valued representation of anti-commutation relations 
(\ref{anti_commuting}). This can be done by the following choice 
for $ \psi, \,  \bar{\psi} $:
\beq 
\label{matrix_representation_fermions}
\psi =  \left[ \begin{array}{ll}
  0 & 0   \\
  1 & 0 \\
\end{array} \right] \equiv \sigma_{-}, \; \; \; 
\bar{\psi} =  \left[ \begin{array}{ll}
  0 & 1   \\
  0 & 0 \\
\end{array} \right] \equiv \sigma_{+}
\eeq 
Using this in Eq.(\ref{Lagrangians_SUSY}) and making the Legendre transform to 
proceed from the Lagrangian to a Hamiltonian, we obtain the SUSY Hamiltonian $ \mathcal{H} $:
\bea
\label{matrix_H_A}
\mathcal{H} = 
  - \frac{\sigma^4}{2} \frac{\partial^2}{
\partial y^2} + \frac{1}{2} \left(\frac{\partial V}{\partial y}
\right)^2 + \frac{\sigma^2}{2} \sigma_3 \frac{\partial^2 V}{\partial y^2} = 
\left[ \begin{array}{ll}
  \mathcal{H}_{+} & 0   \\
  0 & \mathcal{H}_{-} \\
\end{array} \right], \; \; \;
\sigma_3 = \left[ \begin{array}{ll}
  1 & 0   \\
  0 & -1 \\
\end{array} \right], where.
\eea
$ \sigma_3 $ is the Pauli matrix.  This is the Hamiltonian of the Euclidean supersymmetric (SUSY) quantum mechanics of Witten
\cite{Witten_SUSY}, see also below in Appendix D.  
The two FP Hamiltonians (\ref{Hamilton_FP}) describing the 
forward and backward dynamics are therefore combined together within the SUSY approach.
Fore more details on supersymmetric path integrals, see e.g. \cite{Junker}.

\setcounter{equation}{0}
\def\thesection{C}	
\def\theequation{\thesection.\arabic{equation}}

\section*{Appendix C: The Fokker-Planck equation and non-analyticity}
\label{sect:Appendix_C}


The partition function $ Z  $ for the FPE equation is given by the normalization factor in the steady state solution (\ref{steady_state_Str}):
\beq
\label{Z}
Z = \int_{0}^{\infty}  \exp \left[ - U(x) \right] dx =  \int_{0}^{\infty}  x^{ \frac{ 2\theta}{\sigma^2} - \nu} \exp \left\{ 
- \frac{2}{\sigma^2} \left[  \kappa x  + \frac{g}{2} x^2  \right] \right\} dx. 
\eeq
When $ \mbox{Re} \left( \frac{ 2\theta}{\sigma^2} + 1 - \nu \right)  > 0 $ and $  \mbox{Re} \left( g/\sigma^2 \right) > 0 $, this integral is known analytically, see 
\cite{Gradshtein}, Eq.(3.462)):
\beq
\label{part_fun}
Z =   \left( \frac{2g}{\sigma^2} \right)^{-\frac{z}{2}}
\Gamma \left( z \right) e^{ \frac{\kappa^2}{2 g \sigma^2}} 
D_{- z} \left( \frac{\kappa}{\sigma^2}  \sqrt{ \frac{2 \sigma^2}{g}} \right) , \; \; \; z \equiv \frac{2\theta}{\sigma^2} + 1 - \nu, 
\eeq
where $ D_{-z}(\cdot) $ stands for a parabolic cylinder function (\cite{Gradshtein}, Eq.(9.240)) 
\beq
\label{cylinder}
D_{-z}(x) = 2^{\frac{1}{4} - \frac{z}{2}} W_{ \frac{1}{4} - \frac{z}{2}, - \frac{1}{4}} \left( \frac{x^2}{2} \right) 
= 2^{\frac{-z}{2}} e^{ - \frac{x^2}{2} }
\left[ \frac{\sqrt{\pi}}{\Gamma \left( \frac{1+z}{2} \right)} \Phi \left(  \frac{z}{2}, \frac{1}{2}; \frac{x^2}{2} \right)
-  \frac{\sqrt{2 \pi} x}{\Gamma \left( \frac{z}{2} \right)} \Phi \left( \frac{1+z}{2}, \frac{3}{2}; \frac{x^2}{2} \right)
\right].
\eeq
Here $ W_{p,q}(x) $ is a Whittaker function, and $ \Phi(\alpha, \gamma, x) $ is a 
confluent hypergeometric function, also 
known as the Kummer function\footnote{Other notations 
for this function are $ M(a,b,x)$ and $ _{1}F_{1}(a;b;x) $, see \cite{Gradshtein}.}, 
which is defined as a sum of the following infinite series:
\beq
\label{Hypergeom}
\Phi(\alpha, \gamma; x) = 1 + \frac{\alpha}{\gamma} \frac{x}{1!} + \frac{\alpha (\alpha + 1)}{\gamma (\gamma +1) } \frac{x^2}{2!} + 
  \frac{\alpha (\alpha + 1)(\alpha + 2)}{\gamma (\gamma +1)(\gamma+2) } \frac{x^3}{3!} + \ldots.
\eeq 
Importantly, the Kummer function $ \Phi(\alpha, \gamma, x) $ is 
an {\it analytic} (holomorphic) function of $ x $,  therefore the only singularities of the partition function (\ref{part_fun}) in a complex plane of $ \theta $ are due to singularities 
of gamma functions entering Eqs.(\ref{part_fun}) and (\ref{cylinder}).

It is known that for real values of $ z $, the parabolic cylinder function $ D_{z}(x) $ has $ [ z + 1 ] $ real-valued roots, where $ [ z+1 ] $ stands for a largest natural number that is less than $ z + 1 $ if it exists, and zero if it does not \cite{BE}.  According to the Lee-Yang theory of phase transitions, physically observed phase 
transitions correspond to accumulation of non-analytic points of the free energy $ F \equiv \log Z $ on a real positive axis for parameter $ z $ used as an external control parameter driving the phase transition  (see e.g. \cite{Itzykson_Druffe}, Sect. 3.2). 

These non-analytic points of the free energy $ F $ correspond to zeros of the partition function  $ Z $. The latter can have zeros on a real positive axis of 
parameter $ \phi $ due to the presence of the  parabolic cylinder function $ D_{-z}(x) $  in Eq.(\ref{part_fun}).  It should have $ [-z+1] $ real valued zero, therefore
the first zero arises when $ -z + 1 \geq 1 $, or $ z \leq 0 $. Recalling the definition of $ z = \frac{2\theta}{\sigma^2} + 1 - \nu  $ in Eq.(\ref{part_fun}), we conclude that
a phase transition according to the Lee-Yang theory should happen for the following values of parameter $ \theta = r_f - c + \phi $:
\beq
\label{phase_trans_LY}
 \frac{2\theta}{\sigma^2} \leq \nu - 1  \; \; \;  \Leftrightarrow \phi \leq  \left(\nu - 1  \right) \frac{\sigma^2}{2} + c - r_f. 
 \eeq
  


 \def\thesection{D}	
\setcounter{equation}{0}
\def\theequation{\thesection.\arabic{equation}}

\section*{Appendix D: Supersymmetry in the Fokker-Planck dynamics}
\label{sect:Appendix_SUSY}

\subsection{Supersymmetry (SUSY)}
\label{sect:SUSY}

The Schr{\"o}dinger equation (\ref{SE}) for the FPE possesses hidden supersymmetry (SUSY) \cite{Brown}, \cite{Marchesoni}, \cite{vanKampen} that makes it mathematically identical to supersymmetric quantum mechanics (SUSY QM) of Witten \cite{Witten_SUSY}. 

Supersymmetry of the problem is rooted in the fact that the Hamiltonians (\ref{Hamiltonian}) can be factorized into two 
first-order operators as follows:
\beq
\label{factorization}
\mathcal{H}_{-} = \mathcal{A}^{+} \mathcal{A}, \; \; \; 
\mathcal{H}_{+} = \mathcal{A} \mathcal{A}^{+}
\eeq
where
\beq
\label{A}
\mathcal{A} = \frac{1}{\sqrt{2}} \left[ \sigma^2 \frac{\partial}{\partial y} +  \frac{\partial V}{\partial y} \right] , \; \; \; 
\mathcal{A}^{+} = \frac{1}{\sqrt{2}} \left[ - \sigma^2 \frac{\partial}{\partial y} +  \frac{\partial V}{\partial y} \right].  
\eeq 
Note that the Hamiltonian $ \mathcal{H}_{-} $ transforms into 
the partner Hamiltonian $ \mathcal{H}_{+} $ if we flip the sign of 
the potential $ V(y) \rightarrow - V(y) $. They can be paired in the following
matrix-valued Hamiltonian:
\bea
\label{matrix_H}
\mathcal{H} = 
 \left[ \begin{array}{cc}
  \mathcal{H}_{+} & 0   \\
  0 & \mathcal{H}_{-} \\
\end{array} \right]
= 
 \left[ \begin{array}{cc}
  \mathcal{A} \mathcal{A}^{+}  & 0   \\
  0 & \mathcal{A}^{+} \mathcal{A} \\
\end{array} \right].
\eea
This is the Hamiltonian of the Euclidean supersymmetric quantum mechanics of Witten
\cite{Witten_SUSY}. It can also be represented in a form that involves fermion (anti-commuting) fields $ \psi_t, \psi_t^{+} $, in addition to the conventional boson (i.e., commuting) field 
$ y_t $, see Appendix \ref{sect:Appendix_A_SUSY}. Alternatively, two purely boson Hamiltonians 
$ \mathcal{H}_{-} = \mathcal{A}^{+} \mathcal{A} $ and 
$ \mathcal{H}_{+} = \mathcal{A} \mathcal{A}^{+} $ can be thought of as representing two different fermion sectors of the model. The potential
$ V(y) $ is referred to in the context of SUSY models as the superpotential. For a brief review of SUSY quantum mechanics, see Appendix \ref{sect:Appendix_A_SUSY}, while a more complete 
treatment can be found e.g. in \cite{Junker}.

Instead of representation in terms of operators $ \mathcal{A}, \, \mathcal{A}^{+} $,
we can equivalently express the Hamiltonian (\ref{matrix_H}) in terms of supercharges 
\bea
\mathcal{Q}_1 = 
\frac{1}{\sqrt{2}}\left[ \begin{array}{cc}
    0 & \mathcal{A}    \\
   \mathcal{A}^{+} & 0  \\
\end{array} \right], \; \; \; 
\mathcal{Q}_2 = 
\frac{i}{\sqrt{2}}\left[ \begin{array}{cc}
    0 & -\mathcal{A}    \\
   \mathcal{A}^{+} & 0  \\
\end{array} \right]
\eea
This gives
\beq
\label{H_Q12}
\mathcal{H} = 2 \mathcal{Q}_1^2 = 2 \mathcal{Q}_2^2 
= \mathcal{Q}_1^2 +  \mathcal{Q}_2^2.
\eeq
This means that the Hamiltonian $ \mathcal{H} $ commutes with both 
supercharges, i.e. $ [\mathcal{H},\mathcal{Q}_1] \equiv \mathcal{H} \mathcal{Q}_1
- \mathcal{Q}_1 \mathcal{H}  = 0 $, and 
$ [\mathcal{H},\mathcal{Q}_2] = 0 $, therefore 
the supercharges $ \mathcal{Q}_1, \mathcal{Q}_2 $ are constants in time. Furthermore,
Eq.(\ref{H_Q12}) shows that eigenvalues of both Hamiltonians 
$ \mathcal{H}_{\pm} $ are non-negative, with zero being the lowest 
possible eigenvalue.

Due to the factorization property (\ref{factorization}), if $ \Psi_n^{-} $ is an eigenvector of $ \mathcal{H}_{-} $ with an eigenvalue $ E_n^{-} > 0 $ (where
$ n = 1, 2, \ldots $), than the state $ \Psi_n^{+} \equiv \left(E_n^{-} \right)^{-1/2} \mathcal{A}  \Psi_n^{-} $ will be an eigenstate of the SUSY partner Hamiltonian  $ \mathcal{H}_{+} $ with the same eigenvalue (energy) $ E_n^{-} $ (the factor $ \left(E_n^{-} \right)^{-1/2} $ is introduced here for a correct normalization.) This is seen from the following transformation
\beq
\label{SUSY_H}
\mathcal{H}_{+}  \Psi_n^{+} 
= \left(E_n^{-} \right)^{-1/2} \mathcal{A} \mathcal{A}^{+} \mathcal{A}   \Psi_n^{-}  
= \left(E_n^{-} \right)^{-1/2} \mathcal{A}  \mathcal{H}_{-} \Psi_n^{-} 
= \left(E_n^{-} \right)^{-1/2} \mathcal{A} E_n^{-} \Psi_n^{-} 
= E_n^{-} \Psi_n^{+}, \; \; n = 1, 2, \ldots,  \nonumber
\eeq
which means that all eigenstates of  spectra of $ H $, except 
possibly for a 'vacuum' state with energy $ E_0^{-} = 0 $, should be degenerate in energy with eigenstates of the SUSY partner Hamiltonian  $ \mathcal{H}_{+} $ \cite{Witten_SUSY}. Such zero-energy ground state would be unpaired, while all higher states would be doubly degenerate between the SUSY partner Hamiltonians 
$ \mathcal{H}_{\pm} $:
\bea
\label{SUSY_relations}
&&  \mathcal{H}_{-} \Psi_0^{-} = \mathcal{A}  \Psi_0^{-} = 0, \; \;   E_0^{-} = 0 \nonumber \\
&& \Psi_{n+1}^{-} = \left(E_n^{+} \right)^{-1/2}
\mathcal{A}^{+} \Psi_{n}^{+}, \; \; \; 
\Psi_{n}^{+} = \left(E_{n+1}^{-} \right)^{-1/2}
\mathcal{A} \Psi_{n+1}^{-}, \; \; \; n = 0, 1, \ldots \\
&& E_{n+1}^{-} = E_{n}^{+}, \; \; \; n = 0, 1, \ldots 
\nonumber 
\eea
For more details on supersymmetry in quantum mechanics, see e.g. \cite{Weinberg_book}, \cite{Junker}, \cite{Morano}.    

The existence or non-existence of a zero-energy ground state
$ E_0^{-} = 0 $ has to do with supersymmetry being unbroken or 
spontaneously broken.
In scenarios with spontaneous breaking of SUSY, supersymmetry is a symmetry of a Hamiltonian but not of a ground state of that Hamiltonian. On the other hand, an unbroken SUSY is characterized by the existence of a normalizable ground state $ \Psi_0 $ with strictly zero energy $ E_0 = 0 $, while for a spontaneously broken SUSY the energy of the ground state is larger than zero \cite{Witten_SUSY}:
\bea
\label{Unbroken_SUSY}
\mbox{Unbroken SUSY:}  &&  A  \Psi_0 = 0  \cdot \Psi_0  = 0 \; \; (E_0 = 0) \nonumber \\
\mbox{Spontaneously broken SUSY:}  &&  A  \Psi_0  = E_0   \Psi_0 , \; \; E_0 > 0. 
\eea
For SUSY to be unbroken, the derivative of the superpotential 
$ V'(y) = \partial V/ \partial y $ should have different signs at $ y = \pm \infty $, which means that that it should have an odd number of zeros at real values 
of $ y $. 
In our case, the superpotential is 
\beq
\label{Superpotential}
V(y)  = 
- \bar{\theta} y
+ \kappa e^y + \frac{1}{2} g e^{2y}. 
\eeq
For $ y \rightarrow \infty $, we have $ V'(y) \rightarrow \infty $,
while the behavior at $ y \rightarrow - \infty $ depends on the sign of $ \bar{\theta} $. If $ \bar{\theta} > 0 $,
we have $ \lim_{y \rightarrow - \infty}  V'(y) < 0 $, therefore 
SUSY is unbroken. 

On the other hand, if $ \bar{\theta} < 0 $,
SUSY would be spontaneously broken. Note that this is the same 
condition under which defaults through barrier penetration 
in the Langevin dynamics become possible. This is not coincidental,
and is due to the fact that the ground state 
of $ \mathcal{H}_{-} $ with zero energy
turns out to be non-normalizable due to a decay via tunneling.    


Indeed, due to the factorization property (\ref{factorization}), 
the formal zero-energy solution of the 
SE $ \mathcal{H}_{-} \Psi_0^{(-)} = E_0^{(-)} \Psi_0^{(-)} $ with 
zero energy $ E_0^{(-)} = 0 $ is given by a solution of a simpler
equation $ A \Psi_0  = 0 $. The solution of the latter equation is 
\beq
\label{formal}
 \Psi_0^{-}  = C  e^{ -  \frac{1}{\sigma^2} V(y)},
\eeq
where $ C $ is a normalization constant. Note that the square of this solution coincides with the equilibrium FPE density (\ref{p_s_y}). Therefore, it will be square-integrable for the same choice of parameters for which  the FPE density (\ref{p_s_y}) is integrable. In particular, as tunneling becomes possible
for $ \bar{\theta} < 0 $, the zero-energy ground state (\ref{formal}) ceases to be a normalizable state. This leads to a spontaneous breaking of SUSY.

We therefore obtain the following conditions for unbroken and broken SUSY in our setting:
\bea
\label{Unbroken_SUSY_QED}
\mbox{Unbroken SUSY:}  && \bar{\theta} > 0, \; \; \; E_0 = 0   \nonumber \\
\mbox{Spontaneously broken SUSY:}  && \bar{\theta} < 0, \; \; \;  E_0 > 0. 
\eea
 
For certain types of potentials, it is often the case 
that while the potential $ U_{-} $ is bistable, the potential $ V_{+} $ has a single minimum. In this case, SUSY can help to solve the problem of tunneling for 
potential $ U_{-} $ by a simpler problem of finding a ground state for the SUSY partner potential $ U_{+} $ \cite{Brown}.
As suggested by Fig.~\ref{fig:QM_SUSY_potentials}, our case is exactly of this sort.


 
%

\subsection{SUSY at work: the decay rate calculation}
\label{sect:SUSY_at_work}
As discussed above, defaults in the Langevin dynamics correspond to a noise-induced tunneling in the Langevin drift potential $ V(y) $, which becomes possible under the same constraint $ \bar{\theta} < 0 $. This leads to 
metastability of a candidate zero-energy ground state (\ref{formal}), which technically shows up as divergence of a normalization factor $ C $ for such a solution.  

In the quantum mechanical SUSY approach, this means that the zero-energy state 
of QM Hamiltonian $ \mathcal{H}_{-} $ does not exist, and the spectrum starts with the first eigenvalue (energy) $ E_1^{-} > 0 $.
However, supersymmetry relations (\ref{SUSY_relations}) still apply for higher states $ n > 0 $. They can be used to replace a hard 
problem of computing the first level $ E_1^{-} > 0 $ for 
$ \mathcal{H}_{-} $ by a simpler problem of computing the ground state energy $ E_0^{+} $ of the SUSY partner Hamiltonian 
$ \mathcal{H}_{-} $ \cite{Brown}. As indicated in 
Eq.(\ref{spectral_FPE}), the decay rate of a metastable state in the original problem is given by $ E_1^{-}/\sigma^2 $.

We therefore want to compute the lowest eigenvalue $ E_1^{-} > 0 $
of the QM Hamiltonian $ \mathcal{H}_{-} $ when parameters are 
such that $ \bar{\theta} < 0 $, SUSY is spontaneously broken, and a normalizable ground state with $ E_0^{-} = 0 $ does not exists.

The candidate ground-state solution of $ \mathcal{H}_{+} $ 
is easy to compute from the equation $ \mathcal{A}^{+} 
\Psi_0^{+} = 0 $, whose solution can be obtained by flipping the 
sign of $ V(y) $ in Eq.(\ref{formal}):
\beq
\Psi_0^{+}(y) \sim \frac{1}{\Psi_0^{-}(y)} \sim  \exp \left[   \frac{1}{\sigma^2} V(y) \right].
\eeq
However, this cannot be a right zero energy solution because it is not normalizable due to divergence at $ y \rightarrow \infty $.
Following \cite{Keung_1988, Gangopadhyaya_1993} ((see also \cite{Junker} for a similar derivation) consider the following ansatz:
\beq
\label{SUSY_ansatz}
\Psi(y) = 
\left\{ \begin{array}{cc}
   \frac{1}{\Psi_0^{-}(y)} \int_{y}^{\infty} dz 
\left[ \Psi_0^{-}(z) \right]^2 & \text{for} \; y > 0   \\
    \frac{1}{\Psi_0^{-}(-y)} \int_{-y}^{\infty} dz 
\left[ \Psi_0^{-}(z) \right]^2 & \text{for} \; y < 0  \\
\end{array} \right.
\eeq
Here $ \Psi_0^{-}(y) $ is assumed to be normalized, while  
$ \Psi(y) $ is not normalized.
One can easily check that $ \Psi(y) $ is continuous at $ y = 0 $
with $ \Psi(0) = \frac{1}{2 \Psi_0^{-}(0)} $, and that
$ \mathcal{H}_{+} \Psi(y) = 0 $ for $ y \neq 0 $.

However, $ \Psi(y) $ defined in Eq.(\ref{SUSY_ansatz}) is not an eigenstate of $ \mathcal{H}_{+} $ because its derivative has 
a discontinuity at $ y = 0 $:
\beq
\label{discontinuity}
\left. \lim_{\varepsilon \rightarrow 0} \frac{d \Psi(y)}{dy} \right|_{\varepsilon} 
- \left. \frac{d \Psi(y)}{dy} \right|_{ - \varepsilon} = 
\lim_{\varepsilon \rightarrow 0} 2 \varepsilon 
\left. \frac{d^2 \Psi(y)}{dy^2} \right|_{y=0} =
- 2 \Psi_0^{-}(0).
\eeq 
Instead, $ \Psi(y) $ can be seen as a ground state wave function
of the singular Hamiltonian $ \mathcal{H}_{s} $ given by
\beq 
\label{singular_H}
\mathcal{H}_{s} = \mathcal{H}_{+} - 2 \sigma^4 
 \left[ \Psi_0^{-}(0) \right]^2 \delta(y)
\equiv \mathcal{H}_{+} - \delta  \mathcal{H},
\eeq 
that differs from  $ \mathcal{H}_{+} $ by the $ \delta $-function
term $ \delta  \mathcal{H} $. Indeed, with this Hamiltonian we obtain 
\beq 
\label{Normalizaton_SUSY_H_p}
\int dy  \Psi(y) \left[\mathcal{H}_{s} - E_0^s \right]
 \Psi(y) = \int dy  \Psi(y)\mathcal{H}_{s}
 \Psi(y) = 0, 
 \eeq 
as a result of the fact that $ \mathcal{H}_{s} \Psi(y) = 0 $ for 
$ y \neq 0 $, while finite contributions coming from the discontinuity of the first derivative (\ref{discontinuity}) and 
the additional $ \delta $-function
term cancel each other:
\beq 
\label{cancellation}
- \frac{\sigma^4}{2}\int_{-\varepsilon}^{\varepsilon}
 \Psi(y)  \frac{d^2 \Psi(y)}{dy^2} dy - 
\int_{-\varepsilon}^{\varepsilon} \left[ \Psi(y) \right]^2 
\delta  \mathcal{H} dy  =
\frac{\sigma^4}{2} - \frac{\sigma^4}{2} = 0.
\eeq 
Equivalently, we can express $ \mathcal{H}_{+} $ in terms 
of $ \mathcal{H}_{s} $:
\beq 
\label{H_singular_H}
\mathcal{H}_{+} 
= \mathcal{H}_{s} + 
 2 \sigma^4 \left[ \Psi_0^{-}(y^{\star}) \right]^2 
\delta(y) = \mathcal{H}_{s} + \delta \mathcal{H}.
\eeq 
Therefore, the ground-state eigenvalue and the eigenfunction of 
 $ \mathcal{H}_{+} $ can be found using quantum mechanical perturbation theory (see e.g. \cite{Landau_QM}).
 To this end, we treat the additional term $ \delta \mathcal{H} $ as a perturbation  around 
the exactly solvable singular Hamiltonian $ \mathcal{H}_{s} $.
To the first order in perturbation $ \delta \mathcal{H} $, the 
change of energy $ E_0^{+} = E_1^{-} $ is 
\beq   
\label{E_0_pt}
E_0^{+} 
= 
\frac{\int_{-\infty}^{\infty} 
 \Psi^2(y) \delta \mathcal{H}  dy }{ 
\int_{-\infty}^{\infty} 
 \Psi^2(y) dy 
 } = 
 \frac{ 2 \sigma^4 \left[ \Psi(0) \right]^2 \left[ \Psi_0^{-}(0) \right]^2 }{ 
 \int_{-\infty}^{\infty} 
 \Psi^2(y) dy 
 } 
 \simeq \frac{\sigma^4}{2} 
 \frac{1}{\int_{-\infty}^{\infty} e^{- \frac{2 V}{\sigma^2}} dy  
 \int_{-\infty}^{\infty} e^{ \frac{2 V}{\sigma^2}}dy }. 
 \eeq
 Here in the last equality we assumed that the barrier is high, and therefore the ground state wave function is concentrated around the 
 minimum of $ V(y) $ where the exact form of $ V(y) $ is replaced by its approximation around this minimum. In other words, tunneling is neglected when we compute the energy splitting formula 
(\ref{E_0_pt}), similarly to how tunneling is computed in quantum mechanics (\cite{Landau_QM}, Sect. 50).
The constant $ C $ is then determined from the normalization condition (under the same approximation for $ V(y) $)
 \beq 
 \label{norm_cond_C}
 C^2 \int e^{ - \frac{2 V(y)}{\sigma^2}} dy = 1, \nonumber 
 \eeq
 while $ \Psi(y) $ can be approximated as follows:
 $ \Psi(y) \simeq 1/ \Psi_0^{-}(y) = 1/C \exp\left[ 
 \frac{V(y)}{\sigma^2} \right] $ 

While the main contribution to the first integral in the denominator
of Eq.(\ref{E_0_pt}) comes from the region around the minimum 
$ y_{\star} $ of $ V(y) $, the second integral is determined by a vicinity of the maximum $ y^{\star} $ of $ V(y) $\footnote{See 
Sect. 5.10.1 in \cite{FPE_Risken} for a different, FPE-based derivation of the the Kramers relation that leads to the same expression as in the last form in Eq.(\ref{E_0_pt}).}. Therefore,
we expand $ V(y) $ for the first and second integrals as follows:
\bea 
\label{second_order_expansions}
V(y) =
&& \hskip-0.5cm 
V(y_{\star}) + \frac{1}{2} V''(y_{\star}) ( y - y_{\star})^2 + \ldots \nonumber \\
V(y) =
&& \hskip-0.5cm 
V(y^{\star}) - \frac{1}{2} \left| V''(y^{\star}) \right| ( y - y^{\star})^2 + \ldots
\eea 
Using these approximations and performing the Gaussian integration, 
we finally reproduce the Kramers escape rate (\ref{Kramers_rate}) for 
$ r = E_0^{+}/ \sigma^2 $:
\beq 
\label{Kramers_escape_from_SUSY}
r = \frac{\sqrt{ V''(y_{\star}) \left| V''(y^{\star}) \right| }}{2 \pi} \exp \left[ - \frac{2}{\sigma^2} (V(y^{\star}) - V(y_{\star}) ) \right].
\eeq 
Therefore, we demonstrated that the classical Kramers escape rate relation can be obtained using methods of SUSY \cite{Brown, Keung_1988, Gangopadhyaya_1993, Junker}. While the SUSY-based approach may appear more involved than the FPE-based derivation given in Appendix A, its merit is that it can also be relatively easily extended to more difficult multivariate settings, unlike a 1D FPE-based derivation. The SUSY-based approach can also be used to compute corrections to the Kramers relation by including higher-order terms in perturbation theory with the singular potential (\ref{H_singular_H}).

\section*{Appendix E: Numerical Experiments}
The calibrated model parameters used for the results presented in Table \ref{tab:gbm_vs_qed} and Figure \ref{fig:axp_hazard} are listed in Tables \ref{table:theta} to \ref{table:g} below. The regularization parameter $\lambda_1=10$.
\begin{table}[!ht]

\resizebox{\columnwidth}{!}{%
\begin{tabular}{|c||cccccccc|}
\hline
&2010&2011&2012&2013&2014&2015&2016&2017\\
\hline
AXP&-1.6485&-1.5676&-1.0493&-0.3662&-1.0471&-0.598&-0.5643&-0.3563\\
BA&-0.849&-2.9781&-5.1422&-0.2435&-1.5068&-1.2719&-1.0095&-0.1468\\
CAT&-0.3343&-1.0226&-1.5175&-2.2737&-0.7972&-0.8616&-0.6487&-0.145\\
CSCO&-0.8932&-1.1666&-1.5277&-0.5791&-0.357&-0.58&-0.505&-0.3194\\
DIS&-0.7575&-0.5451&-0.2447&-0.2088&-0.2583&-0.2893&-0.6341&-0.7271\\
GS&-2.4782&-1.3811&-1.8919&-2.5805&-1.5015&-1.2657&-0.4338&-1.6921\\
HD&-0.8224&-1.0859&-0.3191&-0.5457&-0.3212&-0.4408&-0.7454&-0.2871\\
IBM&-1.0861&-0.5214&-1.0973&-0.8287&-0.8506&-0.8356&-0.7457&-0.9878\\
JNJ&-1.6686&-0.9216&-0.9047&-0.2556&-0.2501&-0.3198&-0.3485&-0.3611\\
JPM&-1.7856&-1.1131&-1.1225&-1.1016&-1.7978&-1.2017&-0.373&-0.5089\\
MCD&-0.5357&-0.3366&-0.7486&-0.5831&-0.5362&-1.1937&-1.1201&-0.2493\\
PFE&-1.1815&-1.0927&-0.9629&-0.971&-0.598&-0.4283&-0.479&-0.8217\\
PG&-1.8812&-1.6221&-1.2492&-0.9082&-0.5462&-0.3916&-0.5895&-0.6015\\
UNH&-2.0736&-0.7407&-1.5875&-0.4215&-0.2763&-0.5445&-0.3412&-0.1818\\
VZ&-1.0618&-1.91&-0.6247&-0.8863&-1.3047&-1.9708&-2.099&-1.9217\\
WMT&-1.6267&-1.4643&-0.5503&-0.9253&-0.3357&-0.3345&-1.1477&-0.2785\\
\hline
\end{tabular}
}
\caption{Calibrated $\theta $ without a signal.}
\label{table:theta}
\end{table}

\begin{table}[!ht]

\resizebox{\columnwidth}{!}{%
\begin{tabular}{|c||cccccccc|}
\hline
&2010&2011&2012&2013&2014&2015&2016&2017\\
\hline
AXP&0.0318&0.0343&0.0446&0.0634&0.014&0.0271&0.0315&0.0227\\
BA&0.1344&0.0705&0.0387&0.2705&0.0221&0.027&0.0476&0.1582\\
CAT&0.1196&0.0449&0.0347&0.0147&0.02&0.0311&0.0635&0.1079\\
CSCO&0.0274&0.0345&0.026&0.0285&0.0361&0.0165&0.0225&0.0263\\
DIS&0.0261&0.024&0.0374&0.0291&0.019&0.0154&0.0115&0.0128\\
GS&0.0391&0.1093&0.1144&0.0311&0.0302&0.0374&0.1406&0.0205\\
HD&0.0389&0.0263&0.0641&0.0224&0.0235&0.0135&0.0099&0.0246\\
IBM&0.0127&0.0296&0.0114&0.0138&0.0174&0.0229&0.0337&0.0129\\
JNJ&0.0093&0.0158&0.0138&0.0261&0.0129&0.0114&0.0119&0.0122\\
JPM&0.0258&0.0537&0.0677&0.0401&0.0146&0.029&0.0909&0.0399\\
MCD&0.0321&0.0354&0.0095&0.008&0.0104&0.0115&0.0102&0.0317\\
PFE&0.0188&0.0293&0.0297&0.0122&0.0106&0.0131&0.016&0.0108\\
PG&0.0087&0.0103&0.0165&0.0131&0.014&0.0111&0.0092&0.009\\
UNH&0.0419&0.08&0.0328&0.0494&0.049&0.0165&0.0346&0.038\\
VZ&0.0353&0.0162&0.044&0.0346&0.0168&0.0149&0.0132&0.0176\\
WMT&0.0104&0.0128&0.0293&0.0104&0.0126&0.0147&0.0121&0.0445\\
\hline
\end{tabular}
}
\caption{Calibrated $\sigma$ without a signal.}
\label{table:sigma}
\end{table}

\begin{table}[!ht]

\resizebox{\columnwidth}{!}{%
\begin{tabular}{|c||cccccccc|}
\hline
&2010&2011&2012&2013&2014&2015&2016&2017\\
\hline
AXP&-4.9464&-4.2287&-2.6757&-0.8325&-1.6036&-1.2054&-1.4339&-0.7496\\
BA&-3.9613&-10.3299&-15.9398&-1.0578&-2.7674&-2.3009&-2.0436&-0.3095\\
CAT&-1.2435&-2.0677&-2.7605&-4.587&-1.5402&-2.1883&-1.8305&-0.3762\\
CSCO&-1.7808&-3.1219&-4.0273&-1.4105&-0.8867&-1.1334&-0.986&-0.5363\\
DIS&-3.1716&-1.9204&-0.9338&-0.5624&-0.4799&-0.4253&-1.0195&-1.1143\\
GS&-4.6009&-3.1547&-5.8105&-5.5987&-3.16&-2.5091&-1.1926&-2.8655\\
HD&-4.1336&-4.537&-1.1913&-1.3327&-0.682&-0.7047&-1.0644&-0.4021\\
IBM&-2.3178&-1.0711&-1.8213&-1.4064&-1.7304&-2.0507&-2.0223&-2.5079\\
JNJ&-4.9281&-2.753&-2.5928&-0.653&-0.5053&-0.6166&-0.6092&-0.5533\\
JPM&-4.7618&-3.1582&-3.655&-2.7041&-3.6737&-2.4153&-0.8341&-0.754\\
MCD&-1.6211&-0.913&-1.5545&-1.2129&-1.1428&-2.5677&-2.0836&-0.4917\\
PFE&-3.2431&-2.8809&-2.2334&-1.8416&-1.1353&-0.8022&-0.9759&-1.5258\\
PG&-4.3961&-3.8013&-2.9515&-1.8452&-1.0844&-0.781&-1.1096&-1.1261\\
UNH&-11.0125&-3.473&-5.7253&-1.4552&-0.7599&-0.9805&-0.5603&-0.2335\\
VZ&-4.5126&-6.2924&-2.0807&-2.4225&-2.3045&-3.3418&-3.3013&-3.256\\
WMT&-3.7853&-3.5079&-1.2669&-1.8059&-0.6569&-0.6679&-2.4905&-0.6368\\
\hline
\end{tabular}
}
\caption{Calibrated $\kappa$ without a signal.}
\label{table:kappa}
\end{table}

\begin{table}[!ht]

\resizebox{\columnwidth}{!}{%
\begin{tabular}{|c||cccccccc|}
\hline
&2010&2011&2012&2013&2014&2015&2016&2017\\
\hline
AXP&3.7041&2.8461&1.6995&0.4664&0.6134&0.6051&0.9075&0.3931\\
BA&4.5167&8.9223&12.337&0.9479&1.2692&1.0384&1.0297&0.1441\\
CAT&1.11&1.0411&1.252&2.312&0.7425&1.3847&1.2797&0.2278\\
CSCO&0.8851&2.0819&2.6494&0.8562&0.5472&0.5526&0.4801&0.2242\\
DIS&3.3125&1.6859&0.8832&0.3762&0.2222&0.156&0.4095&0.4265\\
GS&2.1306&1.7729&4.4156&3.033&1.66&1.24&0.7853&1.212\\
HD&5.1744&4.73&1.0941&0.8118&0.3607&0.2811&0.3797&0.1402\\
IBM&1.2355&0.5482&0.7554&0.5961&0.8786&1.2554&1.3665&1.5904\\
JNJ&3.6373&2.054&1.8562&0.415&0.2547&0.2969&0.2659&0.2117\\
JPM&3.1696&2.2257&2.9556&1.6542&1.8757&1.2114&0.4552&0.2777\\
MCD&1.2216&0.6152&0.8064&0.6305&0.6084&1.3799&0.9683&0.2409\\
PFE&2.2225&1.8951&1.2921&0.8726&0.5384&0.375&0.4962&0.7079\\
PG&2.5674&2.2259&1.7419&0.9365&0.5376&0.3889&0.5218&0.5266\\
UNH&14.5879&4.0245&5.1527&1.2451&0.5163&0.4409&0.2286&0.0741\\
VZ&4.7781&5.1793&1.723&1.6505&1.0168&1.4157&1.2974&1.3782\\
WMT&2.201&2.0996&0.7268&0.8806&0.3209&0.3327&1.3503&0.3602\\
\hline
\end{tabular}
}
\caption{The calibrated $g$ without a signal.}
\label{table:g}
\end{table}


\clearpage
\begin{thebibliography}{99}
\bibitem{Amihud_2005} Y.~Amihud, H.~Mendelson, and L.H.~Pedersen, "Liquidity and Asset Prices", {\it Foundations and Trends in Finance}, 2005, vol. 1, no. 4, pp. 269-364. 
\bibitem{Bachelier} L.~Bachelier, "ThÃ©orie de la speculation", {\it Annales Scientifiques de L?Ãcole Normale SupÃ©rieure}, {\bf 17}, 21-86 (1900). (English translation by A. J.~Boness in P.H.~Cootner (Editor): {\it The Random Character of Stock Market Prices}, p. 17?75. Cambridge, MA: MIT Press (1964).
\bibitem{BE} H.~Bateman and A.~Erd{\'e}lyi, {\it Higher Transcendental Functions: Volume 2}, McGrau-Hill Book Company (1953).  
\bibitem{BS} F.~Black and M.~Scholes, "The Pricing of Options and Corporate Liabilities", Journal of Political Economy, 
Vol. 81(3),  637-654, 1973.
\bibitem{BC}  J.P.~Bouchaud and R.~Cont, "A Langevin Approach To Stock Market", {\it The European Physical Journal B}, {\bf 6}(4), 543-550 (1998). 
\bibitem{Bouchaud_book} J.P.~Bouchaud and M.~Potters, {\it Theory of Financial Risk and Derivative Pricing}, second edition, Cambridge University Press (2004). 
\bibitem{Brown} M.~Bernstein and L.S.~Brown, "Supersymmetry and the Bistable Fokker-Planck Equation", {\it Physical Review Letters}, {\bf 52} (22), 1933-1935 
(1984).
\bibitem{Boyd_2017} S.~Boyd, E, Busetti, S.~Diamond, R.N.~Kahn, K.~Koh, P. Nystrup, and J.~Speth, "Multi-Period Trading via Convex Optimization", {\it Foundations and Trends in Optimization}. Vol. XX, no. XX, 1-74 (2017).
\bibitem{Coleman_Callan} C.G.~Callan and S.~Coleman, "Fate of the False Vacuum.II. First Quantum Corrections", {\it Phys. Rev. D}, {\bf 16}(6), 1762-1768 (1977). 
\bibitem{Coleman_book} S.~Coleman, {\it Aspects of Symmetry. Selected Erice Lectures}, Cambridge University Press (1988).
\bibitem{Dash} J.~Dash,  {\it Quantitative Finance and Risk Management: a Physicist's Approach}, World Scientific, (2004).
\bibitem{Dixit} A.~Dixit and R.~Pindyck, {\it Investment Under Uncertainty}, Princeton University Press, Princeton NJ (1994).
 \bibitem{Duffie} D.~Duffie, "Black, Scholes and Merton - Their Central Contributions to Economics" (1997).
 \bibitem{Duffie_Singleton} D.~Duffie and K.J.~Singleton, {\it Credit Risk}, Princeton Series in Finance (2003).
 \bibitem{Dyson} F.J.~Dyson, "Divergence of Perturbation Theory in Quantum Electrodynamics", {\it Physical Review}, {\bf 85}, 631 (1952). 
\bibitem{EY} C.O.~Ewald and Z.~Yang, "Geometric Mean Reversion: Formulas for the Equilibrium Density and Analytic Moment Matching", {\it University of 
St. Andrews Economics Preprints} (2007). 
\bibitem{EK} H.~Ezawa and J.R.~Klauder, "Fermions without Fermions", {\it Progress of Theoretical Physics}, {\bf 74}(4), 904-915 (1985).
\bibitem{fama2012}
Eugene~F. Fama and Kenneth~R. French.
\newblock Size, value, and momentum in international stock returns.
\newblock {\em Journal of Financial Economics}, 105(3):457--472, 2012.
\bibitem{Feigelman_Tsvelik} M.V.~Feigelman and A.M.~Tsvelik, "Hidden Supersymmetry of Stochastic Dissipative Dynamics",  Sov. Phys. JEPT, {\bf 56} (4), 823-830 (1982).
\bibitem{Gangopadhyaya_1993} A.~Gangopadhyaya, P.~Panigrahi, and 
U.~Sukhatme, "Supersymmetry and Tunneling in an Asymmetric Double Well", {\it Physical Review A}, {\bf 47}(4), 2720-2724 (1993).
\bibitem{Gardiner} Gardiner, {\it Handbook of Stochastic Methods}  (1996). 
\bibitem{Gozzi_83} E.~Gozzi, "Functional-Integral Approach to Parisi-Wu Stochastic Quantization: Scalar Theory", {\it Phys. Rev. D}, {\bf 28}(8), 1922-1929 (1983). 
\bibitem{Feynman_QM} R.P.~Feynman and A.R.~Hibbs, {\it Quantum Mechanics and Path Integrals}, Dover (2010).
\bibitem{Gradshtein} I.S. Gradshtein and I.M. Ryzhik, {\it Table of Integrals, Series, and Products}, Fifth Edition, Academic Press (1994).   
\bibitem{Horsthemke} W.~Horsthemke and R.~Lefever, {\it Noise-Induced Transitions: Theory and Applications in Physics, Chemistry and Biology}, Springer (1984).
\bibitem{IHIF} I.~Halperin and I.~Feldshteyn, "Market Self-Learning of Signals, Impact and Optimal Trading: Invisible Hand Inference with Free Energy.
(or, How We Learned to Stop Worrying and Love Bounded Rationality)", https://papers.ssrn.com/sol3/papers.cfm?abstract\_id=3174498.
\bibitem{Hanggi_1986} P.~Hanggi, "Escape from a Metastable State", {\it Journal of Statistical Physics}, {\bf 42} (1/2) 105-148 (1986).

\bibitem{Heston93aclosed-form}
S.~L. Heston.
\newblock A closed-form solution for options with stochastic volatility with
  applications to bond and currency options.
\newblock {\em Review of Financial Studies}, 6:327--343, 1993.

\bibitem{Hinrichsen} H.~Hinrichsen, " Nonequilibrium Critical Phenomena and Phase Transitions into Absorbing States", {\it Advances in Physics}, {\bf 49}(7) (2000). 
\bibitem{Itzykson_Druffe} C.~Itzykson and J.M.~Druffe, {\it Statistical Field Theory. Volume I: From Brownian Motion to Renormalization and Lattice Gauge Theory},
Cambridge University Press (1989).
\bibitem{Junker} G.~Junker, {\it Supersymmetric Methods in Quantum and Statistical Physics}, Springer 1996.
\bibitem{Keung_1988} W.Y.~Keung, E.~Kovacs, and U.~Sukhatme, 
"Supersymmetry and Double Well Potentials", {\it Phys. Rev. Lett.}
, {\bf 60}, p. 41 (1988).
\bibitem{Landau_mechanics} L.D.~Landau and E.M.~Lifschitz, {\it Mechanics}, Elsevier (1980).
\bibitem{Landau_QM}  L.D.~Landau and E.M.~Lifschitz, {\it Quantum Mechanics}, Elsevier (1980).
\bibitem{Landau} L.D.~Landau and E.M.~Lifschitz, {\it Statistical Physics}, Elsevier (1980).
\bibitem{Langevin} P.~Langevin, "Sur la Th{\'e}orie du Mouvement Brownien", {\it Comps Rendus Acad. Sci.} (Paris) 146, 530-533 (1908).
\bibitem{LI} A.V.~Lopatin and L.B.~Ioffe, "Instantons in the Langevin Dynamics: an Application to Spin Glasses", {\it Phys. Rev. B} {\bf 60} (9), 6412 (1999).
\bibitem{Marchesoni} F.~Marchesoni, P.~Sodano, and M.~Zannetti, "Supersymmetry and Bistable Soft Potentials", {\it Physical Review Letters}, {\bf 61} (10), 1143-1146 (1988).
\bibitem{MW} J.~Mathews and R.L.~Walker, {\it Mathematical Methods of Physics}, Second Edition, Addison-Wesley Publishing (1969). 
1964.
\bibitem{Merton_74} R.~Merton, "On the Pricing of Corporate Debt: the Risk Structure of Interest Rates", {\it Journal of Finance}, {\bf 29}, 449-470 (1974).
\bibitem{Merton} R.~Merton, "Theory of Rational Option Pricing", Bell Journal of Economics and Management Science, 
Vol.4(1), 141-183, 1974.  
\bibitem{Merton_growth} R.~Merton, "An Asymptotic Theory of Growth Under Uncertainty", {\it Review of Economic Studies}, {\bf 42} (3), 375-393 (1975).
\bibitem{Milnikov_2008} G.~Mil'nikov and H.~Nakamura, "Tunneling Splitting and Decay of Metastable States in Polyatomic Molecules: Invariant Instanton Theory", {\it Phys. Chem. Chem. Phys.} {\bf 10}, 
1374-1393 (2008).
\bibitem{Milnikov_book} H.~Nakamura and G.~Mil'nikov, {\it Quantum Mechanical Tunneling in Chemical Physics}, CRC Press (2013).
\bibitem{Morano} M.~Morano, {\it Instantons and Large $ N $: An Introduction to Non-Perturbative Methods in QFT}, Cambridge University Press (2015). 
\bibitem{Munoz_1998} M.A.~Munoz, "Nature of Different Types of Absorbing States", {\it Physical Review E}, {\bf 57}(2), 1377 (1998).
\bibitem{FPE_Risken} H.~Risken, {\it The Fokker-Plank Equation}, 
Springer (1989).
\bibitem{Samuelson} P.~Samuelson, "Rational theory of warrant pricing", {\it Industrial Management Review}, {\bf 6} (Spring), 13-32 (1965).
\bibitem{CAPM} W.F.~Sharpe, "Capital asset prices: A theory of market equilibrium under conditions of risk", {\it Journal of Finance}, {\bf 19} (3), 425?442 (1964).
\bibitem{Schmittmann} B.~Schmittmann and R.K.P.~Zia, {\it Statistical Mechanics of Driven Diffusiive Systems: Vol 17: Phase Transitions and Critical Phenomena}, 
Ed. C.~Domb and J.L.~Lebowitz, Academic Press (1995). 
\bibitem{Sornette} D.~Sornette, "Stock Market Speculations: Spontaneous Symmetry Breaking of Economic Valuation", {\it Physica A}, {\bf 284} (1-4), 355-375 (2000).
\bibitem{Sornette_book} D.~Sornette, {\it Why Stock Markets Crash}, Princeton University Press (2003). 
\bibitem{Van_Broeck_1997} C.~Van den Broeck, J.M.R.~Parrondo, R.~Toral, and R.~Kawai, "Nonequilibrium Phase Transitions Induced by Multiplicative Noise",
{\it Phys. Rev. E}, {\bf 55} (4), 4084-4094 (1997). 
\bibitem{vanKampen} N.G.~Van Kampen, {\it Stochastic Processes in Physics and Chemistry}, North-Holland (1981).
\bibitem{Weinberg_book} E.~Weinberg, {\it Classical Solutions in Quantum Field Theory: Solitons and Instantons in High Energy Physics}, Cambridge University Press  (2012).
\bibitem{Witten_SUSY} E.~Witten, "Dynamical Breaking of Supersymmetry", {\it Nuclear Physics B} {\bf 188}(3-5), 513-554 (1981).
\bibitem{Zinn-Justin-QFT} C.~Zinn-Justin, {\it Quantum Field Theory and Critical Phenomena}, Fourth Edition, Clarendon Press, Oxford (2002).


\end{thebibliography}
\end{document}